\documentclass[12pt,a4paper,twoside]{report}

\usepackage{type1cm}

\usepackage{amsmath}
\usepackage{amsfonts}
\usepackage{color}
\usepackage{hyperref}

\usepackage{microtype}

\usepackage{graphics}
\usepackage[section]{placeins}
\usepackage{subcaption}

\DeclareCaptionSubType*{figure}

\usepackage{emptypage}

\let\originaleqref\eqref
\renewcommand{\eqref}{eq.~\originaleqref}

\usepackage{listings}

\usepackage{slashed}
\usepackage{xfrac} 
\usepackage{xspace}
\usepackage{empheq}

\usepackage{geometry}
\usepackage{pdflscape}

\newcommand{\Axiloop}{\texttt{Axiloop}\xspace}
\newcommand{\C}{\mathcal{C}}

\newcommand{\Ca}{C_A}
\newcommand{\Cf}{C_F}
\newcommand{\D}[0]{\mathrm{d}}
\newcommand{\Dm}[0]{\mathrm{d}^m}

\newcommand{\FeynCalc}[0]{\texttt{FeynCalc}}

\newcommand{\Gh}[0]{\hat{\Gamma}}

\newcommand{\HERWIG}[0]{\texttt{HERWIG++}\xspace}
\newcommand{\KRKMC}[0]{\texttt{KRKMC}\xspace}
\newcommand{\Li}[0]{\mathrm{Li_2}}

\newcommand{\MCNLO}[0]{\texttt{MC@NLO}\xspace}

\newcommand{\NPV}[0]{{\mathrm{NPV}}}

\renewcommand{\P}[0]{\mathbb{P}}
\newcommand{\POWHEG}[0]{\texttt{POWHEG}\xspace}
\newcommand{\PS}{\Phi}
\newcommand{\PV}[0]{{\mathrm{PV}}}

\newcommand{\Ph}[0]{\hat{P}}

\newcommand{\Pgg}{P_{gg}}

\newcommand{\Pqq}[0]{P_{qq}}
\newcommand{\Pythia}[0]{\texttt{Pythia}\xspace}
\newcommand{\SHERPA}[0]{\texttt{SHERPA}\xspace}
\newcommand{\Tracer}[0]{\texttt{Tracer}\xspace}

\newcommand{\Qr}[0]{Q_\eps^\mathrm{r}}
\newcommand{\Qv}[0]{Q_\eps^\mathrm{v}}

\newcommand{\Tf}{T_F}

\newcommand{\UnitTest}[0]{\texttt{UnitTest}\xspace}
\newcommand{\Wb}[0]{W_B}

\newcommand{\Wirq}{W_{\mathrm{ir}}^k}

\newcommand{\Wirrq}{W_{\mathrm{ir}^2}^k}

\newcommand{\Wr}[0]{W_R}
\newcommand{\Wu}{W_{\mathrm{N}}}

\newcommand{\Wuvk}{W_{\mathrm{uv}}^q}
\newcommand{\Wuvp}{W_{\mathrm{uv}}^p}
\newcommand{\Wuvq}{W_{\mathrm{uv}}^k}
\newcommand{\Wz}[0]{W_Z}

\newcommand{\Wzq}[0]{W_0^k}

\newcommand{\abs}[1]{\lvert #1 \rvert}
\newcommand{\as}{\alpha_\mathrm{s}}

\newcommand{\aspi}{\left(\frac{\as}{2\pi}\right)}

\newcommand{\code}[1]{\colorbox{lstgray1}{\lstinline|#1|}}

\newcommand{\del}[0]{\delta}
\newcommand{\dpp}[0]{\delta_{+}}
\newcommand{\eps}[0]{\epsilon}
\newcommand{\eir}[0]{\epsilon_\mathrm{ir}}
\newcommand{\euv}[0]{\epsilon_\mathrm{uv}}

\newcommand{\ggd}{gg-d}
\newcommand{\gs}{g_\mathrm{s}}

\newcommand{\kn}{\sdot{k}{n}}
\newcommand{\knp}{k_{+}}
\newcommand{\lnp}{l_{+}}
\newcommand{\mf}{\mu_{f}}
\newcommand{\mr}{\mu_{r}}

\newcommand{\nl}{\sdot{l}{n}}
\newcommand{\pgg}[0]{p_{gg}}
\newcommand{\pn}{\sdot{p}{n}}
\newcommand{\pnp}{p_{+}}
\newcommand{\pqq}[0]{p_{qq}}
\newcommand{\qn}{\sdot{q}{n}}

\newcommand{\rnp}{r_{+}}
\newcommand{\sdot}[2]{\,#1\!\cdot\!#2}
\renewcommand{\sl}[1]{\slashed #1}

\newcommand{\vect}[1]{\bar{#1}}
\newcommand{\veps}{\varepsilon}

\newcommand{\cqq}{$\mathrm{c}$}
\newcommand{\dgg}{$\mathrm{d}_{gg}$}
\newcommand{\dqq}{$\mathrm{d}_{qq}$}
\newcommand{\eqq}{$\mathrm{e}$}
\newcommand{\fqq}{$\mathrm{f}$}
\newcommand{\gqq}{$\mathrm{g}$}

\newcommand{\subeq}[1]{ \tag{\thechapter.\arabic{equation}#1} }

\usepackage[
  backend=bibtex,
  doi=false,
  firstinits=true,
  maxnames=3,
  style=alphabetic
]{biblatex}
\bibliography{bibliography.bib}
\renewbibmacro{in:}{}

\renewcommand{\em}{\it}

\usepackage[utf8]{inputenc}
\usepackage[OT4]{fontenc}
\usepackage{textcomp}

\begin{document}
  \lstset{                          %
    aboveskip=4\medskipamount,
    basicstyle=\small\ttfamily,     
    belowskip=4\medskipamount,
    captionpos=b,                   
    frame=lines,
    framesep=6pt,
    language=Mathematica,
    rulesep=4pt,
  }

  \begin{titlepage}
    \hfill IFJPAN-IV-2014-4

    \centering
    \vspace*{0.5cm}
        
    \begin{huge}\bfseries
      Higher-Order Corrections in\\[0.25\baselineskip]
      QCD Evolution Equations and\\[0.25\baselineskip]
      Tools for Their Calculation
    \end{huge}

    \vspace{1.5cm}
        
    A thesis submitted to\\[0.25\baselineskip]
    \begin{large}
      The Henryk Niewodnicza\'nski\\[0.15\baselineskip]
      \uppercase{Institute of Nuclear Physics}\\[0.15\baselineskip]
      \uppercase{Polish Academy of Sciences}\\
    \end{large}
    \vspace{0.8cm}
    \includegraphics[width=0.15\textwidth]{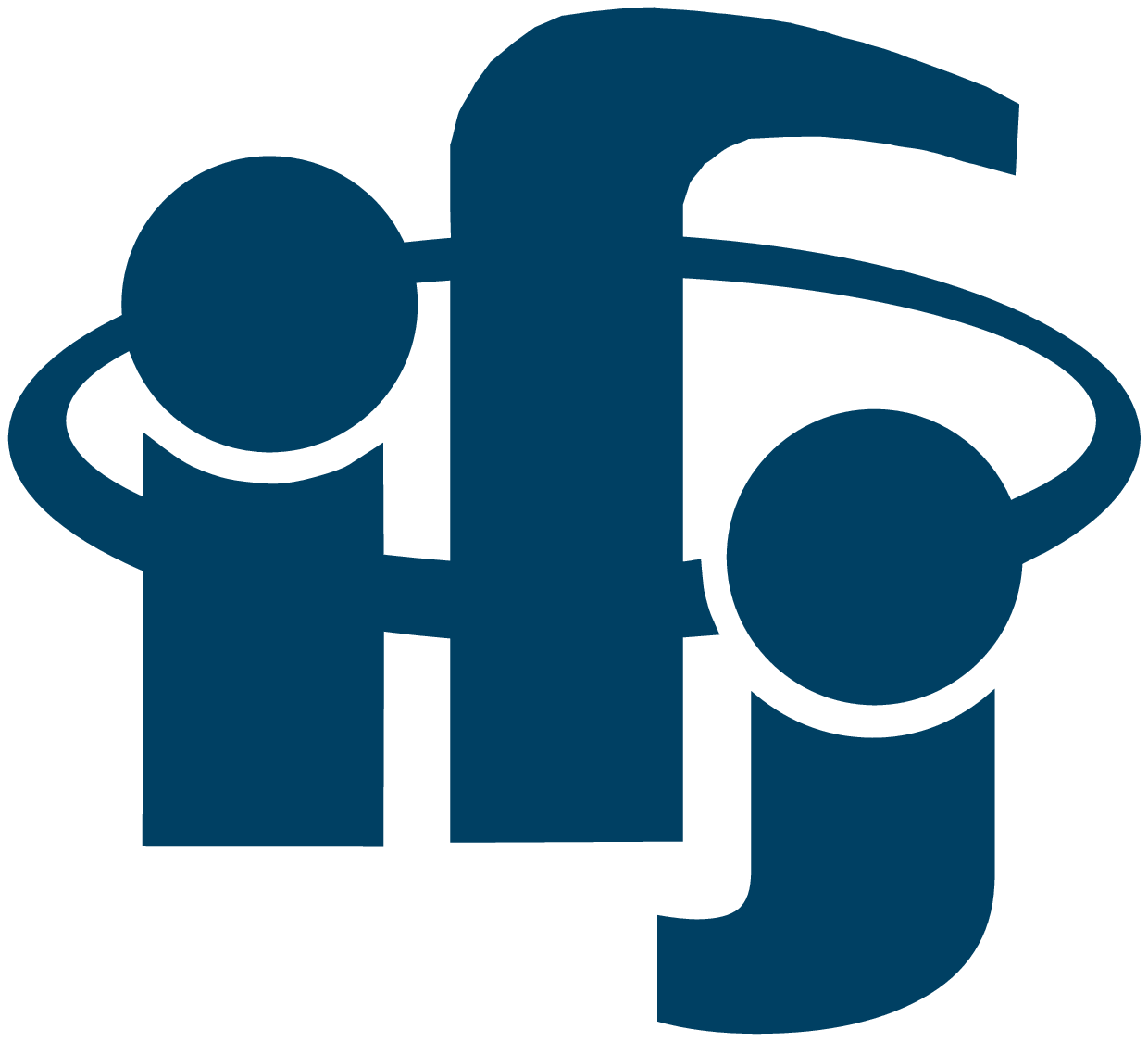}

    \vspace{1.2cm}

    for the degree of\\[0.25\baselineskip]
    Doctor of Philosophy

    \vspace{1.0cm}
        
    presented by\\[0.25\baselineskip]
    \textbf{\large Oleksandr Gituliar}

    \vspace{1.0cm}
        
    written under supervision of\\[0.35\baselineskip]
    {\large Prof. Maciej Skrzypek}\\[0.25\baselineskip]
    {\large Dr. Aleksander Kusina}\\

    \vfill

    Cracow, Poland\\
    2014
  \end{titlepage}
  \cleardoublepage
  \setcounter{page}{1}

  \chapter*{Abstract}
In this thesis we calculate the NLO one-loop virtual contributions to the
QCD DGLAP splitting functions in a form suitable for Monte Carlo simulations.
We use the standard technique based on the factorization properties of mass
singularities in the light-cone axial gauge \cite{EGMPR79,CFP80} but
we propose a modification to this approach by introducing a New Principal Value
prescription \cite{GJKS14} in which we use the PV prescription for
regularization of {\em all} singularities in the light-cone plus components of
the four-momenta.
The main advantage of the NPV prescription is that exclusive splitting functions
calculated with its help can be used for construction of the Monte-Carlo parton
showers. The reason for this is that in the NPV prescription some of the
higher order poles in dimensional $\epsilon$ parameter
are replaced by the logarithms of a cut-off parameter $\delta$ which has a
geometrical meaning in four dimensions. As a consequence, cancellation of the
higher order poles between real and virtual components is reduced. On the other
hand, at the inclusive level the NPV results agree with the results in the
standard PV prescription which shows compatibility of both approaches.

With the help of the NPV prescription, we calculate virtual one-loop
contributions to the NLO non-singlet splitting function $P_{q\to q}^{NLO}$ and
selected contributions to the singlet $P_{g\to g}^{NLO}$ one. We also discuss
the dependence of the results on the choice of the integration variable related
to the evolution time in Monte Carlo parton showers.
Finally, we present the {\tt Axiloop} package written in Wolfram Mathematica
language, that is dedicated to perform NLO calculations in the axial gauge. Results
presented in this thesis were obtained with the help of the {\tt Axiloop} package. 

The possible continuations of this work include
calculation of the remaining contributions to the singlet splitting function
$P_{g\to g}^{NLO}$ and direct calculation of the two-loop virtual
contributions, which are usually deduced indirectly from the sum rules.

\chapter*{Streszczenie}
W pracy obliczone zostały jednopętlowe wkłady wirtualne do jąder ewolucji DGLAP w QCD
w przybliżeniu NLO w formie pozwalającej na ich zastosowanie w symulacjach typu Monte Carlo.
W tym celu użyta została technika rachunkowa oparta o faktoryzację kolinearną
osobliwości masowych w cechowaniu aksjalnym \cite{EGMPR79,CFP80}. W technice tej
dokonaliśmy modyfikacji wprowadzając regularyzację typu PV dla wszystkich
osobliwości w składowej ``plus'' stożka świetlnego czteropędów.
Główną zaletą nowej regularyzacji, nazwanej NPV, jest to, że otrzymane 
ekskluzywne jądra ewolucji mają formę dogodną do konstrukcji
stochastycznych kaskad partonowych. Jest to spowodowane tym, że w schemacie NPV \cite{GJKS14}
część biegunów wyższego rzędu w parametrze $\epsilon$ (pochodzącym z regularyzacji wymiarowej)
zostaje zastąpionych
przez logarytmy obcięcia $\delta$, które ma dogodną interpretację geometryczną
w czterech wymiarach. W konsekwencji, ograniczone zostaje niekorzystne kasowanie
się biegunów wyższego rzędu pomiędzy przyczynkami realnymi i wirtualnymi.
Z drugiej strony jednak, na poziomie inkluzywnym wyniki w schemacie NPV
reprodukują wyniki w standardowym podejściu PV, co pokazuje kompatybilność obu
metod.

Używając schematu NPV obliczyliśmy jednopętlowe wkłady do niesingletowego
jądra $P_{q\to q}^{NLO}$ oraz część wkładów do singletowego jądra $P_{g\to g}^{NLO}$.
Przedyskutowaliśmy również zależność jąder ewolucji od wyboru
zmiennej całkowania, która jest związana z czasem ewolucji w kaskadach
partonowych. Wreszcie, przedstawiliśmy pakiet {\tt Axiloop}, który został przez
nas napisany w jezyku Wolfram Mathematica w celu prowadzenia obliczeń w
cechowaniu aksjalnym. Prezentowane w pracy rezultaty zostały otrzymane przy
użyciu tego pakietu.

Możliwe dalsze zastosowania formalizmu NPV oraz pakietu {\tt Axiloop} obejmują
w szczególności obliczenie pozostałych przyczynków do jądra $P_{g\to g}^{NLO}$
lub bezpośrednie obliczenie przyczynków dwupętlowych, czysto wirtualnych, które
normalnie są uzyskiwane nie wprost wykorzystując reguły sum.

  \tableofcontents

  \chapter{Introduction}


There are four fundamental forces known up to now responsible for the existence of our universe: gravitation, electromagnetic, weak, and strong.
Gravitation is most accurately described by the General Theory of Relativity published by A.Einstein in 1916.
Remaining three forces are described by the Standard Model of particle physics developed through the second half of 20th century.
Standard Model is a quantum theory and unifies Quantum Electrodynamics (electromagnetism), theory of Salam, Glashow, and Weinberg (weak interactions), and Quantum Chromodynamics (strong nuclear interactions).

This thesis is dedicated to the strong forces and theory of quantum chromodynamics.
Strong forces are responsible for the stability of ordinary matter and confining quarks and gluons into hadrons (such as protons and neutrons).
The unique property of quarks and gluons is a color charge, described with a non-Abelian symmetry group SU(3), and a phenomenon associated with it -- {\it color confinement}, stating that color charged particles can not be isolated and particles occur only as colorless combinations.
Another key property of QCD is {\it asymptotic freedom}, which means that in high-energy limit strong interaction between quarks and gluons diminish or, in other words, that the effective coupling constant of QCD goes to zero at small distance.
Asymptotic freedom was first discovered in the early 1970s by Politzer, Gross, and Wilczek \cite{Pol74,GW73}.
This feature allows perturbative calculation approach to be applied in QCD.

\subsubsection{Experimental Tests of QCD}

Thanks to the operation of the two biggest hadron colliders in the world (but not only), predictions made by QCD can be tested with a great precision.
First of them is Tevatron at Fermilab that was running at center of mass energies up to 2 TeV.
Its main achievement was the discovery of the top quark --- the last fundamental fermion predicted by the Standard Model.
The second machine is the highest-energy particle collider ever made --- Large Hadron Collider in Geneva.
During the first phase of operation in 2009--2013 LHC experiments reached the integrated luminosity of nearly 30 fb$^{-1}$.
Such precise measurements allowed to "see" in 2013 the last missing piece of the Standard Model --- the Higgs boson particle in the mass region about 126 GeV.
In the same year The Nobel Prize in Physics was awarded jointly to Francois Englert and Peter W. Higgs "for the theoretical discovery of a mechanism that contributes to our understanding of the origin of mass of subatomic particles, and which recently was confirmed through the discovery of the predicted fundamental particle, by the ATLAS and CMS experiments at CERN's Large Hadron Collider".

\subsubsection{Perturbative QCD and Factorization}

The non-perturbative effects in QCD were first described by the {\it naive parton model} \cite{BP69,Fey69} with partons associated with quarks and gluons.
It describes a low energy (long-distance) structure of hadrons by means of {\it parton distribution functions} or PDFs.
A more recent approach to better understand hadron structure by representing the hadron distributions as functions of more variables, such as the transverse momentum and spin of the parton, is {\it generalized parton model} \cite{BR05}.

Perturbative calculations in QCD lead not only to the ultra-violet, but also to the mass singularities, which appear due to the (nearly) massless origin of quarks and gluons.
This complication makes application of the perturbative theory to QCD non-trivial.
The solution to this problem is the concept of factorization.
It allows one to split a bare (singular) matrix element into two pieces:
1) a short-distance high-energy hard process which is finite and can be calculated after factorization; and
2) long-distance low-energy effects which contain all the singularities and are responsible for the evolution effects.

The most commonly used factorization scheme is the collinear factorization and DGLAP equations \cite{DGLAP} which describes evolution of parton distribution functions (and fragmentation functions) with the change of factorization scale, which in practical applications is identified with hard process scale $Q$.
Another approach to factorization is the BFKL equation \cite{BFKL} that describes evolution in $x$-variable of the unintegrated gluon densities in a small-$x$ region and its non-linear modification --- the BK equation \cite{Bal96,Kov99}.
Finally, the approach which describes evolution in both those variables is CCFM \cite{CCFM}.

The collinear factorization was proposed by Altarelli and Parisi \cite{AP77}.
They worked out corrections to parton densities, which are closer to physical intuition and obey evolution equations known as the Altarelli-Parisi equations or DGLAP equations, since they were also independently considered by V.N.~Gribov, L.N.~Lipatov \cite{GL72,Lip74}, and Y.L.~Dokshitzer \cite{Dok77}.
A formulation of the factorization properties to all orders in perturbative QCD, was first proved in the context of {\it collinear factorization theorem} in the axial-type gauges in ref. \cite{EGMPR79}.
It separates mass singularities in quark and gluon from short-distance inclusive cross section into long-distance parton distribution and decay functions.
This leads to modification of the naive parton model by including perturbative effects and makes perturbative calculations in QCD safe of mass singularities.

It is also important to mention the refinements to the standard factorization introduced in \cite{CSS85,Bod85}, as well as other factorization theorems e.g. allowing to include effects of heavy quark masses. Such an approach has been formulated in \cite{AOT94,ACOT94} and later proven by Collins in \cite{Col98}.
Note that in this approach even though all the heavy quark masses are retained the parton distributions fulfill the same DGLAP evolution equations as in the standard collinear factorization.
For a comprehensive review of factorization approaches see \cite{Col11}.

There are two different methods, which allow one to calculate the evolution of parton densities at next-to-leading order (NLO).
The first method is based on {\it operator product expansion} (OPE) \cite{OPE} with calculations performed in the Feynman gauge and results obtained in the Mellin moment space.
The second method, based explicitly on the factorization properties of mass singularities \cite{EGMPR79} was developed by Curci, Furmanski, and Petronzio \cite{CFP80,FP80}.
What distinguish this technique from the OPE method is the use of a light-cone axial gauge ($n_\mu A^\mu=0$, $n^2=0$) \cite{Lei87,BNS91,Lei94}.
Despite that such an approach increases technical complexity, it works directly in the momentum space and leads to the physical interpretation which is very close to the intuitive parton picture.


The main complication that light-cone gauge introduces is $1/\sdot{l}{n}$ factor in the gluon propagator, which gives rise to "spurious poles", gauge-specific singular terms.
Although these singularities have to cancel in gauge-invariant quantities, one has to apply some regularization prescription in order to be able to evaluate individual diagrams.
The choice done in \cite{CFP80} was the {\it principal value} (PV) prescription.
Later on some theoretical inconsistencies with the canonical quantization procedure in the light-cone gauge were pointed out \cite{BDLS85,MR94}.
Formally performed canonical quantization leads to the {\it Mandelstam--Leibbrandt} (ML) prescription \cite{ML}.
In-depth study of NLO calculations in this scheme has been done by Heinrich \cite{Hei98,HK98}.
The ML prescription, although better justified, leads to calculations significantly more complicated than the PV one.
The NLO results have been obtained in ML scheme in \cite{Hei98,BHKV98}.

Finally, it is worth to mention the most recent state-of-the-art calculations of the next-to-next-to-leading order (NNLO) corrections to the splitting functions \cite{MVV04a,MVV04b,MMV06,MV08,AMV12}.

In this work we propose a modified version of the PV prescription -- a {\it new principal value} (NPV) prescription \cite{GJKS14}.
This new approach leads to a similar treatment of some singularities in Feynman and axial-type integrals which in turn results in simplification of the structure of singularities in both real and virtual NLO splitting functions on graph-by-graph basis.
Moreover, exclusive splitting functions calculated in the NPV prescription are better suited for Monte-Carlo parton shower simulations \cite{JKSS11,JKPSS13}.

\subsubsection{Parton shower}

Up to this point our discussion was related to the splitting functions at the inclusive level, i.e. with momenta of the final-state particles integrated out.
Such splitting functions find their application in solutions of the DGLAP evolution equations for parton distribution and fragmentation functions used in calculations of numerous inclusive experimental observables.
Another approach to perturbative calculations in QCD is based on stochastic technique known as {\it parton shower}.
Parton shower solutions are implemented in QCD Monte-Carlo programs, often called {\em event generators}.
This approach to solving QCD is more powerful as it allows to compare directly with measured results involving complicated experimental acceptances.

In the structure of parton shower we can distinguish two components: perturbative and non-perturbative.
The first one describes the "core" of the process, i.e. the hard scattering and the shower of emitted partons.
This part is rigorously based on the quantum field theory and can be viewed as a stochastic way of solving the QCD Lagrangian.
This solution combines hard matrix element and shower of initial- and final-state partons.
Hard matrix element is calculated perturbatively to a fixed order.
On the contrary, the shower must re-sum dominant contributions to the infinite order.
Thus, resummation is based on the mentioned above evolution equations.
However, the fundamental problem is that Monte-Carlo requires fully exclusive information on the emitted partons, whereas the evolution equations provide only inclusive (i.e. partly integrated) description.
To date this has limited the simulation accuracy of the parton shower to the improved leading order only.

The second, non-perturbative, part involves modeling of the long-distance phenomenon, like hadronization, which currently can not be calculated from the first principles.

General-purpose Monte-Carlo event generators like \Pythia \cite{MSS06,MSS08}, \HERWIG \cite{Herwig08}, and \SHERPA \cite{Sherpa09}, provide fully exclusive simulation of hadronic high-energy collisions.
They play an essential role in QCD modeling, in data analysis, where they are used together with detector simulation to provide a realistic estimate of the detector response to collision events, and in the planning of new experiments, where they are used to estimate signals and backgrounds in high-energy processes.
Event generators are built from several components and simulate the following subprocesses: structure of initial-state hadrons, {\em initial-state radiation}, hard process, multi-particle interactions, resonance decay, {\em final-state radiation}, hadronization, and others depending on the model under consideration.

There are also specialized event generators aimed to improve precision of the perturbative QCD calculations by means of the stochastic methods.
In particular, \MCNLO \cite{FW02} event generator implements NLO corrections in hard processes on top of the LO parton shower.
Similar approach is achieved with \POWHEG \cite{Nas04,FNO07} program.
Additionally the more recent versions of these two programs:
{\tt aMC@NLO} \cite{FFHMP11a,FFHMP11b} and {\tt POWHEG BOX} \cite{ANOR10}
allow for more automatized inclusion of new processes in the same framework.
Other recent and ongoing developments include:
\texttt{GR@PPA} project \cite{FIKK03,OK12} which includes some NLO effects in the cascade;
\SHERPA project \cite{Sherpa09} allowing to use different LO and NLO matching and merging techniques
including \texttt{MEPS@NLO} \cite{HKSS12,GHKSS13}, a technique of combining next-to-leading order
parton-level calculations of varying jet multiplicity and parton showers;
the \texttt{GENEVA} project \cite{BTT08a,BTT08b,ABBHT12} combining higher-order resummation of
large Sudakov logarithms with NLO matrix-element corrections and parton showers;
and the \texttt{MINLO} prescription \cite{HNZ12,HNOZ12} for assigning scales in NLO computations.
Effort going in slightly different direction is pursued in \texttt{DEDUCTOR} shower \cite{NS12,NS14},
where effects due to spin, color and heavy quark masses are being included in the LO shower.

\subsubsection{The \KRKMC project}

Let us summarize the current situation in the QCD Monte-Carlo parton shower programs: the "standard" generators, developed in mid 80s, are constructed on the basis of improved LO shower of partons and LO matrix elements.
The specialized solutions developed in early 2000s, upgraded the precision of matrix element to NLO, retaining the LO shower.
The LO approximation in the shower is one of the factors limiting precision of the current MC generators.
It is a big deficiency compared to the fixed order calculations, and it can turn out to be insufficient for the forthcoming precision measurements at LHC at 14 TeV, preventing us from taking full advantage of the LHC data.
In this context, a different, novel approach to the QCD parton showers has been proposed within the \KRKMC project \cite{GJKPSS11,JJKPS12,JKPSS13}.
The goal of this project is to include NLO corrections in the exclusive form in both the partonic cascade and in the hard matrix element.
It uses the formalism of collinear factorization described earlier, but modifies it in such a way that the evolution becomes fully exclusive.
It requires in particular recalculation of the splitting functions.
This has already been partly done.
Namely, the real part of the exclusive splitting functions has been calculated in a way suitable for NLO event generators in \cite{JKSS11}.
This contribution turned out to be different from the one known from the literature \cite{Hei98} and the difference was related to the treatment of the soft singularities.

\subsubsection{The goal and results of the thesis}

To complete the calculation of splitting functions for the \KRKMC project and to resolve the above issue of soft singularities one needs to compute the virtual contributions in a consistent way, and define the new regularization scheme for the soft singularities,
and this is the goal of the presented thesis.

For this purpose we have developed a new regularization scheme --- New Principal Value (NPV) prescription \cite{GJKS14,GJKS13,GS13} --- which extends the standard PV prescription introduced in \cite{CFP80}.
With the help of the NPV prescription we have calculated the complete NLO $\Pqq$ splitting function as well as some contributions to the $\Pgg$ splitting function.
This was sufficient to show that NPV correctly reproduces the fully inclusive standard PV results \cite{CFP80,EV96} and that it is consistent with the real contributions of \cite{JKSS11}.
With an eye to the MC application, we give an exhaustive list of exclusive and inclusive results for all contributing graphs and their sums.
We also discuss different choices of the integration variable, related to the issue of evolution time variable defining the ordering used in MC algorithms (we have not found such a discussion in the literature).

%


\subsubsection{Outline}

The organization of the remaining part of this thesis is following:

In Chapter \ref{ch:2} we introduce the factorization theorem and a generalized ladder expansion
approach originally developed in \cite{EGMPR79,CFP80}.
A basic formalism for calculating parton densities is introduced followed by the overview
of DGLAP evolution equations and some of their properties.

In Chapter \ref{ch:calc} a practical framework for calculating NLO corrections to the one-loop splitting functions is described.
We provide notation and definitions for the main quantities to be calculated in the following chapters, like exclusive and inclusive parton densities, ultra-violet counter-terms, etc.
We also overview standard approaches to regularizing singularities inside loop integrals (PV and dimensional regularizations) and introduce alternative NPV regularization prescription.

In Chapter \ref{ch:4} complete results for the non-singlet NLO splitting functions in the
NPV regularization scheme are listed and compared with the original results
\cite{CFP80,Hei98}.

In Chapter \ref{ch:5} we describe \Axiloop package written in Wolfram Mathematica language and dedicated to the calculations discussed in this thesis.

In Chapter \ref{ch:6} we summarize our results and provide possible directions they can be extended.

In Appendix \ref{ch:feynman-rules} we list Feynman rules for the light-cone gauge, and
in Appendix \ref{ch:integrals} we provide a complete list of one-loop integrals used for the calculations in the NPV prescription.

  \chapter{Factorization and Evolution Equations in QCD}
\label{ch:2}

QCD describes interaction of quarks and gluons (partons) which carry color charge.
However, the observable particles -- hadrons are bound states of quarks and gluons, and because of confinement we can not observe free partons.
Additionally because of asymptotic freedom formation of these colorless bound states is described by non-perturbative physics which we have limited knowledge about.
Factorization enables us to separate the high energy (short distance) phenomena that are described perturbatively form the low energy (long scale) phenomena that are non-perturbative giving us an indispensable tool for performing calculations within QCD. 

Factorization states that the cross section (or other observable quantity) can be expressed as a convolution of a short distance coefficient function, describing matrix element for hard scattering of partons (that can be calculated perturbatively using Feynman diagrams), and the long distance contribution describing the structure of the initial hadron, namely the parton distribution function or PDF.
In the case of the DIS process it can be written as:
\begin{equation}
\label{eq:fact-common}
  \sigma = \sum_{i} \int \frac{dx}{x}\; f_i(x,\mu) \, \hat{\sigma}_i(x,\alpha_S(\mu),Q^2/\mu^2)
         + O(\Lambda_{\text{QCD}}/Q)
\end{equation}
where $f$ is a PDF and $\hat{\sigma}$ is the coefficient function.

In the above formula we can see that $f$ and $\hat{\sigma}$ depend on an additional energy scale $\mu$, which is referred to as the factorization scale.
This dependence reflects the fact that the procedure of factorization is not uniquely defined, and particular choice of the value of the scale $\mu$ defines the separation point between the hard process and the non-perturbative PDF.
The scale $\mu$ is not physical and we know that no physical observable (like cross-section) can depend on it, and this dependence needs to cancel order by order in perturbative calculations.

The crucial feature of the separation provided by factorization is that the short distance coefficient function is free from the mass singularities and calculable perturbatively, whereas all the non-perturbative effects are separated in the PDF.
Moreover, only the coefficient functions depend on the considered process (e.g. DIS, DY), the parton distributions are universal (in MS-like schemes).
This means that the non-perturbative PDFs can be extracted from one experiment and then used in calculation for another one.
Additionally, we will see that parton distributions fulfill the DGLAP evolution equation which means that if we know PDFs at one scale we can calculate them at any other scale.


In the following we review collinear factorization describing the intermediate steps leading to the final results obtained in \cite{EGMPR79} and \cite{CFP80}.
We start by introducing the {\em generalized ladder expansion} allowing for the reorganization of the perturbative series in section~\ref{sec:ladder}, then in section~\ref{subsec:proj} we introduce projection operators crucial for obtaining the full factorization (decoupling subtracted hard matrix element from PDFs).
In section~\ref{sec:factorization} we formulate factorization and provide final formulas used later in the thesis.
Finally, in Section~\ref{sec:dglap} we provide details about DGLAP equations describing evolution of PDFs.

We keep here close relation to the formulation of the collinear factorization given in \cite{CFP80} and adopt the same notation.

It is worth mentioning that there are also other types of factorization theorems e.g. implementing the so called $k_T$-factorization, for a detailed overview of factorization in QCD we refer to \cite{Col11}.

    \section{Generalized Ladder Expansion}\label{sec:ladder}


In this and next section we will follow arguments presented in \cite{CFP80} and previously proved in \cite{EGMPR79}.
Following these works we will use the DIS process as an example for our considerations.
We will consider squared matrix elements, that are represented as cut Feynman diagrams.

\begin{figure}[ht]
  \centering
  \includegraphics[scale=1.0]{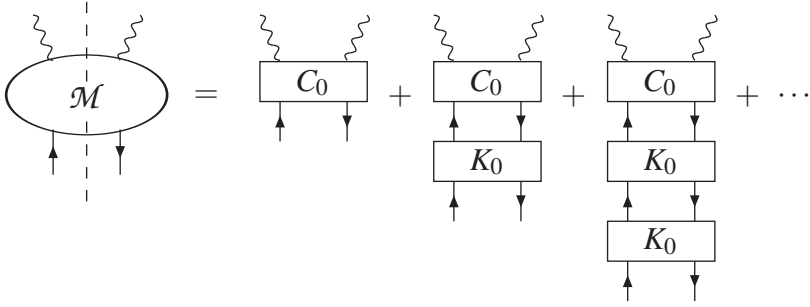}
  \caption{Generalized Ladder Expansion of the squared DIS matrix element.}
  \label{fig:gle}
\end{figure}

In \cite{EGMPR79} authors have shown that in the axial gauge a squared matrix element, for instance describing the DIS process, can be reorganized in terms of 2-particle irreducible (2PI) kernels $C_0$ and $K_0$ in such a way that $C_0$ is free from mass singularities and $K_0$ contains all of them.
This reorganization of the perturbative series is referred to as the {\em generalized ladder expansion} (GLE), and in the case of DIS can be written as:
\begin{equation} \label{eq:m-series}
  M = C_0 \: \left(1 + K_0 + K_0^2 + \cdots \right) = C_0 \: \frac{1}{1-K_0} \equiv C_0 \cdot \Gamma_0 \text{.}
\end{equation}
The $C_0$ kernel represents the hard interaction part, and series of the $K_0$ kernels, contained in $\Gamma_0$, represents ladder part of the expansion that later gives rise to parton densities.
Formula (\ref{eq:m-series}) is represented graphically in Fig.~\ref{fig:gle}, note that the external lines connecting $C_0$ and $K_0$ kernels represent full 4-momentum integration, and kernels themselves are sums of cut diagrams.
By definition every $K_0$ ($C_0$) kernel contains full propagators of the upper lines but does not contain the lower lines, and as was shown by \cite{EGMPR79}, in the axial gauge mass singularities originate from the integration over the connecting lines, so as long as we do not perform this integration kernels are free of the collinear divergences.

The GLE formula of \eqref{eq:m-series} already exhibits properties of the full factorization formulas like \eqref{eq:fact-common}, however, the 2PI kernels are still connected by spinor indices and 4-momentum integration.
This can be clearly seen when we explicitly write all the indices and momenta in $C_0$ and $K_0$ kernels.
Then $C_0$ becomes $C_{\alpha\alpha'}^{\mu\mu'}(p,q)$ and $K_0$ becomes $K_{\alpha\alpha'}^{\beta\beta'}(p,k)$, and a product of two kernels (as in the series in \eqref{eq:m-series}) is given by:
\begin{equation}
      C_{\alpha\alpha'}^{\beta\beta'}(k_1,k_2)
      = 
      \sum_{\gamma\gamma'}
      \int \frac{\Dm r}{(2\pi)^m}
      A_{\alpha\alpha'}^{\gamma\gamma'}(k_1,r)
      B_{\gamma\gamma'}^{\beta\beta'}(r,k_2).
\end{equation}
Additionally we provide here explicit definitions of the shorthand notation for contraction of kernels' indices as used in \cite{CFP80} and in the next section;
in the case of spinor indices we have:
$$ [ \slashed{k} A \quad \text{denotes} \quad \sum_{\alpha\alpha'} \slashed{k}^{\alpha\alpha'} A^{\beta\beta'}_{\alpha\alpha'} \text{,}$$
$$ A \slashed{k} ] \quad \text{denotes} \quad \sum_{\beta\beta'} A^{\beta\beta'}_{\alpha\alpha'} \slashed{k}_{\beta\beta'} \text{,}$$
and in the case of vector indices:
$$ [ g A \quad \text{denotes} \quad \sum_{\alpha\alpha'} g^{\alpha\alpha'} A^{\beta\beta'}_{\alpha\alpha'} \text{,}$$
$$ A \: d(k) ] \quad \text{denotes} \quad \sum_{\beta\beta'} A^{\beta\beta'}_{\alpha\alpha'} \: d_{\beta\beta'}(k) \text{.}$$

To proceed with the factorization procedure we need to introduce projection operators that will decouple the 2PI kernels in spinor indices and in 4-momentum integration, but before doing so, it is beneficial to introduce parametrization of 4-momenta in terms of light-cone variables.

\subsection{Momenta Parametrization}
\label{subsec:momenta}

In the light-cone gauge it is convenient to use the following parametrization for the in- and outgoing momenta $p$ and $k$:%
\footnote{Note that parametrization of \eqref{eq:def-q} is equivalent to the Sudakov decomposition.}
\begin{align} \label{eq:def-p}
  & p = (P,\vect{0},P) \text{,} \quad \text{where $p^2=0$,}
  \\ \label{eq:def-q}
  & k = \left( xP + \frac{k^2-k_\perp^2}{4xP}, \vect{k}_\perp, xP - \frac{k^2-k_\perp^2}{4xP} \right) \text{,} \quad \text{where $k^2<0$,}
  \\
  & k_\perp = (0,\vect{k}_\perp,0) \text{,} \quad \text{so that $k_\perp^2 = - \vect{k}_\perp^2 < 0$,}
\end{align}
and the light-cone gauge-fixing vector
\begin{equation}
  n = \left(\frac{\sdot{p}{n}}{2P}, \vect{0}, - \frac{\sdot{p}{n}}{2P} \right) \text{,} \quad \text{where $n^2=0$.}
\end{equation}
We also introduce a plus notation, so that $k_+ = \sdot{k}{n}$, where $k$ is an arbitrary four-vector.
Additionally we define
\begin{equation}
  x=\frac{\kn}{\pn}
  \text{.}
\end{equation}
that can be interpreted as a momentum fraction of momentum $p$ carried by $k$.\\


\subsection{Projectors}
\label{subsec:proj}

As argued above, the ``raw'' factorization formula (\ref{eq:m-series}), obtained using GLE, can be fully factorized by introduction of an appropriate projection operator.
Such an operator needs to:
\begin{itemize}
  \item decouple $C_0$ and $\Gamma_0$ in spinor indices;
  \item extract the singular part of $\Dm k$ integrals (in terms of $\eps$ poles), and decouple $C_0$ and $\Gamma_0$ in momentum space, leaving only a one-dimensional integration over the light-cone $x$ variable.
\end{itemize}


The action of the projector $\P$ is specified by a product of two operators $\P_{F/G}$ and $\P_\eps$, so that
\begin{equation}
  \P = \P_{F/G} \otimes \P_\eps
  \text{,}
\end{equation}
where the operator acting on fermion lines $\P_F$ is defined as
\begin{equation}\label{eq:pq}
  A \; \P_F \; B = A \slashed{k} ] \Big[ \frac{\slashed{n}}{4\kn} B
  \text{.}
\end{equation}
The operator acting on gluon lines $\P_G$ is
\begin{equation} \label{eq:pg}
  A \; \P_G \; B = A \: d(k) \: \frac{1}{2(1+\eps)} \Big] \big[- g B
\end{equation}
with
\begin{equation*}
  d_{\mu\nu}(k) = -g_{\mu\nu} + \frac{k_\mu n_\nu + n_\mu k_\nu}{\kn}
  \text{.}
\end{equation*}
The operator $\P_\eps$ sets $k^2=0$ on its left side and extracts the pole part in the $\D k^2/k^2$ integral from its right side.

    \section{Factorization} \label{sec:factorization}

With the help of the projection operator $\P$ the ``raw'' factorization formula of \eqref{eq:m-series} can be reorganized in the following way:
\begin{equation}
  M = \left( C_0 \frac{1}{1-(1-\P)K_0} \right) \otimes \left( \frac{1}{1-\P K} \right)
    \equiv C \otimes \Gamma
  \text{,}
\end{equation}
where $\Gamma$ contains all the mass singularities and $C$ is free of them.
Additionally $C$ and $\Gamma$ are now coupled only by a one dimensional convolution integral.%
\footnote{\label{foot:conv}Convolution is defined here in a standard way:
    \begin{equation*}
    f \otimes g(x) = \int_0^1 dz_1 dz_2 f(z_1) g(z_2) \delta(x-z_1z_2)
                   = \int_x^1 \frac{dz}{z} f(x/z) g(z)
                   = \int_x^1 \frac{dz}{z} f(z) g(x/z)
    \end{equation*}.}
The kernels $1/(1-(1-\P)K_0)$ and $1/(1-\P K)$ are defined by series expansion, however, there is an important difference in the action of $(1-\P)$ and $\P$ operators in these series.
In the first case the $(1-\P)$ operator acts on the full expression on the right giving the following expansion:
\begin{equation}
  \frac{1}{1-(1-\P)K_0} \equiv 1 + (1-\P)K_0 + (1-\P)(K_0(1-\P)K_0) + \cdots
  \text{,}
\end{equation}
Whereas in the second case operator $\P$ acts on the immediate/nearest $K$ on the right leading to:
\begin{equation}
  \frac{1}{1-\P K} \equiv 1 + \P K + (\P K)(\P K) + \cdots
  \text{,}
\end{equation}
where $K$ is defined as:
\begin{equation}
  K = K_0 \frac{1}{1-(1-\P)K_0}
  \text{.}
\end{equation}

The final result in the case of quark structure function is the following:
\begin{align*}
  \tilde{F}^{(i)}_i \left(\frac{Q^2}{\mf^2},x,\as,\frac{1}{\eps} \right)
  &=
  \frac{1}{2} \left[ M^{(i)} \slashed{p} \right]\\
  &=
  \int_0^1 \D y \; C^{(i)} \left( \frac{Q^2}{\mf^2},y,\as \right)
  \int_0^1 \D z \; \Gamma_S \left( z,\as,\frac{1}{\eps} \right)
  \del(x-yz)
  \text{,}
\end{align*}
where
\begin{align*}
  & C^{(i)} = \frac{1}{2} \left[ C_0^{(i)} \frac{1}{1-(1-\P)K_0} \slashed{k} \right]_{k^2=0}
  \text{.}
\end{align*}
The parton density $\Gamma_S$ can be written as convolution of $Z_F$ and $\Gh_S$ from fully virtual diagrams and diagrams where at least one internal line is cut respectively:
\begin{equation} \label{eq:gh1}
  \Gamma_S\left(x,\as,\frac{1}{\eps}\right) = \; Z_F\left(x,\as,\frac{1}{\eps}\right) \: \Gh_S \left(x,\as,\frac{1}{\eps}\right)
  \text{,}
\end{equation}
where
\begin{equation} \label{eq:gb}
  \Gh_S \left(x,\as,\frac{1}{\eps}\right) = \delta(1-x) + x \int \frac{\D^mk}{(2\pi)^m} \; \delta\left(x-\frac{\kn}{\pn}\right) \left[ \frac{\sl{n}}{4 \kn} \; \frac{K}{1-\P K} \; \sl{p} \right]
  \text{.}
\end{equation}

Note that $\tilde{F}^{(i)}_i$ is the structure function of a quark.
If we want to calculate hadronic structure function (e.g. for a proton), instead of using the partonic density $\Gamma_S$, we need to use the hadronic one.
The hadronic (``renormalized'') parton density $f_i$ (used earlier in \eqref{eq:fact-common}) is constructed by convoluting $\Gamma_S$ with the ``bare''  density of a parton inside a hadron.
In the case of quark we have
\begin{equation}
\label{eq:PDF}
f_q(x,\mu_f) = \Gamma_S \otimes q_0(x),
\end{equation}
where $q_0$ is the bare quark density.
The bare densities feature mass singularities that cancel exactly the mass singularities from $\Gamma_S$ leaving a finite parton distribution $f_i$ of \eqref{eq:fact-common}, that can be obtained by fitting experimental data.

For the purpose of calculating the next-to-leading order virtual corrections to the DGLAP splitting functions, that we perform in Section~\ref{ch:calc}, we write explicitly the NLO contribution to \eqref{eq:gb} exploiting additional features specific for these corrections.
First, we take contributions from the color factor $\C$ and coupling constant $\as^2$ out of the kernel.
Second, since after renormalization we put $p^2=(p-k)^2=0$, then $k^2$ is the only quantity with mass dimension the integrand of \eqref{eq:gb} depends on.
Thus we can write this integrand  as a product of dimensionless exclusive parton densities $W_R^{(2)}$ and $k^2$ to the corresponding power.
Finally, a one-particle real phase space is denoted as $\PS(k)$ and is defined in Section~\ref{subsec:1PS}.
After described transformations we get the following expression:

\begin{equation} \label{eq:gh2}
  \Gh_S^{(2)}\left(x,\as,\frac{1}{\eps}\right)
  =
  \C
  \aspi^2
  \frac{1}{\mf^{2\eps}}
  \int \frac{\D\,\PS(k)}{\abs{k^2}}
  W_{R}^{(2)}\left(x,\eps,\frac{k^2}{\mr^2}\right)
  \text{,}
\end{equation}
where $\mf$ is the factorization scale which we write here separately from the accompanying $(k^2)^\eps$ factor contained in the real phase space.


    \section{The DGLAP Evolution Equations}
\label{sec:dglap}

The (``renormalized'') parton distribution functions (PDFs) of \eqref{eq:PDF} cannot be calculated perturbatively and currently they need to be obtained by fitting experimental data.
However the evolution of PDFs with the factorization scale is purely perturbative and described be the DGLAP evolution equation~\cite{DGLAP}.
The general form of the DGLAP equation is as follows:
\begin{equation} \label{eq:dglap}
  \frac{\partial}{\partial \ln\mu^2} f_i\left(x,\mu^2\right)
  =
  \sum_{j=g,q,\bar{q}}
  \int_x^1 \frac{\D z}{z} P_{ij}\left(\frac{x}{z},\as(\mu^2)\right) f_j(z,\mu^2)
\end{equation}
or if we introduce convolution notation defined in footnote \ref{foot:conv}:
\begin{equation*}
  \frac{\partial}{\partial \ln\mu^2} f_i\left(x,\mu^2\right)
  =
  \sum_{j=g,q,\bar{q}}
  P_{ij} \otimes f_j
  \text{,}
\end{equation*}
where $i,j=g,q,\bar{q}$ goes over all types of partons.

The evolution variable $\mu$ is a formal parameter -- the factorization scale, however, when we go to practical applications, at the end, we need to choose a particular value for it.
Typically it will be associated with the scale of the considered hard process e.g. in the case of DIS with 4-momentum transfer $Q^2$.
In the case of parton shower Monte Carlo programs, where the evolution is done by the MC program, it needs to be associated with certain kinematical variables, typical choices are e.g.\ virtuality ($Q^2=|k^2|$), transverse momentum or rapidity.
Additional discussion connected to this choice is provided in Section~\ref{subsec:evolTime}.

The evolution kernels governing DGLAP evolution can be calculated perturbatively and we will use the following notation for their expansion in terms of the strong coupling:
\begin{equation}
  P_{ij}(x,\as) = \aspi P_{ij}^{(1)}(x) + \aspi^2 P_{ij}^{(2)}(x) + O(\as^3)
  \text{.}
\end{equation}
%
To calculate these kernels we need to compute appropriate terms in the expansion of the parton density $\Gamma_S$ of \eqref{eq:gh1}, as can be seen below:
\begin{equation*}
  \Gamma_S\left(x,\as,\frac{1}{\eps}\right)
  =
  \delta(1-x)
  +
  \frac{1}{\eps}
  \left(
    \aspi P^{(1)}(x)
    +
    \frac{1}{2} \aspi^2 P^{(2)}(x)
    +
    \cdots
  \right)
  +
  O\left(\frac{1}{\eps^2}\right)
  \text{,}
\end{equation*}

\begin{equation*}
  \Gh_S\left(x,\as,\frac{1}{\eps}\right)
  =
  \delta(1-x)
  +
  \frac{1}{\eps}
  \left(
    \aspi \Ph^{(1)}(x)
    +
    \frac{1}{2} \aspi^2 \Ph^{(2)}(x)
    +
    \cdots
  \right)
  +
  O\left(\frac{1}{\eps^2}\right)
  \text{,}
\end{equation*}

\begin{equation*}
  Z\left(x,\as,\frac{1}{\eps}\right)
  =
  1
  +
  \frac{1}{\eps}
  \left(
    \aspi \xi^{(1)}(x)
    +
    \frac{1}{2} \aspi^2 \xi^{(2)}(x)
    +
    \cdots
  \right)
  +
  O\left(\frac{1}{\eps^2}\right)
  \text{.}
\end{equation*}
For instance the next-to-leading evolution kernel will be given by:
\begin{equation*}
  P^{(2)}(x) = \delta(1-x) \xi^{(2)}(x) + \Ph^{(2)}(x)
  \text{.}
\end{equation*}

Taking into account the following symmetries in the splitting functions \cite{FP82} (resulting from the flavor symmetry and charge conjugation invariance):
\begin{align*}
  & P_{gq_i} \; = \; P_{g\bar{q}_i} = P_{gq}
  \\
  & P_{q_ig} \; = \; P_{\bar{q}_ig} = \frac{1}{2n_f}P_{qg}
  \\
  & P_{q_iq_j} \; = \; P_{\bar{q}_i\bar{q}_j} = \del_{ij} P_{qq}^v + P_{qq}^s
  \\
  & P_{q_i\bar{q}_j} \; = \; P_{\bar{q}_iq_j} = \del_{ij} P_{q\bar{q}}^v + P_{q\bar{q}}^s
  \text{,}
\end{align*}
a system of $2n_f+1$ integro-differential equations (\ref{eq:dglap}) can be reorganized so that the singlet quark density
\begin{equation}
  f_s = \sum_{i=1}^{n_f} (f_i + \bar{f}_i)
\end{equation}
together with the gluon density satisfy the following evolution equations:
\begin{equation}
  \frac{\partial}{\partial \ln\mu^2}
    \begin{pmatrix}
      f_s
      \\
      f_g
    \end{pmatrix}
  =
    \begin{pmatrix}
      P_{qq} & P_{qg}
      \\
      P_{gq} & P_{gg}
    \end{pmatrix}
    \otimes
    \begin{pmatrix}
      f_s
      \\
      f_g
    \end{pmatrix}
  \text{,}
\end{equation}
where
\begin{equation}
  P_{qq} = P_{qq}^v + P_{q\bar{q}}^v + n_f (P_{qq}^s + P_{q\bar{q}}^s)
  \text{.}
\end{equation}
In a similar way, the non-singlet superpositions
\begin{equation} \label{eq:f-ns}
  f_{ij}^\pm = (f_i \pm \bar{f}_i) - (f_j \pm \bar{f}_j)
  \qquad \text{and} \qquad
  f_v = \sum_{i=1}^{n_f} (f_j - \bar{f}_j)
\end{equation}
evolve independently of $f_g$ and each other in the following way:
\begin{equation} \label{eq:p-ns}
  \frac{\partial}{\partial \ln\mu^2} f_{ij}^\pm = P_\pm \otimes f_{ij}^\pm
  \qquad \text{and} \qquad
  \frac{\partial}{\partial \ln\mu^2} f_v = P_v \otimes f_v
  \text{,}
\end{equation}
where the non-singlet splitting functions are defined as
\begin{equation} \label{eq:p-ns2}
  P_\pm = P_{qq}^v \pm P_{q\bar{q}}^v
  \qquad \text{and} \qquad
  P_v = P_{qq}^v - P_{q\bar{q}}^v + n_f (P_{qq}^s - P_{q\bar{q}}^s)
  \text{.}
\end{equation}

The detailed calculation of the next-to-leading order virtual contributions to the non-singlet splitting function $P_{qq}^s$ (with one loop and one real emission) are given in Chapter~\ref{ch:calc}.
The corresponding calculation for the two-real contributions have been done in \cite{JKSS11}.
These calculations allow to obtain the non-singlet inclusive splitting functions of \eqref{eq:p-ns}.
Moreover, the unintegrated (exclusive) splitting functions can be used for the Monte-Carlo simulations of the parton shower \cite{JKPSS13}.

  \chapter{Calculation of Virtual Splitting Functions at NLO}
\label{ch:calc}

In this chapter we describe a complete technique for calculating space-like non-singlet virtual splitting functions at the next-to-leading order (NLO).
We use the method proposed by Curci, Furmanski, and Petronzio \cite{CFP80}, see also \cite{Hei98} for more details.
Our goal is to obtain exclusive splitting functions suitable for Monte-Carlo simulations which are consistent with recently obtained real contributions to NLO splitting functions \cite{JKSS11,Kus11}.
For that purpose we modify the approach of \cite{CFP80} and instead of the Principal Value (PV) prescription, we introduce a New Principal Value (NPV) prescription \cite{GJKS14} for regularizing infra-red singularities in the light-cone gauge.
To ensure consistency and correctness of the NPV prescription and of the obtained results we calculate ultra-violet counter-terms and inclusive splitting functions and compare them with the standard PV inclusive results available in \cite{CFP80,Hei98}.

    \section{Non-Integrated Parton Distribution}
\label{sec:3.1}

\begin{figure}[ht]
  \centering
  \includegraphics[width=3.0cm]{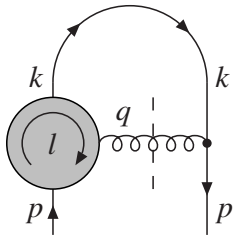}
  \caption{A general topology of $K^{(1+1)}$, one-real-one-virtual 2PI kernels, considered in this work.}
  \label{fig:blob}
\end{figure}

As it was described in Section \ref{sec:ladder}, every matrix element can be expressed as a generalized ladder expansion.
Building blocks of such a ladder are two-particle-irreducible (2PI) cut Feynman diagrams.
The projector operators acting on quark or gluon lines connecting these diagrams transform a generalized ladder expansion into the convolution of scalar objects which are coupled only by the $x$-integral.
We name such an object the {\it non-integrated parton distribution} and define it for the non-singlet case as follows
\begin{equation} \label{eq:wu-general}
  \Wu^{(n_r+n_v)} \left(p,k,q_1,\ldots,q_{n_r},l_1,\ldots,l_{n_v},\eps \right) = x \left[ \frac{\sl{n}}{4 \kn} K^{(n_r+n_v)} \sl{p} \right]
  \text{,}
\end{equation}
where (see fig.~\ref{fig:blob} for the special case of $n_r=n_v=1$):
  1) $p$ is an initial momentum of the incoming leg;
  2) $k$ is a final momentum of the outgoing leg;
  3) $q_i$ are momenta of the real legs depicted as cut lines;
  4) dependence on the virtual momenta $l_i$ arises in cases when there are loops inside a cut amplitude;
  5) $\eps$-dependence is dictated by the fact that calculations are done in $m=4+2\eps$ dimensions.

In this work we consider only NLO kernels with one loop momentum $l$ and one real momentum $q$, i.e. $n_r=n_v=1$, (see fig. \ref{fig:blob}).
In this special case we have
\begin{equation}\label{eq:wu-nlo}
  \Wu^{(1+1)} \left(p,k,q,l,\eps \right) = x \left[ \frac{\sl{n}}{4 \kn} K^{(1+1)} \sl{p} \right]
  \text{.}
\end{equation}

In contrast to other kinds of splitting functions defined later in this work, the non-integrated splitting functions have no physical interpretation, because of the dependence on unphysical virtual momenta $l_i$.
Definitions in eqs. (\ref{eq:wu-general}--\ref{eq:wu-nlo}) are introduced for convenience of the notation as a starting point for the further calculations.

    \section{Virtual-Momentum Integration}
\label{sec:loop-int}

In the consecutive step we need to perform integration over the momentum $l$ in \eqref{eq:wu-nlo}.
The result of such a virtual-momentum integration we name {\it exclusive bare parton density} and define it as follows:
\begin{equation}\label{eq:Wb}
  \Wb^{(1+1)}\left(x,\del,\eps,\frac{k^2}{\mr^2}\right)
  =
  \int \frac{\D^ml}{\mr^{2\eps}} \;
  \Wu^{(1+1)}
  =
  x
  \int \frac{\D^ml}{\mr^{2\eps}} \;
  \left[
    \frac{\sl{n}}{4 \kn} K^{(1+1)} \sl{p}
  \right]
  \text{,}
\end{equation}
where the renormalization scale $\mr$ is shown explicitly.
The index "B" stands for "bare", as this function needs to be renormalized.

\subsection{Singularities and Regularization}
\label{subsec:regularization}

A general structure of the non-integrated parton density suggests that integral in \eqref{eq:Wb} is singular.
A closer analysis leads to the following three types of singularities: 1) ultra-violet, 2) infra-red, and 3) spurious.

Singularities of the first type, {\it ultra-violet} (UV), are common to quantum field theories, like QCD (see \cite{tHV73}).
There are general methods for renormalizing the UV singularities, nevertheless, at first they need to be regularized and we use dimensional regularization for this purpose.
In order to separate them from the other singularities we will mark the corresponding poles as $1/\euv$, where "uv" is a logical marker, i.e. numerically $\euv = \eps$.
 
The second type of singularities, {\it infra-read} (IR), arise because gluons are massless and we consider the limit of massless quarks.
They are related either to softness or to collinearity of the emitted parton.
As explained in the previous chapter, the helpful feature of these singularities is that in the axial gauge they originate only from the integration over the momenta connecting $K_0$ kernels, i.e.\ momentum $k$ in fig.~\ref{fig:blob}.
Nonetheless separate components of $K_0$ contain the "internal" IR singularities which arise from loop-momenta integration.
In the standard approach of \cite{CFP80} these singularities are regularized by the dimensional regularization method.
However, in our approach we use the Principal Value regularization for some of these singularities.
Such a modification of the standard CFP method we call a {\em New Principal Value} scheme.

The last, {\it spurious}, singularities are artifacts of the light-cone gauge.
They result from the axial denominators $1/\nl$ and as unphysical must cancel in the final results.
In the intermediate steps they also require regularization and it is done by the PV prescription.
One should mention that their presence is a source of complication of the renormalization procedure, as the renormalization constants start to depend on vector $n$.
As a consequence determination of these constants is more involved, as described in Section~\ref{sec:renormalization}.
Let us note, that the singularities regularized by means of PV prescription manifest themselves as powers of the logarithm of the regulator $\del$.

\subsubsection{Dimensional Regularization}

A general approach is to regularize IR and UV singularities with the help of dimensional regularization technique.
The idea is to switch from four to an arbitrary number of space-time dimensions.
This technique is well developed and has solid quantum field theory foundation.
In this work we set $m=4+2\eps$.

Let us make here a comment on the UV versus IR singularities.
Consider the following integral:
\begin{equation}\label{eq:int-uv}
  \int \Dm l \frac{1}{l^2 (l+p)^2} = \frac{i}{(4\pi)^{2+\eps}} (p^2)^\eps \left( -\frac{1}{\euv} + 2\right)
  \text{.}
\end{equation}
It is singular in $\eps \to 0$ limit and the pole is of the UV type.
More complicated integrals may lead not only to UV, but also to IR poles, when a massless limit is considered.
The important question is how to distinguish between those two types of poles.
To answer this question let us analyze integrals from Appendix \ref{sec:md-integrals}.
The right hand side of these integrals yet before integration over Feynman parameters in some places contains $\eps$ poles.
Those poles arise from the integration over $\Dm l$ when $\abs{l}\to\infty$, which is well defined for $\eps < 0$.
Therefore they are of ultra-violet origin and we label them with $\euv$ symbol.
Remaining poles, arise from integration over the Feynman parameters.
These integrals are well defined for $\eps > 0$.
Therefore poles resulting from theses integrals are labeled as infra-red with $\eir$ symbol.
Alternatively one can argue that UV poles arise independently of the kinematic configurations we impose on the vectors an integral depends on.
In contrast, IR poles arise only in cases when we put some of those external vectors on shell.

An interesting example that demonstrates this mechanism is \eqref{eq:int-f-2}.
One can easily see that in the case of a three-point integral ($\alpha=3$), the pole in front of a metric tensor $g^{\mu\nu}$ is labeled as UV.
Remaining terms may also contain poles in $\eps$ (after integration over Feynman parameters), but those are of the IR origin.
This can be seen in the final form of integral \eqref{eq:int-f-2}, see \eqref{eq:r36} for a specific kinematic configuration $p^2=(p-k)^2=0$.

\subsubsection{New Principal Value Regularization}

In the light-cone axial gauge gluon propagator has the following form
\begin{equation}
  \frac{i}{l^2 + i\veps} \left(-g^{\mu\nu} + \frac{l^\mu n^\nu + n^\mu l^\mu}{\nl} \right)
  \text{,}
\end{equation}
which features the $1/\nl$ term, that can lead to spurious singularities.
These singularities have to be 
regulated somehow during the intermediate steps of the calculations.
The method proposed in the original paper \cite{CFP80} uses a well-known Principal Value prescription:
\begin{equation}
  \frac{1}{\nl} \to \left[\frac{1}{\nl}\right]_{PV} = \frac{\nl}{(\nl)^2 + \del^2 (\pn)^2}
  \text{.}
\end{equation}
Parameter $\del$ is an infinitesimal regulator and $p$ is an external reference momentum.
It is important to note that $\del$ has a geometrical meaning and can be directly implemented in Monte-Carlo simulations.

Now comes the crucial observation: on top of the light-cone propagator, there are also other sources of $(\nl)^{-1}$ singularities, due to the phase space or ${(l^2+i\veps)^{-1}}$ part of the propagator.
In the dimensional regularization approach they are automatically taken care of by the $\eps$ parameter.
This however introduces in the same graph both $\ln^2{\del}$ and $1/\eps^2$ terms.
Let us illustrate this point with the formula (\ref{eq:int-npv-3}) for the three-point integral.
This formula is organized in such a way, that the $\D \lnp$ part is left unintegrated to the very end, where $\lnp = \nl$.
The function $f(\lnp)$ represents the axial-type denominators and can be a source of $\ln{\del}$ singularities.
On the other hand there are also $(1-y)^{-1+2\eps}$ terms, which are singular in the plus variable $y=\lnp/\pnp$ but they are regulated dimensionally, leading to $1/\eps$ poles.
As a result, for example in the $S_0$ form-factor of \eqref{eq:int-S0} (axial scalar integral) we have all types of singularities: $1/\eps^2$, $I_1$, $1/\eps$, and $I_0$; i.e.
\begin{equation}
  S_0^{PV} = \frac{1}{\eir^2} + \frac{I_0 - \ln{x}}{\eir} + I_1 - I_0 \ln{x} - 2 \Li(1) - 2 \Li(1-x) - \frac{\ln^2{x}}{2}
  \text{,}
\end{equation}
where $I_0$ and $I_1$ stand for
\begin{equation}
\begin{split}
  I_0 = & \int_0^1 \D x \frac{x}{x^2 + \del^2} = -\ln{\del} + O(\del)
  \text{,}
  \\
  I_1 = & \int_0^1 \D x \frac{x \ln x}{x^2+\del^2} = -\frac{1}{2} \ln^2{\del} - \frac{1}{4}\Li(1) + O(\del)
  \text{.}
\end{split}
\end{equation}

To simplify this situation we propose a New Principal Value (NPV) prescription in which {\em all} singularities in plus variables are regulated by $\del$ as in the PV scheme.
In the case of \eqref{eq:int-npv-3} this means that we apply PV also to the ${(1-y)^{-1+2\eps}}$ term.
This is done as follows
\begin{equation}
  \Dm l\; y^{-1+\eps} \to \Dm l \, \biggl[\frac{1}{y}\biggr]_{PV} \Bigl(1+\eps\ln{y} +\eps^2 \frac{1}{2}\ln^2{y} +\dots\Bigr) \text{,} \qquad y = \frac{\lnp}{\pnp}
 \label{plusbar}
  \text{,}
\end{equation}
i.e.\ if needed, we keep higher-order terms in $\eps$ and the limit $\eps \to 0$ is taken before the $\del \to 0$ limit.
In this NPV prescription the above mentioned form-factor $S_0$ takes the form given in \eqref{eq:S0}, which we quote here
\begin{flalign}
  S_0^{NPV} = & - \frac{3 I_0 - \ln{x} + \ln(1-x)}{\eir} - 5 I_1 + I_0 \ln{x} + 2 I_0 \ln(1-x) \nonumber \\
              & + 2 \Li(1-x) + \frac{\ln^2{x}}{2} + \frac{\ln^2(1-x)}{2} + \Li(1)
  \text{.}
\end{flalign}
As we see, the $1/\eps^2$ pole has been converted into $I_0/\eps$ and $I_1$ terms.

Such a modification of the standard PV prescription has also some drawbacks.
Namely, it makes Feynman integrals more complicated, because they start to depend on the auxiliary vector $n$.
For example, the three-point scalar integral with form-factor $R_0$ of \eqref{eq:int-R0}, in the NPV prescription equals
$$ R_0^{NPV} = - \frac{2 I_0 + \ln(1-x)}{\eir} - 4 I_1 + 2 I_0 \ln(1-x) + \frac{\ln^2(1-x)}{2} \text{,} $$
whereas in the PV scheme it reads
$$ R_0^{PV} = \frac{1}{\eir^2} - \frac{\pi^2}{6} \text{.} $$
The complete list of integrals in the NPV prescription is given in Appendix~\ref{sec:int-ir}.

Let us conclude this section with some general comments.
As already mentioned in the introduction, the PV method was proposed in \cite{CFP80} as an efficient, but "phenomenological" prescription.
The main argument supporting it was that the spurious singularities are unphysical and as such {\em must} cancel in the final result.
The proposed NPV scheme follows the same philosophy: the IR-plus singularities also cancel in final expressions as proven in \cite{EGMPR79}.
Therefore we find it justified to treat all of them in the same way with the PV regulator.
In turn, apart from the simplification of the calculations,
``trading'' $\eps$ poles for a geometrical regulator $\del$ makes it possible
to use the resuting NPV splitting functions in a stochastic simulation of a praton cascade.

\subsection{Exclusive Bare Parton Density}

To proceed with the virtual-momentum integration in \eqref{eq:Wb} we first employ regularization techniques of section~\ref{subsec:regularization}.
At this point we assume that $\eps<0$ in order to treat the UV singularities.
The result of the $\Dm l$ integration, which we refer to as the {\it exclusive bare parton density}, can be parametrize in the following way:%
\begin{multline} \label{eq:Wb-series}
  \Wb^{(1+1)}\left(x,\del,\eps,\frac{k^2}{\mr^2}\right)
  =
  -i g^4 \: \C \: \Qv \: \frac{1}{\abs{k^2}} \;
  \Bigg(
  \left(
    \frac{\Wirrq}{\eir^2}
    +
    \frac{\Wirq}{\eir}
    +
    \frac{\Wuvq}{\euv}
    +
    \Wzq
  \right)
  \left(\frac{\abs{k^2}}{\mr^2}\right)^{\eps}
  \\ +
  \frac{\Wuvk}{\euv}
  \left(\frac{q^2}{\mr^2}\right)^{\eps}
  +
  \frac{\Wuvp}{\euv}
  \left(\frac{p^2}{\mr^2}\right)^{\eps}
  \Bigg)
  \text{,}
\end{multline}
where form-factors $W$ depend on $x=\knp/\pnp$ and can also include singularities regularized by means of the geometrical PV or NPV prescription manifesting themselves as powers of $\ln\delta$.
We discarded IR form factors in front of $p^2$ and $q^2=(p-k)^2$ terms.
Those terms vanish in the IR limit, i.e. when put on shell.
On the other hand, we keep UV form factors for all the momenta.
They are needed to build UV counter-term, which is done before analytic continuation from UV to IR domain.
We also explicitly showed the renormalization scale $\mr$ in \eqref{eq:Wb-series}.
We keep it different from the $\mf$ factorization scale.

Note that in the $\PV$ prescription used in \cite{Hei98} term $\Wirrq$ is non-zero for some graphs, while in the NPV approach it always vanishes.

    \section{Renormalization} \label{sec:renormalization}

In this step we renormalize UV singularities.
For this purpose, at first we construct a {\it renormalization constant} $Z^{(1)}$ by dividing out the leading order parton density $W_R^{(1+0)}$ from the UV pole part of \eqref{eq:Wb-series}
\begin{equation}\label{eq:Z}
  Z^{(1)}(x,\del) = \frac{\Wuvk + \Wuvp + \Wuvq}{W_{R,0}^{(1+0)}(x)}
  \text{.}
\end{equation}
Note that the sum of $W$ form-factors in this expression is proportional to the $\eps^0$-part of a LO parton density:
\begin{equation} \label{eq:Wr11}
  \Wr^{(1+0)}(x,\eps) = W_{R,0}^{(1+0)}(x) + \eps \; W_{R,1}^{(1+0)}(x)
  \text{.}
\end{equation}
For the non-singlet case that we consider it reads
\begin{equation} \label{eq:Wr11qq}
  W_{R,\mathrm{qq}}^{(1+0)}(x,\eps) = \frac{1+x^2}{1-x} + \eps \; \xi \; (1-x)
  \text{,}
\end{equation}
where the parameter $\xi=1$ is introduced to track the contribution from a finite part of the UV counter-term.
This somewhat complicated procedure of calculating renormalization constants by extracting them from the whole graph is a consequence of the use of the axial gauge, see \cite{CFP80,Hei98} for further discussion.

We proceed with defining the {\it ultra-violet counter-term} $\Wz$ as follows:
\begin{equation}\label{eq:ZF}
  \Wz^{(1+1)}(x,\del,\eps)
  =
  -i \: g^4 \: \Qv \: \frac{1}{\abs{k^2}} \;
  \frac{Z^{(1)}(x,\del) \; \Wr^{(1+0)}(x,\eps)}{\euv}
  \text{,}
\end{equation}
which is proportional to the complete LO exclusive splitting function with a color factor and all the $\eps$-terms included.


\subsection{Exclusive parton density}

At this point we obtained all the components needed to define {\it exclusive (renormalized) parton density} $W_R$.
We subtract the counter-term of \eqref{eq:ZF} from the exclusive bare parton density of \eqref{eq:Wb-series}.
The resulting object is UV finite (see discussion in Section~\ref{sec:factorization}), so we can analytically continue to the IR domain, i.e. from $\eps < 0$ to $\eps > 0$.
Then we can set $p$ and $q=p-k$ on-shell, i.e. $p^2=q^2=0$.
This results in vanishing of terms proportional to $(p^2)^\eps$ and $(q^2)^\eps$.
We obtain:
\begin{equation}\label{eq:def-wr}
  W_{R}^{(1+1)}\left(x,\del,\eps,\frac{k^2}{\mr^2}\right)
  =
  \lim_{\substack{p^2 \to 0 \\ q^2 \to 0}} 
  \left(
  W_{B}^{(1+1)}\left(x,\del,\eps,\frac{k^2}{\mr^2}\right)
  -
  W_{Z}^{(1)}\left(x,\del,\eps\right)
  \right)
  \text{.}
\end{equation}
Explicitly substituting expressions defined in the previous sections we get
\begin{multline}\label{eq:wbr}
  \Wb^{(1+1)} - \Wz^{(1+1)}
  =
  -i \: g^4 \: \C \: \Qv \: \frac{1}{\abs{k^2}} \;
  \Biggl\{
  \left(
    \frac{\Wirrq}{\eps^2}
    +
    \frac{\Wirq}{\eps}
    +
    \frac{\Wuvq}{\eps}
    +
    \Wzq
  \right)
  \left(\frac{\abs{k^2}}{\mr^2}\right)^{\eps}
  \\ +
  \frac{\Wuvk}{\eps}
  \left(\frac{q^2}{\mr^2}\right)^{\eps}
  +
  \frac{\Wuvp}{\eps}
  \left(\frac{p^2}{\mr^2}\right)^{\eps}
  -
  \frac{Z^{(1)} \; \Wr^{(1+0)}}{\eps}
  \Biggr\}
  \text{,}
\end{multline}
which in the IR limit leads to
\begin{multline}\label{eq:def-wRen}
  \Wr^{(1+1)}\left(x,\del,\eps\right)
  =
  -i \: g^4 \: \C \: \Qv \: \frac{1}{\abs{k^2}} \;
  \Biggl\{
  \left(
    \frac{\Wirrq}{\eps^2}
    +
    \frac{\Wirq}{\eps}
    +
    \frac{\Wuvq}{\eps}
    +
    \Wzq
  \right)
  \left(\frac{\abs{k^2}}{\mr^2}\right)^{\eps}
  \\ -
  \frac{\Wuvq + \Wuvp + \Wuvk}{\eps} \: \frac{\Wr^{(1+0)}}{W_{R,0}^{(1+0)}}
  \Biggr\}
  \text{.}
\end{multline}
Note, that at this point all poles are of the IR type, even if they originate from the UV counter-term.
This is so, because we set $p^2=q^2=0$, as discussed at length in \cite{CFP80}.

    \section{Real-Momentum Integration}

\subsection{One-particle phase space}
\label{subsec:1PS}

We are interested in the configuration with one on-shell particle in the final state, which is the case of the considered here virtual diagrams.
Taking into account definitions of Section \ref{subsec:momenta} the one-particle phase space element has the form
\begin{equation}
  \int \D \PS(k) =
  \int \frac{\Dm k}{(2\pi)^m} \; \delta\left(x - \frac{\kn}{\pn}\right)
  \int \Dm q \; 2\pi \; \delta^{+}(q^2) \; \delta^m(p-k-q)
  \text{.}
\end{equation}
Performing trivial integration over $\Dm q$ and taking into account that
\begin{equation} \label{eq:q2}
  q^2 = (p-k)^2 = \frac{k_\perp^2 - (1-x)k^2}{x}
  \text{,}
\end{equation}
which leads to
\begin{equation}
  \delta^{+}(q^2) = x \; \delta(\abs{k_\perp^2} - (1-x)\abs{k^2})
  \text{,}
\end{equation}
we get
\begin{equation}
  \int \D \PS(k) =
  \int \frac{\Dm k}{(2\pi)^m} \; \delta\left(x - \frac{\kn}{\pn}\right)
  2 \pi \; x \; \delta(\abs{k_\perp^2} - (1-x)\abs{k^2})
  \text{.}
\end{equation}
Since the integrand has no angular dependence, we find
\begin{equation}
  \int \Dm k = \Omega_{m-2} \; \int \: \D x \; \D \abs{k^2} \; \D \abs{k_\perp^2} \; \frac{\abs{k_\perp^2}^{\eps}}{4 x}
  \text{,}
\end{equation}
where
\begin{equation}
  \Omega_{m-2} = \frac{2 \pi^{1+\eps}}{\Gamma(1+\eps)}
\end{equation}
is the surface of a hypersphere in $m-2$ dimensions. 

A final expression for the one-particle real phase space in $m=4+2\eps$ dimensions reads
\begin{equation} \label{eq:def-ps}
  \int \D \PS(k) = \Qr \; (1-x)^{\eps} \int_0^{Q^2} \D \abs{k^2} \abs{k^2}^{\eps}
  \text{,}
\end{equation}
where
\begin{equation}\label{eq:qr}
  \Qr = \frac{1}{(4\pi)^{2+\eps}} \frac{1}{\Gamma(1+\eps)}
\end{equation}
and the upper integration limit is denoted as $Q^2$.
It is a dummy parameter in the construction of the inclusive splitting function,
as the final result does not depend on it.

\subsection{Integration variables}
\label{subsec:evolTime}


The variable we have chosen in the parametrization of \eqref{eq:def-ps} is closely related to the choice of the Monte-Carlo evolution time in the construction of the parton shower.
The common choices are virtuality (invariant mass) of the virtual quark, transverse momentum, and rapidity of the emitted real gluon.
The complete inclusive splitting functions should not depend on this choice, however at the exclusive level some of the distributions can differ.

In order to describe all the above three choices we define a new, generic, integration variable
\begin{equation}
  a_\sigma^2
  =
  \frac{\abs{k_\perp^2}}{(1-x)^\sigma}
  =
  \frac{\abs{k^2}}{(1-x)^{\sigma-1}}
  \text{,}
\end{equation}
where parameter $\sigma$ corresponds to the choices defined below:
\begin{equation}
\label{eq:sigma}
  \sigma = \left\{
    \begin{array}{l l l}
      0  \text{ -- transverse momentum,}
      \\
      1 \text{ -- virtuality,}
      \\
      2 \text{ -- rapidity,}
    \end{array}
    \right.
\end{equation}
since from \eqref{eq:q2} we have $q_\perp^2 = k_\perp^2 = (1-x)k^2$ and $k_\perp^2/(1-x)^2 = q_\perp^2/(1-x)^2 \sim q_+q_-/q_+^2 = q_-/q_+$.

Taking into account such a parametrization, the one-particle phase space of \eqref{eq:def-ps} can be re-expressed as
\begin{equation}
  \int \D \PS(a_\sigma)
  =
  \Qr \; (1-x)^{\sigma (1+\eps) - 1} \int_0^{Q^2} \D a_\sigma^2 (a_\sigma^2)^{\eps}
  \text{.}
\end{equation}

For the calculation of inclusive NLO splitting functions with one loop the following integral needs to be evaluated:
\begin{align} \label{eq:ps1-zeta}
& \frac{1}{\mf^{2\eps} }
   \int \frac{\D\PS(k)}{\abs{k^2}} \left(\frac{\abs{k^2}}{\mr^2}\right)^{\zeta \eps}
  \; = \;
  (1-x)^{(1-\sigma)(1-\zeta\eps)}
  \frac{1}{\mf^2}
  \int \frac{\D\PS(a_\sigma)}{a_\sigma^2}
  \left(\frac{a_\sigma^2}{\mr^2}\right)^{\zeta \eps}
\\ & \qquad = \;
  \Qr \; (1-x)^{\eps (\sigma + \zeta (\sigma - 1))}
  \frac{1}{\mf^{2\eps}}
  \int_0^{Q^2} \D a_\sigma^2 (a_\sigma^2)^{-1+\eps}
  \left(\frac{a_\sigma^2}{\mr^2}\right)^{\zeta \eps}
\\ & \qquad = \;
  \frac{\Qr}{(1+\zeta)\eps} \; (1-x)^{\eps (\sigma + \zeta (\sigma - 1))}
  \bigg(\frac{Q^2}{\mf^2}\bigg)^{\eps}
  \left(\frac{Q^2}{\mr^2}\right)^{\zeta \eps}
  \text{,}
\end{align}
where we introduce one more "marker" $\zeta$ to distinguish: $\zeta = 1$ for the bare part and $\zeta = 0$ for the counter-term part of the renormalized parton density (see \eqref{eq:def-wr}).
In \eqref{eq:ps1-zeta} we explicitly added also the factorization scale $\mf$ related to the real phase-space integration.
It is kept different from the renormalization scale $\mr$, coming from the loop integration.

\subsection{Real-momentum integration}
\label{sec:34-real-int}

Now we are ready to integrate exclusive parton density (\ref{eq:def-wRen}) over the real-momentum phase-space (\ref{eq:ps1-zeta}).
This way we obtain the {\it inclusive parton density}
\begin{multline}
  \Gh^{(1+1)}
  =
  -i g^4 \: \C \: \frac{\Qv}{\mf^{2\eps}}
  \int \frac{\D\PS(k)}{\abs{k^2}}
  \\ \times
  \left\{
    \left(
      \frac{\Wirrq}{\eps^2}
      +
      \frac{\Wirq + \Wuvq}{\eps}
      +
      \Wzq
    \right)
    \left(\frac{\abs{k^2}}{\mr^2}\right)^{\eps}
    -
    \frac{Z^{(1)} \; \Wr^{(1+0)}}{\eps}
  \right\}
\end{multline}

\begin{multline} \label{eq:gh}
  \Gh^{(1+1)}
  =
  \aspi^2 \: \C \: \frac{1}{(4\pi)^{2\eps}} \: \frac{\Gamma(1-\eps)}{\Gamma(1+\eps)} \:
  \frac{1}{4}
  \Biggl\{
    - \frac{Z^{(1)} \; W_{R}^{(1+0)}}{\eps^2}
    \left(
      (1-x)^{\sigma} \frac{Q^2}{\mf^2}
    \right)^\eps
    \\ + \frac{1}{2}
    \left(
      \frac{\Wirrq}{\eps^3}
      +
      \frac{\Wirq + \Wuvq}{\eps^2}
      +
      \frac{\Wzq}{\eps}
    \right)
    \left(
      (1-x)^{2\sigma-1} \frac{Q^4}{\mf^2 \mr^2}
    \right)^\eps
  \Biggr\}
  \text{,}
\end{multline}
where $\C$ is a corresponding color factor.

Next we make $\eps$-expansion and parametrize \eqref{eq:gh} in the following way:%
\begin{equation} \label{eq:gh-series}
  \Gh^{(1+1)}
  =
  \aspi^2 \: \C \: \frac{1}{(4\pi)^{2\eps}} \: \frac{\Gamma(1-\eps)}{\Gamma(1+\eps)} \:
  \frac{1}{4}
    \left\{
      \frac{\Gh_{-3}^{(1+1)}}{\eps^3}
      +
      \frac{\Gh_{-2}^{(1+1)}}{\eps^2}
      +
      \frac{\Gh_{-1}^{(1+1)}}{\eps}
      +
      O(\eps^0)
    \right\}
  \text{,}
\end{equation}
where:
\begin{align} \label{eq:gh-3}
  \Gh_{-3}^{(1+1)} = & \; \frac{\Wirrq}{2}
\text{,}\\ \label{eq:gh-2}
  \Gh_{-2}^{(1+1)}
  = & \; \frac{\Wirrq}{2} \; \ln\left( (1-x)^{2\sigma-1} \frac{Q^4}{\mf^2 \mr^2} \right) + \frac{\Wirq + \Wuvq}{2} - Z^{(1)} \; W_{R,0}^{(1+0)}
\text{,}\\ \label{eq:gh-1}
  \Gh_{-1}^{(1+1)}
  = & \;
    \frac{\Wirrq}{4}\ln^2\left( (1-x)^{2\sigma-1} \frac{Q^4}{\mf^2 \mr^2} \right)
  + \frac{\Wzq}{2} - Z^{(1)} \; W_{R,1}^{(1+0)}
  \\ &
  + \frac{\Wirq + \Wuvq}{2} \; \ln\left( (1-x)^{2\sigma-1} \frac{Q^4}{\mf^2 \mr^2} \right)
  - Z^{(1)} \; W_{R,0}^{(1+0)} \; \ln\left((1-x)^{\sigma} \frac{Q^2}{\mf^2}\right)
 \nonumber
\text{.}
\end{align}

In the following two paragraphs we will consider two special cases of eqs.~(\ref{eq:gh-3}--\ref{eq:gh-1}) for which we apply some additional assumptions that simplify $W$ form-factors.
These assumptions will be proven in explicit calculations later on.
For now we will assume them in order to show the analytic structure of the formulae.

\subsubsection{Topologies (c), (d), and (e)}

For the topologies (c), (d), and (e) we assume the following relation between UV and IR form-factors:
\begin{equation}\label{eq:ir-uv-rel}
  \Wuvk + \Wuvp - \Wirq = 0
  \text{.}
\end{equation}
By inspecting tables (\ref{tb:wc},\ref{tb:wd},\ref{tb:we}) it can be explicitly checked that the above relation is true in the $\NPV$ prescription, however it is not always the case in the $\PV$ prescription.
With the relation (\ref{eq:ir-uv-rel}) formulae (\ref{eq:gh-3}--\ref{eq:gh-1}) simplify considerably:
\begin{align} \label{eq:gh-3-i}
  \Gh_{-3}^{(1+1)}
  = & \; \frac{\Wirrq}{2}
  \text{,}
\\ \label{eq:gh-2-i}
  \Gh_{-2}^{(1+1)}
  = & \;
  - \frac{1}{2} \left( \Wuvq + \Wirq - \Wirrq \; \ln\left( (1-x)^{2\sigma-1} \frac{Q^4}{\mf^2 \mr^2} \right) \right)
  \text{,}
\\ \label{eq:gh-1-i}
  \Gh_{-1}^{(1+1)}
  = & \;
    \frac{1}{2} \; \Wzq - \frac{1}{2} \; Z^{(1)} \; W_{R,0}^{(1+0)} \; \left( \ln\frac{\mr^2}{\mf^2} + \ln(1-x) \right) - Z^{(1)} \; W_{R,1}^{(1+0)}
  \nonumber \\ &
  + \frac{1}{4} \; \Wirrq \; \ln^2\left( (1-x)^{2\sigma-1} \frac{Q^4}{\mf^2 \mr^2} \right)
  \text{.}
\end{align}
In the above we used the fact that thanks to \eqref{eq:ir-uv-rel} also the renormalization constant $Z^{(1)}$ of \eqref{eq:Z} simplifies to
\begin{equation}
  Z^{(1)}(x,\del) = \frac{\Wuvq + \Wirq}{W_{R,0}^{(1+0)}}
  \text{.}
\end{equation}

Additionally for the NPV prescription the form factor $\Wirrq$ vanishes and the formulae (\ref{eq:gh-3-i}--\ref{eq:gh-1-i}) are further simplified to
\begin{align} \label{eq:gh-3-ii}
  \Gh_{-3}^{(1+1)} = & \; 0 \text{,}
\\ \label{eq:gh-2-ii}
  \Gh_{-2}^{(1+1)}
  = & \;
  - \frac{\Wuvq + \Wirq}{2}
  \text{,}
\\ \label{eq:gh-1-ii}
  \Gh_{-1}^{(1+1)}
  = & \;
  \frac{\Wzq}{2} - \frac{\Wuvq + \Wirq}{2} \; \left( 2 \xi \: \frac{(1-x)^2}{1+x^2} + \ln\frac{\mr^2}{\mf^2} + \ln(1-x) \right)
  \text{.}
\end{align}

Number of comments related to eqs.~(\ref{eq:gh-3-i}--\ref{eq:gh-1-ii}) is in order here:
\begin{enumerate}
  \item The $Q^2$ dependence vanishes separately for each component of $\Gh$ in the NPV prescription.
  \item The same holds for the dependence on the integration variable described by the $\sigma$ parameter related to the evolution time.
  \item In the standard PV prescription the above properties hold {\em only} after adding the real and virtual contributions,
        which makes the results difficult to use in the Monte-Carlo applications.
  \item As can be seen from tables (\ref{tb:wc}--\ref{tb:we}) the sum $\Wirq+\Wuvq$ entering the definition of the renormalization constant $Z^{(1)}$ in \eqref{eq:Z} is always proportional to $W_{R,0}^{(1+0)} = (1+x^2)/(1-x)$, so that $Z^{(1)}$ reduces to simple function as expected.
\end{enumerate}

Let us mention that for those topologies a combinatorial factor of 2 should be included.


\subsubsection{Topologies (f) and (g)}

In the case of topologies (f) and (g) only the $\Wuvk$ form-factor is non-zero in both PV and NPV prescription.
This fact leads to the following results:
\begin{align}
  \Gh_{-3}^{(1+1)}
  = & \; 0
  \text{,}
\\
  \Gh_{-2}^{(1+1)}
  = & \; - \Wuvk
  \text{,}
\\
  \Gh_{-1}^{(1+1)}
  = & \; - \Wuvk \; \left( \xi \frac{(1-x)^2}{1+x^2} + \sigma \ln(1-x) + \ln\frac{Q^2}{\mf^2} \right)
  \text{.}
\end{align}

In contrast to the previous case, here neither the $Q^2$-dependence nor the evolution time related $\sigma$-dependence vanishes.
This will happen only after the matching real contributions are added at a fully inclusive level.
The physical and Monte-Carlo interpretations are also more complex because on the one hand these graphs can be seen as related to the running coupling.
On the other hand, the real graphs can be interpreted as parts of the final-state shower.

  \chapter{Results}
\label{ch:4}

\begin{figure}[ht]
  \centering

  \subcaptionbox*{\centerline{(\cqq): $\Cf^2-\frac{1}{2}\Cf\Ca$}}{\includegraphics[scale=1.15]{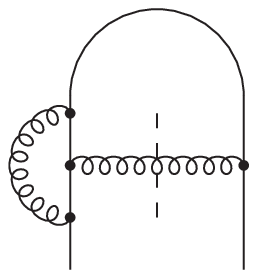}}
  \hspace{2cm}
  \subcaptionbox*{\centerline{(\dqq): $\frac{1}{2}\Cf\Ca$}}{\includegraphics[scale=1.15]{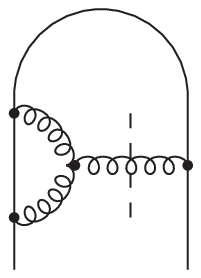}}
  \hspace{2cm}
  \subcaptionbox*{\centerline{(\dgg): $\Ca^2$}\vspace{1cm}}{\includegraphics[scale=1.15]{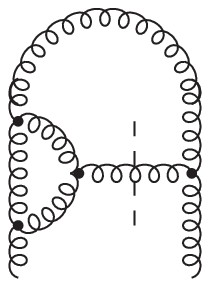}}
  \hspace{2cm}
  \subcaptionbox*{\centerline{\hspace{6mm}(\eqq): $\Cf^2$} }{\includegraphics[scale=1.15]{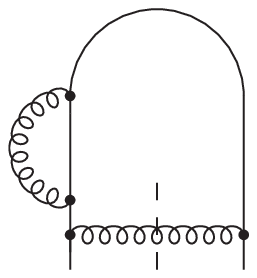}}
  \hspace{2cm}
  \subcaptionbox*{\centerline{(\fqq): $\Cf\Ca$}}{\includegraphics[scale=1.15]{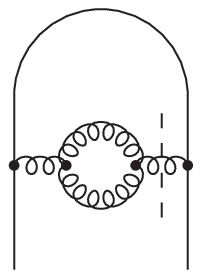}}
  \hspace{2cm}
  \subcaptionbox*{\centerline{(\gqq): $\Cf\Tf$}}{\includegraphics[scale=1.15]{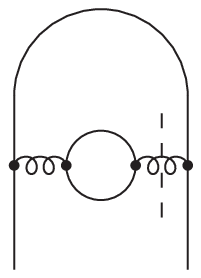}}

  \caption{NLO contributions to the one-loop splitting functions with corresponding color factors.}
  \label{fig:nlo_ns}
\end{figure}

In this chapter we will present results calculated in the NPV prescription for six graphs depicted in fig.~\ref{fig:nlo_ns}.
At first we provide results for each topology separately, we start with graph (c) and discuss it in detail.
For the remaining graphs we present results in the same format and comment only on the issues specific to a given topology.
Later we collect results featuring the same color factors and discuss the resulting distributions.


Presented results were obtained with the help of the \Axiloop package\footnote{The \Axiloop package is described in detail in Chapter \ref{ch:5}.} which we have written in Wolfram Mathematica language as a comprehensive tool for NLO calculations in the light-cone gauge.

\section{Topology (c)}
\label{sec:41}

We start the calculation from the definition of the non-integrated parton density given in \eqref{eq:wu-nlo}.
We use Feynman rules for the light-cone gauge as described in Chapter~\ref{ch:5} and listed in Appendix~\ref{ch:feynman-rules}.
The expression for the non-integrated parton density function of topology (c) in the form of \Axiloop code reads
\begin{lstlisting}
x G[n]/(4k.n)**FP[k]**FV[i1]**FP[l+k]**FV[mu]**FP[l+p]**
  FV[i2]**GP[i1,i2,l]**FPx[p]**GPx[mu,nu,p-k]**FV[nu]**FP[k]
\end{lstlisting}
where one can easily recognize the fermion propagator \lstinline{FP}, gluon propagator \lstinline{GP}, fermion-gluon vertex \lstinline{FV}, external cut lines (spin-averaged density matrices) \lstinline{FPx} and \lstinline{GPx} and fermion projector operator \lstinline{G[n]/(4k.n)}.

The color factor $\C$ can be easily calculated or taken from the literature, e.g. \cite[Table 2]{CFP80} or \cite[Figure 3.6]{Hei98}:
\begin{equation}
  \C^{(c)} = \Cf^2 - \frac{1}{2}\Cf\Ca
  \text{.}
\end{equation}

The next step is to perform trace over the corresponding fermion lines as defined in the projector operator $\P_F$, see \eqref{eq:pq}.
The result is quite long so we do not list it here.
However, this expression can be obtained with the help of the \Axiloop package, see file "NLO-C.ms" in the \Axiloop repository.

We proceed with loop integration as described in Section \ref{sec:loop-int}.
The result corresponding to \eqref{eq:Wb-series} for topology (c) is presented in Listing \ref{lst:c-loop}, where the \lstinline{Qv[r]} stands for the virtual phase-space factor as defined in \eqref{eq:qv} and the form-factors are listed in Appendix~\ref{sec:int-ir}.
Presented result is valid in both PV and NPV schemes, however it depends on the choice of the regularization prescription via form-factors \texttt{B0}, \texttt{B1}, etc.

Symbols \lstinline{eir} and \lstinline{euv} are numerically equal (\lstinline{eir = euv = eps}), however they specify a type of poles for a given form-factor, e.g. \lstinline{T0[eir] = -1/eir + 2}.
They are used to distinguish between the UV and IR \texttt{eps}-poles and can be canceled in the numerator and denominator with \lstinline{eps}, i.e. \lstinline{T0[eir] eps = -1 + 2 eps}, but the difference \lstinline{T0[eir] - T0[euv] = 1/euv - 1/eir} must be kept non-zero at this stage of calculation.

\begin{lstlisting}[caption=Exclusive Bare Parton Density (c).,label=lst:c-loop]
I g^4 / ((1-x)(-k.k)) (
  Qv[p](
    B0[euv] (-2(1+x^2 + (1-x)^2eps)) +
    B1[euv] 2x(x - (1-x)eps) +
    T0[euv] (3-2x^2 + (1-2x^2)eps - 2(1-x)eps^2)) +
  Qv[k](
    P0[euv] (-6(1+x^2 + (1-x)^2eps)) +
    T0[euv] x(4+5x - (2+x)eps + 2(1-x)eps^2) +
    R0[eir] (-2(1+x^2 + (1-x)^2eps)) +
    S0[eir] 2(1+x^2 + (1-x)^2eps) +
    T0[eir] (6-4x+x^2 + (4-8x+3x^2)eps - 2(2-3x+x^2)eps^2)) +
  Qv[q](
    K0[euv] (-2(1-x+x^2 + (1-3x+2x^2)eps)/(1-x)) +
    T0[euv] (1-x)(3-x + (1-3x)eps - 2(1-x)eps^2) +
    V1[euv] 2x^2(x - (1-x)eps) +
    V2[euv] 2x^2(x - (1-x)eps)))
\end{lstlisting}

We can now explicitly expand form-factors in Listing \ref{lst:c-loop} in the NPV prescription using expressions from Appendix \ref{sec:int-ir}.
The results --- form-factors $W$ from \eqref{eq:Wb-series} --- are presented in Table~\ref{tb:wc}.
Let us note that this decomposition into from-factors $W$ in the table and Listing~\ref{lst:c-loop} differs from the one in \cite[eqs.~3.110--3.111]{Hei98}.
It is not unique and we choose different set of integrals during the simplification procedure in \Axiloop.
An explicit example is given in Section~\ref{sec:53}.

The ultra-violet counter-term defined in \eqref{eq:ZF} equals
\begin{equation}
  W_{Z}^{(c)}
  \: = \;
  \as^2 \: \C^{(c)} \: \frac{\Gamma(1-\eps)}{(4\pi)^\eps} \: \frac{1}{\abs{k^2}} \: \frac{1}{\eps} \: (6 - 4\ln{x} - 8I_0) \: P_{qq}
  \text{,}
\end{equation}
where
\begin{equation}
  \Pqq(x,\eps) = \pqq + \eps \: \xi \: (1-x) \qquad \text{and} \qquad  \pqq = \frac{1+x^2}{1-x}
  \text{.}
\end{equation}

Subtracting this counter-term from the bare parton density and putting external momenta on shell (see \eqref{eq:wbr}) we obtain the renormalized parton density as defined in \eqref{eq:def-wRen}:
\begin{align}
  W_{R}^{(c)} = \; &
  \as^2 \: \C^{(c)} \: \frac{\Gamma(1-\eps)}{(4\pi)^\eps} \: \frac{1}{\abs{k^2}} \;
  \biggl\{
    \frac{1}{\eps} (6 - 4\ln{x} - 8I_0) \left( \left(\frac{\abs{k^2}}{\mr^2}\right)^\eps - 1 \right) \Pqq \nonumber \\
    & + \bigg( \pqq \: \left(-14 + 8 \Li(1) + 4 \ln^2{x} + 4 \Li(1-x) + 8 I_0 \ln{x} - 8 I_1\right)  \nonumber\\
    & \hspace{60mm} - (1-x) + (1+x) \bigg) \left(\frac{\abs{k^2}}{\mr^2}\right)^\eps
  \biggr\}
  \text{.}
\end{align}

Finally, after integration over the real momentum $q$, coefficients in the series of \eqref{eq:gh-series} read
\begin{align}
  \Gh_{-3}^{(c)} = & \; 0
  \text{,}
  \\
  \Gh_{-2}^{(c)} = & \; \pqq \left( -3 + 2 \ln{x} + 4 I_0 \right)
  \text{,}
  \\
  \Gh_{-1}^{(c)} = & \;
    \pqq \biggl\{ -7 + 4 \Li(1) + 2 \ln^2{x} + 2 \Li(1-x) + 4 I_0 \ln{x} - 4 I_1
     \nonumber\\ &  \hspace{50mm}+ \left( -3 + 2 \ln{x} + 4 I_0 \right) \left( \ln(1-x) + \ln\frac{\mr^2}{\mf^2} \right) \biggr\}
     \nonumber\\ &  + (1-x)\biggl\{-\frac{1}{2} + (2\xi-1)(-3 + 2 \ln{x} + 4 I_0)\biggr\} + \frac{1}{2} (1+x)
  \text{.}
\end{align}
This result agrees with the standard PV result given in \cite{Hei98} and there is no difference between the PV and NPV schemes.

\begin{table}[h]
  \centering
  \begin{tabular}{|l|c|c|c|c|c|}
    \hline
                           & $\Wuvq$ & $\Wuvp$ & $\Wuvk$ & $\Wirq$ & $\Wzq$
    \\ \hline
    \multicolumn{6}{|c|}{Topology (c)}
    \\ \hline
    $\pqq$                 & $9/2$   & $3/2$   & $0$     & $3/2$   & $-14$
    \\
    $\pqq \; \ln{x}$       & $-6$    & $0$     & $2$     & $2$     & $0$
    \\
    $\pqq \; \ln^2{x}$     & $0$     & $0$     & $0$     & $0$     & $4$
    \\
    $\pqq \;\: \Li(1)$       & $0$     & $0$     & $0$     & $0$     & $8$
    \\
    $\pqq \;\: \Li(1-x)$     & $0$     & $0$     & $0$     & $0$     & $4$
    \\
    $\phantom{(}1-x$       & $-2$    & $0$     & $2$     & $2$     & $5$
    \\
    $(1-x) \; \ln{x}$      & $0$     & $0$     & $0$     & $0$     & $-4$
    \\
    $\phantom{(}1+x$       & $-5/2$  & $3/2$   & $1$     & $5/2$   & $1$
    \\ \hline
    $\pqq \; I_0$          & $-6$    & $-2$    & $0$     & $-2$    & $0$
    \\
    $\pqq\;I_0\:\ln{x}$    & $0$     & $0$     & $0$     & $0$     & $8$
    \\
    $\pqq \; I_1$          & $0$     & $0$     & $0$     & $0$     & $-8$
    \\
    $(1-x) \; I_0$         & $0$     & $0$     & $0$     & $0$     & $-8$
    \\ \hline
  \end{tabular}
  \caption{Form-factors for the topology (c).}
  \label{tb:wc}
\end{table}

\newpage

\section{Topology (\dqq)}
\label{sec:dqq}

As advertised earlier the structure of this and following sections mimics Section~\ref{sec:41}.
Therefore we keep comments to the minimum and concentrate on the results.

The starting expression, the {\em non-integrated parton density}, for the topology (\dqq) in the form of \Axiloop code is presented in Listing~\ref{lst:d-def}.
\begin{lstlisting}[caption=Non-Integrated Parton Density for topology (\dqq).,label=lst:d-def]
x G[n]/(4 k.n)**FP[k]**FV[i1]**FP[l]**FV[i2]**GP[i1,i3,l+k]**
  GP[i2,i4,l+p]**GV[i3,-l-k,i4,l+p,mu,-p+k]**FPx[p]**
  GPx[mu,nu,p-k]**FV[nu]**FP[k]
\end{lstlisting}

The corresponding {\em color factor} for this topology is
\begin{equation}
  \C_{qq}^{(d)} = \frac{1}{2}\Cf\Ca \text{.}
\end{equation}

The {\em exclusive bare parton density} in terms of form-factors of Appendix~\ref{sec:int-ir} is
\begin{lstlisting}[caption=Exclusive Bare Parton Density (\dqq) in \Axiloop.,label=lst:d-ebs]
I g^4/((1-x)(-k.k)) (
  Qv[q] (
    C0[euv] (-6(1+x^2 + (1-x)^2eps)) +
    C1[euv] 4x(1+x) +
    T0[euv] (3-6x-5x^2 + 3(1-x)^2eps)) +
  Qv[p] (
    B0[euv] (-4(1+x^2 + (1-x)^2eps)) +
    B1[euv] 2x(x - (1-x)eps) +
    D0[euv] (-2(1+x^2 + (1-x)^2eps)) +
    T0[euv] (3-2x^2 + (3-2x-2x^2)eps)) +
  Qv[k] (
    E1[euv] (-2x^2(x-(1-x)eps)) +
    E2[euv] (-4x^2(x-(1-x)eps)) +
    E3[euv] (-2x^3(x-(1-x)eps)) +
    P0[euv] (-6(1+x^2 + (1-x)^2eps)) +
    T0[euv] x(2+7x - 3(2-x)eps) +
    R0[eir] 2(1+x^2 + (1-x)^2eps) +
    T0[eir] (6-2x-x^2 + (4-8x+3x^2)eps) +
    U0[eir] 2(1+x^2 + (1-x)^2eps)))
\end{lstlisting}

In Table~\ref{tb:wd} we present form-factors $W$ out of which all the main results can be built.
\begin{table}[h]
  \centering
  \begin{tabular}{|l|c|c|c|c|c|}
    \hline
                           & $\Wuvq$     & $\Wuvp$     & $\Wuvk$     & $\Wirq$     & $\Wzq$
    \\ \hline
    \multicolumn{6}{|c|}{Topology (\dqq)}
    \\ \hline
    $\pqq$                 & $9/2$       & $3/2$       & $0$         & $3/2$       & $-14$
    \\
    $\pqq \; \ln{x}$       & $-6$        & $2$         & $0$         & $2$         & $0$
    \\
    $\pqq \; \ln(1-x)$     & $0$         & $-2$        & $-6$        & $-8$        & $0$
    \\
    $\pqq \; \ln^2{x}$     & $0$         & $0$         & $0$         & $0$         & $4$
    \\
    $\pqq \;\: \Li(1)$     & $0$         & $0$         & $0$         & $0$         & $16$
    \\
    $\pqq \;\: \Li(1-x)$   & $0$         & $0$         & $0$         & $0$         & $-4$
    \\
    $\phantom{(}1-x$       & $-1$        & $0$         & $1$         & $1$         & $7$
    \\
    $(1-x) \; \ln{x}$      & $0$         & $0$         & $0$         & $0$         & $-4$
    \\
    $(1-x) \; \ln(1-x)$    & $0$         & $0$         & $0$         & $0$         & $-8$
    \\
    $\phantom{(}1+x$       & $-7/2$      & $3/2$       & $2$         & $7/2$       & $-1$
    \\ \hline
    $\pqq \; I_0$          & $-6$        & $-4$        & $-6$        & $-10$       & $0$
    \\
    $\pqq \; I_0 \ln{x}$   & $0$         & $0$         & $0$         & $0$         & $8$
    \\
    $\pqq \; I_0 \ln(1-x)$ & $0$         & $0$         & $0$         & $0$         & $8$
    \\
    $\pqq \; I_1$          & $0$         & $0$         & $0$         & $0$         & $-24$
    \\
    $(1-x) \; I_0$         & $0$         & $0$         & $0$         & $0$         & $-16$
    \\
    \hline
  \end{tabular}
  \caption{Form-factors for the topology (\dqq)}
  \label{tb:wd}
\end{table}
As in the case of topology (c), this decomposition into from-factors $W$ differs from the one in \cite[eqs.~3.152]{Hei98}, see Section~\ref{sec:53} for details.

The {\em ultra-violet counter-term} is
\begin{equation}
  W_{Z,qq}^{(d)}
  \: = \;
  \as^2 \: \C_{qq}^{(d)} \: \frac{\Gamma(1-\eps)}{(4\pi)^\eps} \: \frac{1}{\abs{k^2}} \: \frac{1}{\eps} \: (6 - 4\ln{x} - 8 \ln(1-x) - 16 I_0) \: P_{qq}.
\end{equation}

The {\em  exclusive (renormalized) parton density} is
\begin{align}
  & W_{R,qq}^{(d)}
  =
  \as^2 \: \C_{qq}^{(d)} \: \frac{\Gamma(1-\eps)}{(4\pi)^\eps} \: \frac{1}{\abs{k^2}} \:
  \biggl\{
    \frac{1}{\eps} (6 - 4\ln{x} - 8 \ln(1-x) - 16 I_0) \left( \left(\frac{\abs{k^2}}{\mr^2}\right)^\eps - 1 \right) \Pqq \nonumber \\
    & \qquad + \biggl( \pqq \left(-14 + 16 \Li(1) + 4 \ln^2{x} - 4 \Li(1-x) + 8 I_0 \ln{x} + 8 I_0 \ln(1-x) - 24 I_1 \right) \nonumber\\
    & \hspace{70mm} + (1-x) - (1+x) \biggr) \left(\frac{\abs{k^2}}{\mr^2}\right)^\eps
  \biggr\}
  \text{.}
\end{align}

Contributions to the {\em inclusive parton density} are
\begin{align}
  \Gh_{-3,qq}^{(d)} = & \; 0
  \text{,}
  \\
  \Gh_{-2,qq}^{(d)} = & \; \pqq \left( -3 + 2 \ln{x} + 4 \ln(1-x) + 8 I_0 \right)
  \text{,}
  \\
  \Gh_{-1,qq}^{(d)} = & \;
    \pqq \biggl\{ -7 - 2 \Li(1-x) + 2 \ln^2{x} + 8 \Li(1) - 12 I_1 + 4 I_0 \ln{x} + 4 I_0 \ln(1-x) \nonumber \\
     &  + \left( -3 + 2 \ln{x} + 4 \ln(1-x) + 8 I_0 \right) \left( \ln(1-x) + \ln\frac{\mr^2}{\mf^2} \right) \biggr\} \nonumber \\
     &  + (1-x) \biggl\{ \frac{1}{2} + (2\xi-1)(-3 + 2 \ln{x} + 4 \ln(1-x) + 8 I_0) \biggr\} - \frac{1}{2} (1+x)
  \text{.}
\end{align}

This result differs from the standard PV one of \cite{Hei98}.
The most important difference is that the term proportional to $1/\eps^3$, present in the PV prescription, is replaced in NPV by the contributions of $I_0/\eps^2$ and $I_1/\eps$.
This matches exactly the change in the real contribution presented in \cite[eq.~(3.48)]{JKSS11} in which all higher-order poles in $\eps$ are absent.
Once these two contributions (real and virtual) are added, we recover the standard PV result of \cite{CFP80, Hei98}.
This is strong confirmation of the correctness of the proposed NPV scheme.
Together with the similar result for the singlet graph (\dgg), see Section~\ref{sec:dgg}, it demonstrates that NPV is correct for both $\Pqq$ and $\Pgg$ splitting functions.

\newpage

\section{Topology (e)}

The starting expression, the {\em non-integrated parton density}, for the topology (\eqq) in the form of \Axiloop code is presented in Listing~\ref{lst:e-def}.
\begin{lstlisting}[caption=Non-Integrated Parton Density (e).,label=lst:e-def]
x G[n]/(4 k.n)**FP[k]**FV[i1]**FP[k-l]**GP[i1,i2,l]**FV[i2]**
  FP[k]**FV[mu]**FPx[p]**GPx[mu,nu,p-k]**FV[nu]**FP[k]
\end{lstlisting}

The corresponding {\em color factor} for this topology is
\begin{equation}
  \C^{(e)} = \Cf^2
  \text{.}
\end{equation}

The {\em exclusive bare parton density} in terms of form-factors of Appendix~\ref{sec:int-ir} is
\begin{lstlisting}[caption=Exclusive Bare Parton Density (e).]
I g^4/((1-x)k.k) (
  Qv[k] (
    P0[euv] (-8(1+x^2 + (1-x)^2eps)) + 
    T0[euv] 2(3(1+x^2) + 2(1-3x+x^2)eps - (1-x)^2eps^2)))
\end{lstlisting}

\begin{table}[h]
  \centering
  \begin{tabular}{|l|c|c|c|c|c|}
    \hline
                           & $\Wuvq$ & $\Wuvp$ & $\Wuvk$ & $\Wirq$ & $\Wzq$
    \\ \hline
    \multicolumn{6}{|c|}{Topology (e)}
    \\ \hline
    $\pqq$                 & $-6$    & $0$     & $0$     & $0$     & $14$
    \\
    $\pqq \; \ln{x}$       & $8$     & $0$     & $0$     & $0$     & $0$
    \\
    $\pqq \; \ln^2{x}$     & $0$     & $0$     & $0$     & $0$     & $-4$
    \\
    $\pqq \;\: \Li(1)$     & $0$     & $0$     & $0$     & $0$     & $-8$
    \\
    $\phantom{(}1-x$       & $0$     & $0$     & $0$     & $0$     & $-6$
    \\
    $(1-x) \; \ln{x}$      & $0$     & $0$     & $0$     & $0$     & $8$
    \\ \hline
    $\pqq \; I_0$          & $8$     & $0$     & $0$     & $0$     & $0$
    \\
    $\pqq \; I_0 \ln{x}$   & $0$     & $0$     & $0$     & $0$     & $-8$
    \\
    $\pqq \; I_1$          & $0$     & $0$     & $0$     & $0$     & $8$
    \\
    $(1-x) \; I_0$         & $0$     & $0$     & $0$     & $0$     & $8$
    \\
    \hline
  \end{tabular}
  \caption{Form factors for the topology (e).}
  \label{tb:we}
\end{table}

The {\em ultra-violet counter-term} reads
\begin{equation}
  W_{Z}^{(e)} \: = \; \as^2 \: \C^{(e)} \: \frac{\Gamma(1-\eps)}{(4\pi)^\eps} \: \frac{1}{\abs{k^2}} \: \frac{1}{\eps} \: (-6 + 8\ln{x} + 8I_0) \: P_{qq}.
\end{equation}

The {\em  exclusive (renormalized) parton density} is
\begin{multline}
  W_{R}^{(e)}
  \; = \;
  \as^2 \: \C^{(e)} \: \frac{\Gamma(1-\eps)}{(4\pi)^\eps} \: \frac{1}{\abs{k^2}} \;
  \biggl\{
    \frac{1}{\eps} (-6 + 8\ln{x} + 8I_0) \left( \left(\frac{\abs{k^2}}{\mr^2}\right)^\eps - 1 \right) P_{qq}
  \\
    + \pqq \: (14 - 4 \ln^2{x} - 8 \Li(1) - 8 I_0 \ln{x} + 8 I_1)
    \left(\frac{\abs{k^2}}{\mr^2}\right)^\eps
  \biggr\}
  \text{.}
\end{multline}

Contributions to the {\em inclusive parton density} are
\begin{align}
  \Gh_{-3}^{(e)} = & \; 0 \text{,}
  \\
  \Gh_{-2}^{(e)} = & \; \pqq \left( 3 - 4 \ln{x} - 4 I_0 \right) \text{,}
  \\
  \Gh_{-1}^{(e)} = & \;
    \pqq \biggl\{ 7 - 4 \Li(1) - 2 \ln^2{x} - 4 I_0 \ln{x} + 4 I_1
       \nonumber \\ & \hspace{40mm} + \left(3 - 4 \ln{x} - 4 I_0 \right) \left( \ln(1-x) + \ln\frac{\mr^2}{\mf^2} \right) \biggr\}
    \nonumber \\ & + (2\xi-1) (1-x)\left(3 - 4 \ln{x} - 4 I_0 \right).
\end{align}

This result agrees with the standard PV result given in \cite{Hei98} and there is no difference between the PV and NPV schemes.

\newpage

\section{Topology (f)}

The starting expression, the {\em non-integrated parton density}, for the topology (\fqq) in the form of \Axiloop code is presented in Listing~\ref{lst:f-def}.
\begin{lstlisting}[caption=Non-Integrated Parton Density (f).,label=lst:f-def]
x (G[n]/(4k.n))**FP[k]**FV[i1]**FPx[p]**FV[nu]**FP[k]**
  GPx[mu,nu,q] GP[i1,i2,q] GV[i2,-q,i3,-l,i4,l+q] GP[i3,i5,l]
  GP[i4,i6,l+q] GV[i5,l,mu,q,i6,-l-q];
\end{lstlisting}

The corresponding {\em color factor} for this topology is
\begin{equation}
  \C_{qq}^{(f)} = \Cf\Ca
  \text{.}
\end{equation}

The {\em exclusive bare parton density} in terms of form-factors of Appendix~\ref{sec:int-ir} is
\begin{lstlisting}[caption=Exclusive Bare Parton Density (f).]
I g^4/((1-x) k.k) (
   Qv[q] (
     C0[euv] (-16(2(1+x^2) + (3-4x+3x^2)eps
                                  + (1-x)^2eps^2))/(2+eps) + 
     C1[euv] 8x(1+x) +
     T0[euv] 8(2-x+x^2 + 2(1-x)^2eps) +
     T2[euv] (-8(1+x^2 + 2(1-x+x^2)eps
                                  + (1-x)^2eps^2))/(2+eps)))
\end{lstlisting}

\begin{table}[h]
  \centering
  \begin{tabular}{|l|c|c|c|c|c|}
    \hline
                           & $\Wuvq$ & $\Wuvp$ & $\Wuvk$ & $\Wirq$ & $\Wzq$
    \\ \hline
    \multicolumn{6}{|c|}{Topology (f)}
    \\ \hline
    $\pqq$                 & $0$     & $0$     & $-11/3$ & $0$     & $0$
    \\
    $\pqq \; \ln(1-x)$     & $0$     & $0$     & $4$     & $0$     & $0$
    \\
    \hline
    $\pqq \; I_0$          & $0$     & $0$     & $4$     & $0$     & $0$
    \\
    \hline
  \end{tabular}
  \caption{Form factors for the topology (f).}
  \label{tb:wf}
\end{table}

The {\em ultra-violet counter-term} reads
\begin{equation}
  W_{Z}^{(f)} \: = \; \as^2 \: \C^{(f)} \: \frac{\Gamma(1-\eps)}{(4\pi)^\eps} \: \frac{1}{\abs{k^2}} \: \frac{1}{\eps} \: \left(-\frac{11}{3} + 4 \ln(1-x) + 4 I_0\right) P_{qq}
  \text{.}
\end{equation}

The {\em  exclusive (renormalized) parton density} is
\begin{equation}
  W_{R}^{(f)}
  =
  \as^2 \: \C^{(f)} \: \frac{\Gamma(1-\eps)}{(4\pi)^\eps} \: \frac{1}{\abs{k^2}} \: \frac{1}{\eps} \:
    \left(\frac{11}{3} - 4 \ln(1-x) - 4 I_0 \right) P_{qq}
  \text{.}
\end{equation}

Contributions to the {\em inclusive parton density} are
\begin{align}\label{eq:gh-e-3}
  \Gh_{-3}^{(f)} = & \; 0
  \text{,}
  \\
  \Gh_{-2}^{(f)} = & \; \left( \frac{11}{3} - 4 \ln(1-x) - 4 I_0 \right) \; \pqq
  \text{,}
  \\ \label{eq:g1-f}
  \Gh_{-1}^{(f)} = & \;
    \left( \frac{11}{3} - 4 \ln(1-x) - 4 I_0 \right)
    \nonumber \\ & \; \times
     \left( \xi (1-x) + \sigma \; \pqq \ln(1-x) + \pqq \ln\frac{Q^2}{\mf^2} \right)
  \text{.}
\end{align}

In \eqref{eq:g1-f} one can notice the dependence on the $\sigma$-parameter related to the choice of the evolution time.
This is the example of the general situation described in Section~\ref{sec:34-real-int}.
As discussed there, this dependence should vanish once the real and virtual graphs are combined.
The same holds for the graph (\gqq) of the next section.
Note, that in the literature \cite{Hei98} only the virtuality choice is used ($\sigma=1$) and eqs.~(\ref{eq:gh-e-3})--(\ref{eq:g1-f}) result  agrees with the PV one given in \cite{Hei98} for the case of $\sigma = 1$ and $Q^2=\mf^2$.

\newpage

\section{Topology (g)}

The starting expression, the {\em non-integrated parton density}, for the topology (\gqq) in the form of \Axiloop code is presented in Listing~\ref{lst:g-def}.
\begin{lstlisting}[caption=Non-Integrated Parton Density (g).,label=lst:g-def]
x G[n]/(4 k.n)**FP[k]**FV[i1]**FPx[p]**FV[nu]**FP[k]
  FV[i2,Line->f2]**FP[l,Line->f2]**FV[mu,Line->f2]**
  FP[l+q,Line->f2] GP[i1,i2,q] GPx[mu,nu,q]
\end{lstlisting}

The corresponding {\em color factor} for this topology is
\begin{equation}
  \C_{qq}^{(g)} = \Cf \Tf \text{,} \quad \Tf = \frac{1}{2} n_f
  \text{.}
\end{equation}

The {\em exclusive bare parton density} in terms of form-factors of Appendix~\ref{sec:int-ir} is
\begin{lstlisting}[caption=Exclusive Bare Parton Density (g).]
-I g^4/((1-x)(-k.k)) (
  Qv[q] (
    T0[euv] (-4(1+x^2 + (1-x)^2eps)) +
    T2[euv] (8(1+x^2 + (1-x)^2eps))/(2+eps)))
\end{lstlisting}

\begin{table}[h]
  \centering
  \begin{tabular}{|l|c|c|c|c|c|}
    \hline
    & $\Wuvq$ & $\Wuvp$ & $\Wuvk$ & $\Wirq$ & $\Wzq$
    \\ \hline
    \multicolumn{6}{|c|}{Topology (g)}
    \\ \hline
    $\pqq$                 & $0$  & $0$ & $4/3$ & $0$ & $0$
    \\
    \hline
  \end{tabular}
  \caption{Form factors for the topology (g).}
  \label{tb:wg}
\end{table}

The {\em ultra-violet counter-term} reads
\begin{equation}
  W_{Z}^{(g)} = \; \as^2 \: \C^{(g)} \: \frac{\Gamma(1-\eps)}{(4\pi)^\eps} \: \frac{1}{\abs{k^2}} \: \frac{1}{\eps} \: \frac{4}{3} \: P_{qq}.
\end{equation}

The {\em  exclusive (renormalized) parton density} is
\begin{equation}
  W_{R}^{(g)} = \; \as^2 \: \C^{(g)} \: \frac{\Gamma(1-\eps)}{(4\pi)^\eps} \: \frac{1}{\abs{k^2}} \: \frac{1}{\eps} \left(-\frac{4}{3} \right) P_{qq}
  \text{.}
\end{equation}

Contributions to the {\em inclusive parton density} are
\begin{align}
  \Gh_{-3}^{(g)} = & \; 0 \text{,}
  \\
  \Gh_{-2}^{(g)} = & \: -4/3 \; \pqq \text{,}
  \\
  \Gh_{-1}^{(g)} = & \; -4/3 \left( \xi (1-x) + \sigma \; \pqq \ln(1-x) + \pqq \ln\frac{Q^2}{\mf^2} \right).
\end{align}
Note the presence of the $\sigma$ parameter that was discussed in detail in the previous section.

This result agrees with the PV one given in \cite{Hei98} for the case of $\sigma = 1$ and $Q^2=\mf^2$.

\newpage

\section{Topology (\dgg)}
\label{sec:dgg}

The NS results of the previous sections showed that the NPV prescription modified only the graph (\dqq).
It is the only non-singlet topology which in the PV prescription has the triple pole $1/\eps^3$.
On the other hand, the results for the singlet case, given in \cite{Hei98}, show this in that case also only the similar topology (\dgg), shown in Fig.~4.2 exhibits $1/\eps^3$ poles.
Therefore it will be the only one we expect to be modified in the NPV prescription.
Consequently, to verify the NPV prescription also for the singlet case of $P_{gg}$ we have calculated the virtual and real singlet graphs (\dgg) of Fig.~4.2.
The calculation of the virtual one closely followed the scheme described in Section~\ref{sec:41}.
The only differences were the appropriate projection operators of \eqref{eq:pg} and the trace replaced by Lorentz contraction.
The real graph was calculated in \cite{GJKS14} with the method described in \cite{JKSS11, Kus11}.
Below we give various intermediate results for the virtual graph (\dgg), in analogy to Section~\ref{sec:41}.

The {\em non-integrated parton density} in the form of \Axiloop code is
\begin{lstlisting}[caption=Non-Integrated (singlet) Parton Density (\dgg).,label=lst:dgg-def]
x (1-eps)/2 GPx[i1,i11,p] GP[i6,i7,k] ({i7}.{i8}) GP[i8,i9,k]
  GV[i1,-p,i2,p-l,i3,l] GP[i3,i4,l] GV[i4,-l,i6,k,i5,l-k]
  GP[i2,i12,p-l] GV[i12,l-p,i10,k-l,mu,q] GP[i5,i10,l-k]
  GV[i9,-k,nu,-q,i11,p] GPx[mu, nu, q]
\end{lstlisting}

The corresponding {\em color factor} for this topology is
\begin{equation}
  \C_{gg}^{(d)} = \Ca^2
  \text{.}
\end{equation}

The {\em exclusive bare parton density} in terms of form-factors of Appendix~\ref{sec:int-ir} is
\begin{lstlisting}[caption=Exclusive Bare Parton Density (\dgg) in \Axiloop.,label=lst:ebsf-dgg]
-I g^4/((1-x)(-k.k)) (
  Qv[p] (
    B0[euv] (-24(-1+eps^2)(1-x+x^2)^2)/x +
    B1[euv] 2(-1+2x)(-5+4x-4x^2 - eps + (6-4x+4x^2)eps^2) +
    D0[euv] (8(1-eps^2)(1-x+x^2)^2)/x +
    T0[euv] (-2(12-15x+8x^2+4x^3-8x^4 + x(1-2x)eps
            + 2(-6+7x-3x^2-2x^3+4x^4)eps^2))/x +
    T2[euv] 8(1-eps^2)(-2+x)/x) + 
  Qv[k] (
    E1[euv] (-2x(3-6x+5x^2-4x^3 + (1-x^2)eps
            + (-4+6x-4x^2+4x^3)eps^2)) + 
    E2[euv] (-4x(3-6x+5x^2-4x^3 + (1-x^2)eps
            + (-4+6x-4x^2+4x^3)eps^2)) + 
    E3[euv] (-2x^2(3-6x+5x^2-4x^3 + (1-x^2)eps
            + (-4+6x-4x^2+4x^3)eps^2)) +
    P0[euv] (48(1-eps^2)(1-x+x^2)^2)/x + 
    S0[eir] (8(-1 + eps^2)(1-x+x^2)^2)/x + 
    T0[eir] (4(-1+x)(x(-18+16x-3x^2) - 2(-4 + x + x^2)eps
            + (2-x)^2(-2+3x)eps^2))/x +
    T0[euv] 4(-1+eps^2)x(10-11x+11x^2) +
    T2[eir] (8(-1+eps)(-2+x)(-22+6x+3x^2+2x^3
            + 3(-8+x^2+x^3)eps + (-8+x^3)eps^2))/((2+eps)x) +
    T2[euv] (8(1-eps^2)(-3 + (-2+x)eps)x^2)/(2+eps) + 
    U0[eir] (8(-1+eps^2)(1-x+x^2)^2)/x) + 
  Qv[q] (
    C0[euv] (24(1-eps^2)(1-x+x^2)^2)/x + 
    C1[euv] 2(1+x)(-5+6x-5x^2 - (1-x)^2 eps
            + (6-8x+6x^2)eps^2) +
    K0[euv] (2(-4+12x-18x^2+12x^3-5x^4 + (-2+x)x^3eps
            + 2(2-6x+9x^2-5x^3+2x^4)eps^2))/((-1+x)x) +
    T0[euv] (-2(12-33x+35x^2-23x^3+x^4 - x(-1+x)^2(1+x)eps
            + 2(-6+17x-18x^2+11x^3)eps^2)/x) + 
    T2[euv] (8(-1+eps^2)(-2+x)(-1+x)^3)/x + 
    V1[euv] 2x(2-8x+7x^2-4x^3 + x(-2+x)eps
            + 2(-1+5x-4x^2+2x^3)eps^2) + 
    V2[euv] 2x(2-8x+7x^2-4x^3 + x(-2+x)eps
            + 2(-1+5x-4x^2+2x^3)eps^2)))
\end{lstlisting}

\begin{table}[h]
  \centering
  \begin{tabular}{|l|c|c|c|c|c|}
    \hline
                           & $\Wuvq$ & $\Wuvp$ & $\Wuvk$  & $\Wirq$ & $\Wzq$
    \\ \hline
    \multicolumn{6}{|c|}{Topology (\ggd)}
    \\ \hline
    $\pgg$                 & $-22$   & $-22/3$ & $0$      & $-22/3$ & $536/9$
    \\
    $\pgg \; \ln{x}$       & $24$    & $-4$    & $-4$     & $-8$   & $0$
    \\
    $\pgg \; \ln(1-x)$     & $0$     & $4$     & $12$     & $16$    & $0$
    \\
    $\pgg \; \ln^2{x}$     & $0$     & $0$     & $0$      & $0$     & $-16$
    \\
    $\pgg \;\: \Li(1)$     & $0$     & $0$     & $0$      & $0$     & $-48$
    \\
    $1/x$                  & $22$    & $-22/3$ & $-44/3$  & $-22$   & $0$
    \\
    $1$                    & $-22$   & $-2/3$  & $68/3$  & $22$    & $0$
    \\
    $x$                    & $24$    & $-2/3$  & $-70/3$ & $-24$   & $4/3$
    \\
    $x^2$                  & $0$     & $-22/3$ & $22/3$   & $0$     & $0$
    \\ \hline
    $\pgg \; I_0$          & $24$    & $12$    & $12$     & $24$    & $0$
    \\
    $\pgg \; I_0 \ln{x}$   & $0$     & $0$     & $0$      & $0$     & $-32$
    \\
    $\pgg \; I_0 \ln(1-x)$ & $0$     & $0$     & $0$      & $0$     & $-16$
    \\
    $\pgg \; I_1$          & $0$     & $0$     & $0$      & $0$     & $64$
    \\
    \hline
  \end{tabular}
  \caption{Form factors for the topology (\ggd).}
  \label{tb:wdgg}
\end{table}

The {\em ultra-violet counter-term} reads
\begin{equation}
  W_{Z,gg}^{(d)}
  \: = \;
  \as^2 \: \C_{gg}^{(d)} \: \frac{\Gamma(1-\eps)}{(4\pi)^\eps} \: \frac{1}{\eps} \: \frac{8}{\abs{k^2}} \: \left(\frac{11}{3} - 2\ln{x} - 2 \ln(1-x) - 6 I_0 \right) \Pgg.
\end{equation}
where
\begin{equation}
P_{gg} = \frac{(1-x+x^2)^2}{x(1-x)}.
\end{equation}
Note that $P_{gg}$ does not have a part proportional to $\eps$, in contrast to the case of~$P_{qq}$.

The {\em  exclusive (renormalized) parton density} is
\begin{align}
  & W_{R,gg}^{(d)}
  =
  \as^2 \: \C_{gg}^{(d)} \: \frac{\Gamma(1-\eps)}{(4\pi)^\eps} \: \frac{8}{\abs{k^2}} \:
  \biggl\{
    \frac{1}{\eps} \left(-\frac{11}{3} + 2\ln{x} + 2 \ln(1-x) + 6 I_0 \right) \left( 1 - \left(\frac{\abs{k^2}}{\mr^2}\right)^\eps \right) \Pgg \nonumber \\
    & \qquad -
    \left( \Pgg \left(\frac{67}{9} - 6 \Li(1) - 2 \ln^2{x} - 4 I_0 \ln{x} - 2 I_0 \ln(1-x) + 8 I_1 \right) + \frac{x}{6} \right)
    \left(\frac{\abs{k^2}}{\mr^2}\right)^\eps
  \biggr\}
  \text{.}
\end{align}

Contributions to the {\em inclusive parton density} are
\begin{align}
  \Gh_{-3,gg}^{(d)} = & \; 0
  \text{,}
  \\
  \Gh_{-2,gg}^{(d)} = & \; 4 \: \biggl\{ \pgg \left( -\frac{11}{3} + 2 \ln{x} + 2 \ln(1-x) + 6 I_0 \right) \biggr\}
  \text{,}
  \\
  \Gh_{-1,gg}^{(d)} = & \; 4 \: \biggl\{
    \pgg \left( -\frac{11}{3} + 2 \ln{x} + 2 \ln(1-x) + 6 I_0 \right) \left( \ln(1-x) + \ln\frac{\mr^2}{\mf^2} \right)
     \\ &  +
    \pgg \left( -\frac{67}{9} + 6 \: \Li(1) + 2 \ln^2{x} + 4 I_0 \ln{x} + 2 I_0 \ln(1-x) - 8 I_1 \right) - \frac{x}{6}
    \biggr\}
  \text{.} \nonumber
\end{align}

The comments to the above results are identical as given in the case of the non-singlet graph (\dqq), Section~\ref{sec:dqq}:
the $1/\eps^3$ terms vanish in the NPV scheme as compared to the PV results {\em but} once the real and virtual components are added, the NPV and standard PV results of \cite{EV96,Hei98} agree.
This verifies the correctness of the NPV scheme for the $\Pgg$ splitting functions.


\newpage

\section{Color structure $\Cf^2$}

In the following two sections we will present the results in yet another form --- grouped by the color factors.
This is the most practical form of the results.
It also exhibits cancellations of mass singularities.
In this section we will discuss $\Cf^2$ component which is built of topologies (c) and (e), see Fig.~\ref{fig:nlo_ns}.

\begin{table}[h]
  \centering
  \begin{tabular}{|l|c|c|c|c|c|}
    \hline
                           & $\Wuvq$ & $\Wuvp$ & $\Wuvk$ & $\Wirq$ & $\Wzq$
    \\ \hline
    \multicolumn{6}{|c|}{$\Cf^2$}
    \\ \hline
    $\pqq$                 & $-3/2$  & $3/2$   & $0$     & $3/2$   & $0$
    \\
    $\pqq \; \ln{x}$       & $2$     & $0$     & $2$     & $2$     & $0$
    \\
    $\pqq \;\: \Li(1-x)$   & $0$     & $0$     & $0$     & $0$     & $4$
    \\
    $\;1-x$                & $-2$    & $0$     & $2$     & $2$     & $-1$
    \\
    $(1-x) \; \ln{x}$      & $0$     & $0$     & $0$     & $0$     & $4$
    \\
    $\;1+x$                & $-5/2$  & $3/2$   & $1$     & $5/2$   & $1$
    \\ \hline
    $\pqq \; I_0$          & $2$     & $-2$    & $0$     & $-2$    & $0$
    \\ \hline
  \end{tabular}
  \caption{Form-factors for the $\Cf^2$ color structure}
  \label{tb:cf2}
\end{table}

Let us begin with the {\em exclusive parton density} of \eqref{eq:def-wRen}:
\begin{align}
  W_{R}^{(\Cf^2)} &= W_{R}^{(c)} + W_{R}^{(e)} \nonumber \\
  &=
  \as^2 \: \Cf^2 \: \frac{\Gamma(1-\eps)}{(4\pi)^\eps} \: \frac{1}{\abs{k^2}} \;
  \biggl\{
    \frac{1}{\eps} \: 4\ln{x} \left( \left(\frac{\abs{k^2}}{\mr^2}\right)^\eps - 1 \right) \Pqq \nonumber \\
    & + \Big( \pqq \; 4 \: \Li(1-x)  - (1-x) + (1+x) \Big) \left(\frac{\abs{k^2}}{\mr^2}\right)^\eps
  \biggr\}
  \text{.}
\end{align}

This formula is important for the practical applications in stochastic simulations.
After integrating out the real gluon phase space we get contributions to the inclusive parton density:
\begin{align}
  \Gh_{-3,virt}^{(\Cf^2)} = & \; 0 \text{,}
  \\
  \Gh_{-2,virt}^{(\Cf^2)} = & \: -\pqq \: 2 \ln{x} \text{,}
  \\
  \Gh_{-1,virt}^{(\Cf^2)} = & \; \pqq \: \biggl\{  2 \Li(1-x) - 2 \ln{x} \ln(1-x) - 2 \ln{x} \ln\frac{\mr^2}{\mf^2} \biggr\}
  \nonumber \\ & + (1-x) \biggl\{ -\frac{1}{2} - 2 \: (2\xi-1) \ln{x} \biggr\} + \frac{1}{2}(1+x) \text{.}
  \\
  \intertext{For completion we also give a contribution of the corresponding real graphs based on \cite{JKSS11}.}
  \Gh_{-1,real}^{(\Cf^2)} = & \; \pqq \: \biggl\{  -\frac{3}{2}\ln{x} - 2 \Li(1-x) \biggr\} + (1-x) \biggl\{-\frac{9}{2} + 3\ln{x}\biggr\}
  \nonumber  \\ & + (1+x) \biggl\{ -\frac{1}{2} - \frac{5}{2}\ln{x} - \frac{1}{2} \ln^2{x} \biggr\} \text{.}
\end{align}
As we see, the singularities $I_0$ and $I_1$ cancel between graphs, separately for virtual and real components.

\newpage

\section{Color structure $\Cf\Ca$ and $\Cf\Tf$}

In this section we analyze a combined contribution of the $\Cf\Ca$ and $\Cf\Tf$ color structures --- graphs (c), (\dqq), (f), and (g).
(Note that the (c) graph enters the color structure $\Cf\Ca$ with negative
sign).

\begin{table}[h]
  \centering
  \begin{tabular}{|l|c|c|c|c|c|}
    \hline
                           & $\Wuvq$     & $\Wuvp$     & $\Wuvk$     & $\Wirq$     & $\Wzq$
    \\ \hline
    \multicolumn{6}{|c|}{$\Cf\Tf$}
    \\ \hline
    $\pqq$                 & $0$         & $0$         & $4/3$       & $0$         & $0$
    \\ \hline
    \multicolumn{6}{|c|}{$\Cf\Ca$}
    \\ \hline
    $\pqq$                 & $0$         & $0$         & $-11/3$     & $0$         & $0$
    \\
    $\pqq \; \ln{x}$       & $0$         & $1$         & $-1$        & $0$         & $0$
    \\
    $\pqq \; \ln(1-x)$     & $0$         & $-1$        & $1$         & $-4$        & $0$
    \\
    $\pqq \;\: \Li(1)$     & $0$         & $0$         & $0$         & $0$         & $4$
    \\
    $\pqq \;\: \Li(1-x)$   & $0$         & $0$         & $0$         & $0$         & $-4$
    \\
    $\phantom{(}1-x$       & $1/2$       & $0$         & $-1/2$      & $-1/2$      & $1$
    \\
    $(1-x) \; \ln(1-x)$    & $0$         & $0$         & $0$         & $0$         & $-4$
    \\
    $\phantom{(}1+x$       & $-1/2$      & $0$         & $1/2$       & $1/2$       & $-1$
    \\ \hline
    $\pqq \; I_0$          & $0$         & $-1$        & $1$         & $-4$        & $0$
    \\
    $\pqq \; I_0 \ln(1-x)$ & $0$         & $0$         & $0$         & $0$         & $4$
    \\
    $\pqq \; I_1$          & $0$         & $0$         & $0$         & $0$         & $-8$
    \\
    $(1-x) \; I_0$         & $0$         & $0$         & $0$         & $0$         & $-4$
    \\
    \hline
  \end{tabular}
  \caption{Form-factors for the $\Cf\Ca$ color structure}
  \label{tb:cfca}
\end{table}

The ultra-violet counter-term defined in \eqref{eq:ZF} equals
\begin{equation}
\begin{split}
  W_{Z}^{(\Cf\Ca)}
  \: &= 
  -\frac{1}{2} W_{Z}^{(c)} + \frac{1}{2}W_{Z}^{(\text{\dqq})} + W_{Z}^{(f)} + W_{Z}^{(g)}\\
  &=\;
  \as^2 \: \Cf \: \frac{\Gamma(1-\eps)}{(4\pi)^\eps} \: \frac{1}{\abs{k^2}} \: \frac{1}{\eps} \: \left( - \Ca \frac{11}{3} + \Tf \frac{4}{3} \right) P_{qq}.
\end{split}
\end{equation}
In the above equation we can
recognize the $\beta_0$ term contributing to the 1-loop beta
function:
\begin{equation}
\beta_{1\text{loop}}(g) = \frac{g^3}{16\pi^2}\beta_0, \;\;\;\;
          \beta_0= \frac{4}{3}\Tf -\frac{11}{3}\Ca.
\end{equation} 
We will see the $\beta_0$ function in all the remaining formulae of this section as
well.

The {\em exclusive parton density} of \eqref{eq:def-wRen} reads
\begin{align}
  & W_{R}^{(\Cf\Ca)}
  = -\frac{1}{2} W_{R}^{(c)} + \frac{1}{2}W_{R}^{(\text{\dqq})} + W_{R}^{(f)} + W_{R}^{(g)} \nonumber \\
  & \quad =
  \as^2 \: \Cf \: \frac{\Gamma(1-\eps)}{(4\pi)^\eps} \: \frac{1}{\abs{k^2}} \:
  \biggl\{
    \frac{1}{\eps} \left( \Ca \frac{11}{3} - \Tf \frac{4}{3} - 4 \: \Ca \: \big(\ln(1-x) + I_0 \big) \left(\frac{\abs{k^2}}{\mr^2}\right)^\eps \right) \Pqq \nonumber \\
    & \qquad + 4 \: \Ca \biggl( \pqq \Big(\Li(1) - \Li(1-x) + I_0 \ln(1-x) - 2 I_1 \Big) - \frac{x}{2} \biggr) \left(\frac{\abs{k^2}}{\mr^2}\right)^\eps
  \biggr\}
  \text{.}
\label{eq:Gamma1_CACF_excl}
\end{align}

Finally, after integrating out the real gluon phase space, we get contributions to the {\em inclusive parton density}:
\begin{align}
  \Gh_{-2,virt}^{(\Cf\Ca)} = & \; \pqq \: \left\{ -\beta_0
                  - 2\:\Ca I_0^{(1-x)} \right\}
  \text{,}
  \\
  \label{eq:Gamma1_CACF}
  \Gh_{-1,virt}^{(\Cf\Ca)} = & 
  \; \pqq \: \biggl\{\biggl( -\beta_0 - 4\Ca I_0^{(1-x)} \biggr) \biggl(\sigma\ln(1-x) + \ln\frac{Q^2}{\mu_f^2}\biggr)+ 2 \: \Ca I_0^{(1-x)} \ln\frac{\mr^2}{\mf^2} \biggr\}
  \nonumber \\ &
  + \Ca \bigg\{ p_{qq}\biggl( 2\Li(1) - 2\Li(1-x)  -4I_1^{(1-x)}\biggr) - x
 -2 (1-x) I_0^{(1-x)} \bigg\}
    \nonumber \\ &
 - \beta_0(1-x)
  \text{.}
\end{align}
where
\begin{align}
I_0^{v} = I_0 +\ln v,
\;\;\;\;
I_1^{v} = I_1 -I_0\ln v -\frac{1}{2}\ln^2v.
\end{align}

In the above results we can see a number of differences when compared to
the similar formulae for the case of $C_F^2$ color factor. Let us comment on
them:
\begin{itemize}
\item
The results (\ref{eq:Gamma1_CACF_excl})--(\ref {eq:Gamma1_CACF}) have uncanceled soft singularities manifesting themselves as $I_0$ and $I_1$ terms.
These singularities will cancel only after adding the corresponding real contributions.
\item
The same holds for the $\ln({Q^2}/{\mu_f^2})$ terms, which introduce dependence on the upper phase space limit.
\item
The term related to the running coupling  is explicitly shown.
It is proportional to the $\beta_0$ times the LO kernel $P_{qq}$. 
\item
Finally, we note that the results depend on the choice of the integration variable (through the $\sigma$ parameter, see \eqref{eq:sigma}).
Once the real contributions are added this dependence is supposed to vanish.
However, at the exclusive level, relevant for the Monte-Carlo, the distributions differ.
\end{itemize}

\newpage

\section{Inclusive non-singlet case}

We complete presentation of the results with a summary Table~\ref{tb:inclusive} for all (virtual and  real) inclusive non-singlet contributions.
This table is normalized to match \cite[Table~1]{CFP80} in the PV prescription, where $m=4+\eps$ convention is used. 
In order to normalize one should multiply the lines "single poles" by $1/(2\epsilon)$ and the lines "double poles" by $1/(4\epsilon^2)$.
The color factors are shown explicitly for each graph ($\Tf = n_f/2$).
The table is given for the $\sigma=1$ case only, i.e.\ virtuality, as the integration variable. 
By comparing Table~\ref{tb:inclusive} with \cite[Table~1]{CFP80} it is easy to check that the presented NPV results, agree with the corresponding results in the PV scheme for all the inclusive sums of real and virtual graphs.
One should mention here that the double pole terms can be found in the preprint version of \cite{CFP80} available at CERN server (\href{http://cds.cern.ch/record/133945}{http://cds.cern.ch/record/133945}).

\newgeometry{margin=2.5cm} 
\begin{landscape}

\begin{table}
  \centering
  \begin{tabular}{|l||c|c|c||c|c|c||c||c|c|c||c|c|c|}
    \hline
    & \includegraphics[height=1.5cm,trim=0 0 0 -5]{nlo_d_n_v.eps}
    & \includegraphics[height=1.5cm,trim=0 0 0 -5]{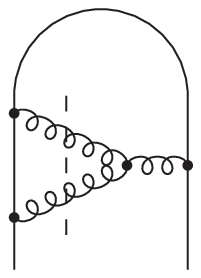}
    & SUM
    & \includegraphics[height=1.5cm,trim=0 0 0 -5]{nlo_c_n_v.eps}
    & \includegraphics[height=1.5cm,trim=0 0 0 -5]{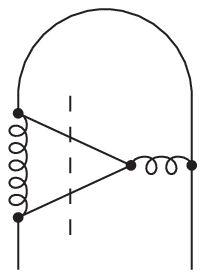}
    & SUM
    & \includegraphics[height=1.5cm,trim=0 0 0 -5]{nlo_e_n_v.eps}
    & \includegraphics[height=1.5cm,trim=0 0 0 -5]{nlo_f_n_v.eps}
    & \includegraphics[height=1.5cm,trim=0 0 0 -5]{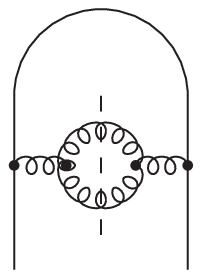}
    & SUM
    & \includegraphics[height=1.5cm,trim=0 0 0 -5]{nlo_g_n_v.eps}
    & \includegraphics[height=1.5cm,trim=0 0 0 -5]{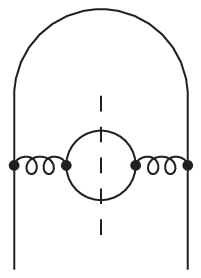}
    & SUM
    \\ \hline
      \multicolumn{1}{|c||}{}
    & \multicolumn{3}{|c||}{$(d):\;\sfrac{1}{2}\:\Cf\Ca$}
    & \multicolumn{3}{|c||}{$(c):\;\Cf^2 - \sfrac{1}{2}\:\Cf\Ca$}
    & \multicolumn{1}{|c||}{$(e):\;\Cf^2$}
    & \multicolumn{3}{|c||}{$(f):\;\sfrac{1}{2}\:\Cf\Ca$}
    & \multicolumn{3}{|c|}{$(g):\;\Cf\Tf$}
    \\ \hline \hline
    \multicolumn{14}{c}{\bf Double poles}
    \\ \hline
    $\pqq$                            & $-6$    & $0$   & $-6$   & $-6$   & $0$    & $-6$    & $6$ & $44/3$  & $-22/3$ & $22/3$  & $-8/3$ & $4/3$   & $-4/3$
    \\ 
    $p_{qq} \; \ln{x}$                & $4$    & $0$    & $4$    & $4$    & $0$    & $4$    & $-8$ & $0$     & $0$     & $0$     & $0$    & $0$     & $0$
    \\ 
    $p_{qq} \; \ln(1-x)$              & $8$    & $0$    & $8$    & $0$    & $0$    & $0$    & $0$  & $-16$    & $8$    & $-8$    & $0$    & $0$     & $0$
    \\
    $p_{qq} \; I_0$                   & $16$   & $0$    & $16$   & $8$    & $0$    & $8$    & $-8$ & $-16$   & $8$     & $-8$    & $0$    & $0$     & $0$
    \\ \hline \hline
\multicolumn{14}{c}{\bf Single poles}
    \\ \hline
    $p_{qq}$                          & $-7$   & $-4$   & $-11$  & $-7$   & $0$    & $-7$   & $7$  & $0$     & $103/9$ & $103/9$ & $0$    & $-10/9$ & $-10/9$
    \\
    $p_{qq} \; \ln{x}$                & $ 0$   & $-3/2$ & $-3/2$ & $0$    & $-3/2$ & $-3/2$ & $0$  & $0$     & $11/3$  & $11/3$  & $0$    & $-2/3$  & $-2/3$
    \\
    $p_{qq} \; \ln(1-x)$              & $-3$   & $8$    & $5$    & $-3$   & $0$    & $-3$   & $3$  & $22/3$  & $-34/3$ & $-4$    & $-4/3$ & $4/3$   & $0$
    \\
    $p_{qq} \; \ln^2{x}$              & $2$    & $-1$   & $1$    & $2$    & $-1$   & $1$    & $-2$ & $0$     & $0$     & $0$     & $0$    & $0$     & $0$
    \\
    $p_{qq} \; \ln{x} \ln(1-x)$       & $2$    & $4$    & $6$    & $2$    & $0$    & $2$    & $-4$ & $0$     & $-4$    & $-4$    & $0$    & $0$     & $0$
    \\
    $p_{qq} \; \ln^2(1-x)$            & $4$    & $-2$   & $2$    & $0$    & $0$    & $0$    & $0$  & $-8$    & $6$     & $-2$    & $0$    & $0$     & $0$
    \\
    $p_{qq} \;\: \Li(1)$              & $8$    & $-2$   & $6$    & $4$    & $0$    & $4$    & $-4$ & $0$     & $-4$    & $-4$    & $0$    & $0$     & $0$
    \\
    $p_{qq} \;\: \Li(1-x)$            & $-2$   & $2$    & $0$    & $2$    & $-2$   & $0$    & $0$  & $0$     & $0$     & $0$     & $0$    & $0$     & $0$
    \\
    $\phantom{(}1-x$                  & $-5/2$ & $3/2$  & $-1$   & $-7/2$ & $-15/2$ & $-11$ & $3$  & $22/3$  & $-4$    & $10/3$  & $-4/3$ & $0$     & $-4/3$
    \\
    $(1-x) \; \ln{x}$                 & $2$    & $0$    & $2$    & $2$    & $0$    & $2$    & $-4$ & $0$     & $0$     & $0$     & $0$    & $0$     & $0$
    \\
    $(1-x) \; \ln(1-x)$               & $4$    & $0$    & $4$    & $0$    & $0$    & $0$    & $0$  & $-8$    & $4$     & $-4$    & $0$    & $0$     & $0$
    \\
    $\phantom{(}1+x$                  & $-1/2$ & $1/2$  & $0$    & $1/2$  & $-1/2$ & $0$    & $0$  & $0$     & $0$     & $0$     & $0$    & $0$     & $0$
    \\
    $(1+x) \; \ln{x}$                 & $0$    & $1/2$  & $1/2$  & $0$    & $-7/2$ & $-7/2$ & $0$  & $0$     & $0$     & $0$     & $0$    & $0$     & $0$
    \\ \hline
\multicolumn{14}{|c|}{Spurious poles}
    \\ \hline
    $p_{qq} \; I_0$                   & $0$    & $8$    & $8$    & $0$    & $0$    & $0$    & $0$  & $0$     & $-4$    & $-4$    & $0$    & $0$     & $0$
    \\
    $p_{qq} \; I_0 \ln{x}$            & $4$    & $4$    & $8$    & $4$    & $0$    & $4$    & $-4$ & $0$     & $-4$    & $-4$    & $0$    & $0$     & $0$
    \\
    $p_{qq} \; I_0 \ln(1-x)$          & $12$   & $-4$   & $8$    & $4$    & $0$    & $4$    & $-4$ & $-8$    & $4$     & $-4$    & $0$    & $0$     & $0$
    \\
    $p_{qq} \; I_1$                   & $-12$  & $4$    & $-8$   & $-4$   & $0$    & $-4$   & $4$  & $0$     & $4$     & $4$     & $0$    & $0$     & $0$
    \\
    $(1-x) \; I_0$                    & $8$    & $0$    & $8$    & $4$    & $0$    & $4$    & $-4$ & $-8$    & $4$     & $-4$    & $0$    & $0$     & $0$
    \\ \hline
  \end{tabular}
  \caption{Contributions to the inclusive splitting function $P_{qq}$ from the real and virtual topologies in the NPV prescription.}
  \label{tb:inclusive}
\end{table}

\end{landscape}
\restoregeometry

  \chapter{The \Axiloop Package}
\label{ch:5}

\definecolor{lstgray1}{rgb}{0.85,0.85,0.85}
\definecolor{lstgray2}{rgb}{0.5,0.5,0.5}
\lstset{                          %
  aboveskip=3\medskipamount,
  backgroundcolor=\color{lstgray1},
  basicstyle=\footnotesize\ttfamily,     
  belowskip=3\medskipamount,
  frame=single,
  frameround=ffff,
  rulecolor=\color{lstgray2},
}

In this chapter we provide documentation for the \Axiloop package, exclusively written for calculating NLO contributions to splitting functions according to the calculation framework discussed in this thesis.


\subsubsection{Program Summary}

\begin{small}
\noindent
{\em Program Title:} \Axiloop                                 \\
{\em Version:} 2.3 (Mar 2014)                                 \\
{\em Licensing provisions:} GNU GPL v3                          \\
{\em Programming language:} Wolfram Mathematica               \\
{\em Computer:} x86-64                                           \\
{\em Operating system:} Linux, Mac OS X, Windows                                       \\
{\em RAM:} 256 MB                                              \\
{\em Number of processors used:} 1                              \\
{\em Keywords:} DGLAP splitting functions, axial gauge, one-loop, light-cone, infra-red, ultra-violet, spurious, principal value, new principal value  \\
{\em External routines/libraries:} \\
\hspace*{1cm} - \texttt{Tracer v1.1.1} (30 Dec 1991) \cite{JL93}           \\
{\em Running time:} 150 sec \\

\end{small}

\newpage

    \section{Overview}

Wolfram Mathematica language provides a set of mathematical routines developed by scientists and software engineers for more than 30 years, which form a solid framework for developing reliable software.
It is also a widespread tool for implementing analytical calculations and algorithms in high-energy physics many of which are open-sourced and freely available, such as \FeynCalc \cite{DM91}, \texttt{FeynArts} \cite{Hah01}, \texttt{LoopTools} \cite{HP99} and others.
Taking that into account, we chose Wolfram Mathematica language for implementing \Axiloop package and distribute it as free software.

The reason we started to develop \Axiloop from scratch is because the light-cone gauge is poorly (if at all) implemented in the existing software.
Moreover, we wanted to have a convenient and easy way to implement custom regularization prescriptions for loop integrals: dimensional regularization, PV and NPV prescriptions as well as their combinations.
Nevertheless, where possible we used third-party code, namely \Tracer package \cite{JL93} for vector and gamma algebras in arbitrary dimensions.
In addition, an independent auxiliary package \UnitTest was written for testing \Axiloop routines.

    \section{Installation}

The installation of \Axiloop package consists of two steps:
\begin{enumerate}
  \item
  Download \Axiloop source code from the official repository by running the following command in your shell (Linux only):
  \begin{lstlisting}[basicstyle=\ttfamily\footnotesize]
curl http://gituliar.org/axiloop/install.sh | sh
  \end{lstlisting}
  or clone a Git repository:
  \begin{lstlisting}
git clone https://github.com/gituliar/Axiloop.git
  \end{lstlisting}
  In both cases, a new folder `\lstinline{Axiloop}' with complete source code will be created in you current working directory.
  \item
  Register \Axiloop package, so that Mathematica Kernel knows its location.
  For that purpose append the following line into `\lstinline{~/.Mathematica/Kernel/init.m}' file in your home directory (where \lstinline{<path-to-axiloop>} is an absolute path to the `\lstinline{Axiloop}' folder, e.g.\ `\lstinline{/home/gituliar/src/}'):
  \newline
  \begin{lstlisting}
AppendTo[$Path, "<path-to-axiloop>"];
  \end{lstlisting}
\end{enumerate}

After these steps are completed successfully you can start Mathematica, load and start working with \Axiloop package, e.g.
\begin{lstlisting}
$ math
Mathematica 9.0 for Linux x86 (64-bit)
Copyright 1988-2013 Wolfram Research, Inc.

In[1]:= << Axiloop`;

In[2]:= Axiloop`$Version                                                                                                                                                                               

Out[2]= Axiloop 2.3 (Mar 2014)
\end{lstlisting}

    \section{User Manual}
\label{sec:53}

\subsection{Vector and Gamma Algebras}

Implementation of the vector and gamma algebras in arbitrary number of space-time dimensions is based on the \Tracer package \cite{JL93} and can be found in \code{GammaTrace.m} and \code{Tracer.m} files.
These files provide the following definitions:
\begin{itemize}
  \item {\em 4-vectors} are denoted with regular symbols, i.e.\ \code{k}, \code{p}, etc.
  \item {\em Vector indices} are denoted as one-element lists, i.e.\ \code{\{mu\}}, \code{\{nu\}}, etc.
  \item {\em $\gamma$-matrix} is denoted with symbol \code{G}, e.g.\ \code{G[\{mu\}]} is equivalent to $\gamma^\mu$ matrix.
  \item {\em Scalar product} of two vectors \code{a} and \code{b} is denoted by \code{S[a,b]} or, using a short "dot" notation, i.e.\ \code{a.b}.
  \item {\em Vector $p^\mu$} is defined as a scalar product, i.e.\ \code{p.\{mu\}}.
  \item {\em Metric tensor $g^{\mu\nu}$} is defined as a scalar product of two indices, i.e.\ \code{\{mu\}.\{nu\}}.
  \item {\em Non-commutative product of $\gamma$-matrices}, e.g.\ $\gamma^\mu \gamma^\nu \gamma^\xi$ can be written as \newline \code{G[\{mu\},\{nu\},\{xi\}]} which is equivalent to the explicit \code{G[\{mu\}]**G[\{nu\}]**G[\{xi\}]} expression.
  \item {\em Einstein summation convention} is assumed, so that any two repeated indices are automatically contracted, e.g.\ \code{k.\{mu\} p.\{mu\}} $\to$ \code{k.p}.
  \item {\em Number of dimensions} is denoted by \code{Global`d} symbol, so that \code{\{mu\}.\{mu\}} equals \code{Global`d}.
  \item {\em Slash-notation} is assumed in places where regular symbols are used instead of indices, e.g.\ \code{G[p,\{mu\}]} is equivalent to $\slashed{p}\gamma^\mu$ expression.
  \item \code{GammaTrace[expr, NumberOfDimensions -> 4 + 2 eps]} function calculates trace of the product of $\gamma$-matrices over all known spinor lines in the arbitrary number of dimensions ($4+2\eps$ by default). The spinor lines are recorded when \code{G[__]} matrix is used.
  \begin{lstlisting}
In[1]:= GammaTrace[G[{mu}]**G[{nu}]]

Out[1]= 4 g_{mu nu}

In[2]:= GammaTrace[G[k]**G[p]**G[k,Line->f2]**G[p,Line->f2]]
               
              2
Out[2]= 16 k.p

In[3]:= GammaTrace[G[k]**G[p]**G[k]**G[p]]
              
             2
Out[3]= 8 k.p  - 4 k.k p.p

  \end{lstlisting}
\end{itemize}

\subsection{Feynman Rules and Projectors}

Feynman rules defined in Appendix~\ref{ch:feynman-rules} are implemented in the \Axiloop with the help of the following routines:
\begin{itemize}
  \item \code{FP[p, Line -> f1]} and \code{FV[mu, Line -> f1]} define a fermion propagator and a quark-gluon vertex.
        Option \code{Line} defines a spinor line which is taken into account when a trace of the $\gamma$-matrices is performed (see usage of \code{GammaTrace} for details).
  \item \code{GP[mu,nu,p]} and \code{GV[i1,p1,i2,p2,i3,p3]} define a gluon propagator and a three-gluon vertex (with outgoing momenta lines).
  \item \code{FPx[p, Line -> f1]} and \code{GPx[mu,nu,p]} define cut (on-shell) propagators.
  \item \code{FPc}, \code{FVc}, \code{GPc}, and \code{GVc} define complex conjugated Feynman rules with arguments as defined for corresponding non-conjugated quantities.
\end{itemize}

Projectors defined in Section~\ref{subsec:proj} are implemented as
\begin{itemize}
  \item \code{PFi[p, Line -> f1]} and \code{PFo[p, Line -> f1]} define fermion projectors for the kernel's incoming and outgoing momenta.
  \item \code{PGi[mu,nu,p]} and \code{PGo[mu,nu]} define gluon projectors for the incoming and outgoing momenta.
\end{itemize}

Using routines of this section we can, for example, define a LO quark-quark kernel $K_{q\to q}^{(1+0)}$ (see eq.~\ref{eq:wu-general}) as
\begin{lstlisting}
x PFi[p]**FVc[nu]**FPc[k]**PFo[k]**FP[k]**FV[mu]**GPx[mu,nu,p-k]
\end{lstlisting}
This expression can be further used as an argument to \code{SplittingFunction} routine, defined later in this section.

\subsection{Loop-momenta integration}

The \code{IntegrateLoop[Wn, l]} function performs one-loop integration of non-integrated parton density \code{Wn} (see \eqref{eq:wu-nlo}) over the loop momentum \code{l}, discussed at length in Section~\ref{sec:loop-int}.

The function \code{IntegratedLoop} accepts the following options:
\begin{itemize}
  \item \code{Prescription -> "NPV"} which defines a prescription for the singularities of axial type.
        Accepted values are \code{"NPV"} (default) for the New Principal Value and \code{"PV"} for the Principal Value prescription correspondingly.
  \item \code{SimplifyNumeratorAndDenominator -> True} which determines if a numerator is canceled with a denominator where possible, to decrease a rank of tensor integrals.
\end{itemize}

Loop integration can be done not only for a non-integrated splitting function, but for any expression which contains Feynman and axial-type denominators (see eqs.~(\ref{eq:IF}--\ref{eq:IA})), for example
\begin{lstlisting}
In[1]:= IntegrateLoop[ 1/(l.l (l+p).(l+p)), l]

Out[1]= {
  {collected, $$[{}, {0, p}, {}]},
  {simplified, $$[{}, {0, p}, {}]},
  {integrated, {{short, Qv[p] T0[euv]},
                   Qv[p]
  {long, 2 Qv[p] - -----}}}}
                    euv
\end{lstlisting}
The term \code{Qv[p]} defines a one-loop phase space factor and is defined in \eqref{eq:qv}.

In the case of vector and tensor integrals \code{IntegrateLoop} can operate in two modes, which are controlled by \code{SimplifyNumeratorAndDenominator} option.
If \code{True} the following simplification rule is applied to the integrand:
\begin{equation}
  \frac{\sdot{l}{k}}{l^2 (l+k)^2 (l+p)^2} \to \frac{1}{2} \left( \frac{1}{l^2 (l+p)^2} - \frac{1}{(l+k)^2 (l+p)^2} - \frac{k^2}{l^2 (l+k)^2 (l+p)^2} \right)
  \text{.}
\end{equation}
This way three-point vector integrals are simplified to a simpler three-point scalar and two-point integrals, for example
\begin{lstlisting}
In[2]:= $Get[
  IntegrateLoop[ l.k/(l.l (l+k).(l+k) (l+p).(l+p)), l,
    SimplifyNumeratorAndDenominator -> True]
  ,
  {"integrated", "short"}
]

          Qv[k] R0[eir]   Qv[p] T0[euv]   Qv[q] T0[euv]
Out[2]= - ------------- + ------------- - -------------
                2               2               2
\end{lstlisting}

In the case of \code{False} value the above simplification is not applied and vector and tensor integrals are done using Passarino-Veltman reduction formulas presented in Appendix~\ref{sec:int-ir}.
\begin{lstlisting}
In[3]:= $Get[
  IntegrateLoop[ l.k/(l.l (l+k).(l+k) (l+p).(l+p)), l,
    SimplifyNumeratorAndDenominator -> False]
  ,
  {"integrated", "short"}
]


        Qv[k] (2 R2[eir] k.k + R1[eir] (k.k + p.p - q.q))
Out[3]= -------------------------------------------------
                              2 k.k
\end{lstlisting}

It is important to stress that the two above strategies for the integrand simplification lead to different $W$ form-factors defined in \eqref{eq:Wb-series}, as seen in the two examples above.
Despite of that the final result in the infra-red limit (when we put \code{p.p = 0} and \code{q.q = 0}) is equivalent in both approaches.

In addition to the above simplification rules applied to Feynman denominators, axial-type denominators can be simplified as well.
In particular, a product of two axial denominators can be expressed as
\begin{equation}
  \frac{1}{\sdot{(l+k)}{n}} \frac{1}{\sdot{(l+p)}{n}}
  \to
  \frac{1}{\sdot{(p-k)}{n}} \left( \frac{1}{\sdot{(l+k)}{n}} - \frac{1}{\sdot{(l+p)}{n}} \right)
  \text{.}
\end{equation}


\subsection{Final-state integration}

The \code{IntegrateLeg[Wr, k, NumberOfDimensions -> 4 + 2 eps]} function integrates renormalized parton density \code{Wr} (see \eqref{eq:def-wRen}) over the outgoing momentum \code{k}, as depicted in Fig.~\ref{fig:blob}.
The integration is performed in an arbitrary number of dimensions defined by \code{NumberOfDimensions} option which equals \code{4 + 2 eps} by default.
The lower and upper limit for the \code{k.k} is considered to be \code{-Q^2} and \code{0} respectively.

\begin{lstlisting}
In[1]:= WrLOqq = 2 g^2 (1+x^2 + (1-x)^2 eps)/((1-x) k.k);

In[2]:= IntegrateLeg[ WrLOqq, k ]

             2   2 eps        2           2
        - 2 g  (Q )    Qr (1+x + eps (1-x) ) (1 + eps Log[1-x])
Out[2]= --------------------------------------------------------
                               eps (1-x)
\end{lstlisting}
The \code{Qr} defines a real-momentum phase-space factor $\Qr$ defined in \eqref{eq:qr}.

\subsection{Splitting function calculation}

The \code{SplittingFunction[Wn, WrLO]} function calculates a splitting function up to the next-to-leading order.
The first argument \code{Wn} is a non-integrated parton density defined in \eqref{eq:wu-nlo}.
The second argument \code{WrLO} represents a corresponding exclusive leading-order parton density, which is needed to build an ultra-violet counter-term (see \eqref{eq:ZF}).
For examples on how to use \code{SplittingFunction} see \texttt{LO-*.ms} and \texttt{NLO-*.ms} files in the \Axiloop repository.
In these files one can find encoded calculations of all the graphs described in previous chapter.

  \chapter{Summary and Outlook}
\label{ch:6}

\subsubsection{Summary}

In this work we discussed calculation of the next-to-leading contributions to the one-loop non-singlet splitting function $P_{qq}$ and selected contributions to the singlet $P_{gg}$ splitting function.
Our goal was to re-calculate these contributions at exclusive level in a way suitable for NLO Monte-Carlo parton shower simulations and fulfilling two requirements:
\begin{itemize}
  \item consistency with the recently calculated exclusive NLO real contributions \cite{JKSS11}; 
  \item consistency with the inclusive NLO results of \cite{CFP80}.
\end{itemize}

The real NS contributions have been calculated in \cite{JKSS11} in such a way that the higher-order poles in $\eps$ were eliminated and the result was suitable for MC implementation.
These results turned out to be different from the standard ones available in the literature \cite{CFP80,Hei98} and the difference was related to the regularization of the soft singularities.
In order to fully understand, define, and justify this new regularization scheme it was  necessary to calculate the virtual contributions in a new way as well.
It was done in this work.

One of the important results of this thesis is the formulation of the New Principal Value (NPV) regularization prescription \cite{GJKS14}.
Results obtained with its help fulfill the above two requirements: are consistent with the real contributions \cite{JKSS11} and the sum of real and virtual ones is in a total agreement with inclusive calculations in standard Principal Value (PV) scheme \cite{CFP80,EV96,Hei98}. 
The main idea of the NPV prescription is the following: it treats {\em all} singularities in light-cone plus variable in the same way --- regularizing them by the PV prescription.
In contrast, the standard prescription uses PV method {\em only} for the axial-type denominators and dimensional regularization for the rest.
We explain it in detail in  Chapter~\ref{ch:calc}, where we also presented a complete procedure for calculated NLO virtual splitting functions in both PV and NPV prescriptions.
We also showed general formulae for inclusive (eqs.~\ref{eq:gh-3}--\ref{eq:gh-1}) exclusive bare (\eqref{eq:Wb-series}) and renormalized (\eqref{eq:def-wRen}) splitting functions as well as ultra-violet counter-terms (eqs.~\ref{eq:Z} and \ref{eq:ZF}).
Finally, we showed how the choice of the integration variable (related to the evolution time in the stochastic methods) influence the results.
We have not found such an explicit result in the literature.

With the help of these general formulae we gave in Chapter~\ref{ch:4} the results for the complete non-singlet $P_{qq}$ splitting function and for the graph (\dgg) to the $P_{gg}$ splitting function in which the NPV results differ from the PV ones.
By comparing the obtained inclusive results with literature \cite{CFP80,JKSS11} we showed correctness of the NPV prescription: after adding all the contributions (real and virtual) the inclusive splitting functions in NPV and standard PV schemes are identical, both for $P_{qq}$ and $P_{gg}$ splitting functions.

The benefits of the new scheme are:
\begin{itemize}
  \item The $1/\eps^3$ and some of $1/\eps^2$ terms are eliminated from all partial contributions.
  As a result, there is no need to cancel them between real and virtual pieces.
  \item Calculation of real corrections is much simpler, as in most cases it can be done in four dimensions. 
  \item This scheme is compatible with the stochastic methods as the regulator
$\del$ has a meaning of geometrical cut-off and, contrary to the dimensional $\eps$,
can be simulated in a Monte-Carlo program.
\end{itemize}

The drawback of the NPV scheme is that Feynman integrals (without axial-type denominator $1/\nl$) start to depend on the auxiliary axial vector $n$ and become  more complicated.

Calculations of the NLO splitting functions in the NPV prescription presented in this thesis were performed with the help of the \Axiloop package \cite{Axiloop}.
The \Axiloop package has been developed by us especially for this purpose, to support the loop integration in the light-cone gauge, and it is also an important result of this thesis.
It is available under the GNU license at \href{http://gituliar.org/axiloop/}{http://gituliar.org/axiloop/}.


The \Axiloop package is written in Wolfram Mathematica language.
It is dedicated to automatic calculation of NLO contributions to the one-loop splitting functions in PV and NPV prescriptions.
In particular, it contains a complete library of integrals in both the NPV and PV schemes (we also provide these integrals in Appendix~\ref{sec:int-ir}). 
Some modules of the package are universal enough to be used for other tasks, e.g.  vector algebra and trace calculation, performing one-loop integration in the light-cone gauge, various simplification algorithms for those integrals and others.

\subsubsection{Outlook}

The natural extension of this work is to calculate remaining contributions to the singlet splitting function $P_{gg}$ in the NPV scheme.
For this set of contributions we found in the literature \cite{FP80,EV96,Hei98} only the inclusive results.
The \Axiloop package is ready to perform these calculations, as all the necessary  integrals and procedures are already implemented there.

Another interesting task is to analyze the fully virtual NLO contributions to splitting functions in the NPV prescription.
These results are routinely obtained from the sum rules.
To our knowledge, their direct calculation has been done in the ML prescription and for selected contributions only \cite{BHKV98}.
The complete calculation in PV-based prescription is still an open challenge.


%
%

\chapter*{Acknowledgments}
\addcontentsline{toc}{chapter}{Acknowledgments}

First of all I would like to thank my advisers, Maciej Skrzypek and Aleksander Kusina, for fruitful discussions and their invaluable experience I could profit from.

I thank to the directorate of IFJ PAN, in particular to the head of the division of theoretical physics Stanisław Jadach, the head of international PhD studies Tadeusz Lesiak, and the director of the institute Marek Jeżabek, for the financial support and the opportunity to live and graduate in Poland.

I also thank to my friends and colleagues from the institute for their support and presence I enjoyed a lot, in particular Amanda Bartkowiak, Andreas van Hameren, Agnieszka Karczmarska, Kamil Klimkiewicz, Marta Kulij, Krzysztof Kutak, Ewelina Lipiec, and Piotr Morawski.

Special thanks go to my friends who opened for me an exciting world of sport climbing, in particular to Piotr Morawski, Piotr Suder, Maciej Wolak, and others who shared with me their motivation and passion.

Finally, I would like to thank to my parents and family for their faith and mental support.

This work is partly supported by the Polish National Science Center grants DEC-2011/03/B/ST2/02632 and  UMO-2012/04/M/ST2/00240.

  \appendix
  \chapter{Feynman Rules for QCD}
\label{ch:feynman-rules}

In this Appendix we collect Feynman rules for the massless QCD in the light-cone gauge \cite[Appendix 3]{BNS91}, used in the \Axiloop package.

The {\em quark propagator}:
\begin{equation}
  \mathrm{FP}(p)
  =
  \frac{i \slashed{p}}{p^2 + i \veps}
  \text{.}
\end{equation}

The {\em gluon propagator} is
\begin{equation}
  \mathrm{GP}_{ab}^{\mu\nu}(p)
  =
  \frac{i \del_{ab} d^{\mu\nu}(p)}{p^2 + i \veps}
  \text{,}
\end{equation}
where the {\em Lorentz tensor} in the light-cone gauge reads
\begin{equation}
  d_{\mu\nu}(p) = - g^{\mu\nu} + \frac{p^\mu n^\nu + n^\mu p^\nu}{\sdot{p}{n}} \quad \text{and} \quad n^2=0
  \text{.}
\end{equation}

The {\em quark-gluon vertex} is
\begin{equation}
  \mathrm{FV}_{a}^{\mu}
  =
  i \gs (2\pi)^m T_a \gamma^\mu
  \text{.}
\end{equation}

The {\em three-gluon vertex} is
\begin{equation}
  \mathrm{GV}_{abc}^{\mu\nu\rho}(p,q,r)
  =
  i \gs f_{abc} \del^m(p+q+r)
  \left(
    g^{\mu\nu}(p-q)^\rho + g^{\nu\rho}(q-r)^\mu + g^{\rho\mu}(r-p)^\nu
  \right)
  \text{.}
\end{equation}

The {\em four-gluon vertex} is
\begin{multline}
  \mathrm{GW}_{abcd}^{\mu\nu\sigma\rho}(p,q,s,r)
  =
  -i\gs^2 \del^m(p+q+s+r)
  (
    f_{eab} f_{ecd} (g^{\mu\rho} g^{\nu\sigma} - g^{\mu\sigma} g^{\nu\rho})
    \\ +
    f_{eac} f_{edb} (g^{\mu\sigma} g^{\rho\nu} - g^{\mu\nu} g^{\rho\sigma})
    +
    f_{ead} f_{ebc} (g^{\mu\nu} g^{\sigma\rho} - g^{\mu\rho} g^{\sigma\nu})
  )
  \text{.}
\end{multline}

In all these rules the convention is that:
\begin{itemize}
  \item all momenta from vertices are outgoing;
  \item each vertex contains a four-momentum conserving $\del$-function;
  \item each internal line is accompanied by the integral over its four-momentum; 
  \item for every closed quark loop an additional factor $-1$ should be included; 
  \item for every group of $k$ vertices playing the same role a symmetry factor $1/k!$ should be included.
\end{itemize}

  \chapter{One-Loop Integrals} \label{ch:integrals}

Such a denominator is singular at some points of phase space and thus should be properly regularized.
For our purposes we choose Principal Value (PV) prescription, though other choices are possible.
Singularities which arise from the axial denominator we call spurious.
They are unphysical, since they have gauge-dependent origin, and should not contribute to the final result.

Below we present two approaches to calculating integrals \ref{eq:IF}--\ref{eq:IA}.
The first one (PV) is rather standard and leads to the known results presented in many papers.
The other one, NPV, is new and was developed in the context of this work.

    \section{Parametrization Techniques}


A technique proposed by R.~Feynman looks as follows:
\begin{equation}
  \frac{1}{D_1 \ldots D_n}
  =
  (n-1)!
  \int_0^1 \D z_1 \ldots \D z_n
    \frac{\del (1 - z_1 - \cdots - z_n)}
         {(z_1 D_1 + \dots + z_n D_n)^n}
  \text{.}
\end{equation}
Such a transformation leads to the expression which can be integrated over the loop-momentum $l$ using standard formulas obtained by G.~t'Hooft [...].

Below we provide explicit parametrization formulas for 2- and 3-point loop integrals, which are frequently used in this work.

\begin{equation}
  \frac{1}{((l+k_1)^2 + i\veps) ((l+k_2)^2 + i\veps)}
     = \int_0^1 \D z \;
       \frac{1}
            {(l^2+2\sdot{l}{k}+M^2 + i\veps)^2}
  \text{,}
\end{equation}
where
\begin{gather}
  k^\mu = z k_1^\mu + (1-z) k_2^\mu \text{,} \quad
  M^2 = z k_1^2 + (1-z) k_2^2 \text{,} \subeq{a} \\
  \text{and} \nonumber \\
  M^2-k^2 = z(1-z)\,(k_1-k_2)^2 \subeq{b}
  \text{.}
\end{gather}

\begin{equation}
  \frac{1}{(l^2 + i\veps) ((l+k_1)^2 + i\veps) ((l+k_2)^2 + i\veps)}
     = \int_0^1 \D z_1 \D z_2\;
       \frac{2 z_1}
            {(l^2+2\sdot{l}{k}+M^2 + i\veps)^2}
  \text{,}
\end{equation}
where
\begin{gather}
  k^\mu = z_1 z_2 k_1^\mu + (1-z_1) k_2^\mu \text{,} \quad
  M^2 = z_1 z_2 k_1^2 + (1-z_1)\,k_2^2 \text{,} \subeq{a} \\
  \text{and} \nonumber \\
  M^2-k^2 = z_1^2 z_2 (1-z_2) k_1^2 + z_1 (1-z_1)(1-z_2) k_2^2 + z_1 z_2 (1-z_1)(k_1-k_2)^2 \subeq{b}
  \text{.}
\end{gather}

    \section{Dimensionally Regularized Integrals} \label{sec:md-integrals}

We list Feynman and axial-type integrals in the arbitrary number of dimensions.
These integrals are useful to properly identify IR and UV singularities as will be explained in the Appendix \ref{sec:int-ir}.

\subsubsection{Feynman Integrals}

Formulas \ref{eq:int-f-0}--\ref{eq:int-f-2} were taken from \cite[Appendix B]{tHV73}.

\begin{align}\label{eq:int-f-0}
  & \int \D^ml \frac{1}{(l^2 + 2\sdot{l}{k} + M^2)^\alpha}
  \; = \;
  \frac{\Gamma\left(\alpha - \frac{m}{2}\right)}{\Gamma(\alpha)}
  \frac{i \pi^{\sfrac{m}{2}}}{(M^2 - k^2)^{\alpha - \sfrac{m}{2}}}
  \text{,}
\\
  & \int \D^ml \frac{l^\mu}{(l^2 + 2\sdot{l}{k} + M^2)^\alpha}
  \; = \;
  \frac{\Gamma\left(\alpha - \frac{m}{2}\right)}{\Gamma(\alpha)}
  \frac{i \pi^{\sfrac{m}{2}}}{(M^2 - k^2)^{\alpha - \sfrac{m}{2}}}
  (- k^\mu)
  \text{,}
\\
  & \int \D^ml \frac{l^\mu l^\nu}{(l^2 + 2\sdot{l}{k} + M^2)^\alpha}
  \nonumber
\\ \label{eq:int-f-2}
  & \qquad \qquad = \;
  \frac{\Gamma\left(\alpha - \frac{m}{2}\right)}{\Gamma(\alpha)}
  \frac{i \pi^{\sfrac{m}{2}}}{(M^2 - k^2)^{\alpha - \sfrac{m}{2}}}
    \left(
      k^\mu k^\nu
    + \frac{g^{\mu\nu}}{2}\frac{M^2-k^2}{\alpha - \sfrac{m}{2} - 1}
    \right)
  \text{.}
\end{align}

\subsubsection{Axial-Type Integrals in the PV Prescription}

Integrals \ref{eq:int-a-0}--\ref{eq:int-a-2} were taken from \cite[eqs. B.29--B.31]{Pok00}.
Note that formulas \ref{eq:int-a-1}--\ref{eq:int-a-2} and higher-order tensor integrals can be obtained by taking the derivative of \eqref{eq:int-a-0} over the momentum $k$.

\begin{align} \label{eq:int-a-0}
  & \int \D^ml \frac{1}{(l^2 + 2\sdot{l}{k} + M^2)^\alpha} \frac{\lnp}{\lnp^2 + \dpp^2}
  \; = \;
  \frac{\Gamma\left(\alpha - \frac{m}{2}\right)}{\Gamma(\alpha)}
  \frac{i \pi^{\sfrac{m}{2}}}{(M^2 - k^2)^{\alpha - \sfrac{m}{2}}}
  \frac{\knp}{\knp^2 + \dpp^2}
  \text{,}
\\ \label{eq:int-a-1}
  & \int \D^ml \frac{l^\mu}{(l^2 + 2\sdot{l}{k} + M^2)^\alpha} \frac{\lnp}{\lnp^2 + \dpp^2}
  \nonumber
\\
  & \qquad = \;
  \frac{\Gamma\left(\alpha - \frac{m}{2}\right)}{\Gamma(\alpha)}
  \frac{i \pi^{\sfrac{m}{2}}}{(M^2 - k^2)^{\alpha - \sfrac{m}{2}}}
  \frac{\knp}{\knp^2 + \dpp^2}
  \left(
    k_\mu
    -
    \frac{1}{2}
    \frac{M^2-k^2}{\alpha-\sfrac{m}{2}-1}
    \frac{\knp^2-\dpp^2}{\knp^2+\dpp^2}
    \frac{n_\mu}{\knp}
  \right)
  \text{,}
\\ \label{eq:int-a-2}
  & \int \D^ml
    \frac{l^\mu l^\nu}{(l^2 + 2\sdot{l}{k} + M^2)^\alpha}
    \frac{\lnp}{\lnp^2 + \dpp^2}
\nonumber \\
  & \qquad = \;
  \frac{\Gamma\left(\alpha - \frac{m}{2}\right)}{\Gamma(\alpha)}
  \frac{i \pi^{\sfrac{m}{2}}}{(M^2 - k^2)^{\alpha - \sfrac{m}{2}}}
  \frac{\knp}{\knp^2 + \dpp^2}
  \bigg(
    k^\mu k^\nu
    +
    \frac{1}{2}
    \frac{M^2-k^2}{\alpha-\sfrac{m}{2}-1}
    g^{\mu\nu}
  \nonumber \\ & \qquad \qquad + \;
    \frac{1}{2}
    \frac{M^2-k^2}{\alpha-\sfrac{m}{2}-1}
    \frac{\knp^2 - \dpp^2}{\knp^2 + \dpp^2}
    \frac{k^\mu n^\nu + n^\mu k^\nu}{\knp}
  \nonumber \\ & \qquad \qquad + \;
    \frac{1}{2}
    \frac{(M^2-k^2)^2}{(\alpha-\sfrac{m}{2}-1)(\alpha-\sfrac{m}{2}-2)}
    \frac{\knp^2 - 3 \dpp^2}{(\knp^2 + \dpp^2)^2}
    n^\mu n^\nu
  \bigg)
  \text{.}
\end{align}

In order to preserve dimensional structure of the axial denominator, we introduced an abbreviation $\dpp=\del\,\pnp$, where $\del$ is a dimensionless PV regulator and $p$ some external reference momentum \footnote{In this work as $p$ we choose a momentum of the incoming quark, see \eqref{eq:def-p}.}.
The "plus" notation for momentum variables stands for $\rnp=\sdot{r}{n}$ as everywhere in this work.

    \section{Integrals in the Infra-Red Region}
\label{sec:int-ir}

In this appendix we present one-loop integrals used to calculate NLO virtual splitting functions.
We define general $n$-point Feynman integrals by
\begin{equation} \label{eq:IF}
  J^{\mu_1 \ldots \mu_s}_F(k_1 \ldots k_n) = 
    \int \D^ml
    \frac{l^{\mu_1} \ldots l^{\mu_s}}{((l+k_1)^2+i\eps) \ldots ((l+k_n)^2+i\eps)}
\end{equation}
and axial-type integrals with denominator $1/\lnp$ by
\begin{equation} \label{eq:IA}
  J^{\mu_1 \ldots \mu_s}_A(k_1 \ldots k_n;r) = 
    \int \D^ml
    \frac{l^{\mu_1} \ldots l^{\mu_s}}
         {((l+k_1)^2+i\eps) \ldots ((l+k_n)^2+i\eps)}
    \frac{1}{\lnp+\rnp}
    \text{.}
\end{equation}
The complete list of all the integrals needed in the standard PV scheme is given in \cite{Hei98}.
As compared to \cite{Hei98} we provide also some additional tensor integrals.
They can be reduced to scalar ones by using Passarino-Veltman reduction method.
We provide results of such a reduction which are valid in both PV and NPV schemes.

In the infra-red region we impose some momenta to go on-shell, so that everywhere in this appendix we assume $p^2=(p-k)^2=0$ and $k^2\neq0$.

\subsection{General Integrals}

General integrals \ref{eq:int-npv-2}--\ref{eq:int-npv-3} were taken from \cite{EV96}.
In these formulas the only integration left to perform is over $\lnp=\sdot{l}{n}$, the plus component of the loop momentum.
That makes these expression useful in calculation of integrals in the NPV scheme.

Two- and three-point integrals read as
\begin{equation} \label{eq:int-npv-2}
  \int \Dm l
    \frac{f(l_{+})}{(l^2+i\veps) ((l-k)^2+i\veps)}
  \\ =
  - \frac{\Qv(k)}{\eps}
  \int_0^1 \D z \: f(l_{+}) z^{\eps} (1-z)^{\eps}
\end{equation}
and
\begin{multline} \label{eq:int-npv-3}
  \int \Dm l
    \frac{f(l_{+})}{(l^2+i\veps) ((l-k)^2+i\veps) ((l-p)^2+i\veps)}
  \\=
  \frac{\Qv(k)}{\eps}
  \frac{1}{\abs{k^2}}
  \bigg(
    \int_0^x \D y f(l_{+}) z^{\eps} (1-z)^{-1+\eps}
    \:_2F_{1}\left( 1-\eps,1; 1+\eps, \frac{z(1-x)}{z-1} \right)
    \\+
    2 \: \frac{\Gamma^2(1+\eps)}{\Gamma(1+2\eps)}
    \int_x^1 \D y f(l_{+}) (1-y)^{-1+2\eps}
  \bigg)
  \text{,}
\end{multline}
where $f$ is an arbitrary function, $m=4+2\eps$, and the rescaled momentum values $x=\knp/\pnp$, $y=\lnp/\pnp$, and $z=y/x=\lnp/\knp$.
The one-loop phase-space factor reads
\begin{equation} \label{eq:qv}
  \Qv(r) = \Qv \; \abs{r^2}^\eps = \frac{i}{(4\pi)^{2+\eps}} \; \Gamma(1-\eps) \; \abs{r^2}^\eps
  \text{.}
\end{equation}

\subsection{Scalar Integrals}

All the scalar integrals needed for the calculations done in Chapter~\ref{ch:4} are listed below.
At first we parametrize integrals in terms of form factors.

\subsubsection{Two-Point Feynman Integrals}

\begin{equation}
  \boxed{J_{F}(0,r) = \Qv(r) \; T_0}
\end{equation}
\begin{flalign}
  T_0 = -\frac{1}{\euv} + 2 &&
\end{flalign}

\subsubsection{Two-Point Axial Integrals}

\begin{equation}
  \boxed{J_{A}(0,k;0) = - \frac{\Qv(k)}{\kn} \; P_0}
\end{equation}
\begin{flalign}
  P_0 = \frac{ -I_0 - \ln{x}}{\euv} - I_1 + I_0 \ln{x} + \frac{\ln^2x}{2} + \Li(1) &&
\end{flalign}

\begin{equation}
  \boxed{J_{F}(p,q;0) = - \frac{\Qv(k)}{\pn} \; E_0}
\end{equation}
\begin{flalign}
  E_0 = \frac{1}{\euv} \frac{\ln(1-x)}{x} &&
\end{flalign}

\subsubsection{Three-Point Feynman Integrals}

\begin{equation} \label{eq:int-R0}
  \boxed{J_{A}(0,k,p) = \frac{\Qv(k)}{\abs{k^2}} \; R_0}
\end{equation}
\begin{flalign}
  R_0^{NPV} = - \frac{2 I_0 + \ln(1-x)}{\eir} - 4 I_1 + 2 I_0 \ln(1-x) + \frac{\ln^2(1-x)}{2} &&
\end{flalign}

\subsubsection{Three-Point Axial Integrals}

\begin{equation} \label{eq:int-S0}
  \boxed{J_{A}(0,k,p;0) = \frac{\Qv(k)}{\abs{k^2}} \frac{1}{\pn} \; S_0}
\end{equation}
\begin{flalign} \label{eq:S0}
  S_0^{NPV} = & - \frac{3 I_0 - \ln{x} + \ln(1-x)}{\eir} - 5 I_1 + I_0 \ln{x} + 2 I_0 \ln(1-x) \nonumber \\
              & + 2 \Li(1-x) + \frac{\ln^2{x}}{2} + \frac{\ln^2(1-x)}{2} + \Li(1) &&
\end{flalign}

\begin{equation} \label{eq:int-u0}
  \boxed{J_{A}(0,k,p;k) = \frac{\Qv(k)}{\abs{k^2}} \frac{1}{\qn} \; U_0}
\end{equation}
\begin{flalign}
  U_0^{NPV} = & - \frac{3 I_0 - \ln{x} + 3 \ln(1-x)}{\eir} - 5 I_1 + I_0 \ln{x} + 2 I_0 \ln(1-x) \nonumber \\
              & - 2 \Li(1-x) + \frac{\ln^2{x}}{2} - \frac{\ln^2(1-x)}{2} + 5 \Li(1) &&
\end{flalign}

\begin{equation} \label{eq:int-w0}
  \boxed{J_{A}(0,k,p;p) = \frac{\Qv(k)}{\abs{k^2}} \frac{1}{\qn} \; W_0}
\end{equation}
It turns out that form factor $W_0$ cancels with corresponding terms from vector and tensor integrals $W_1$ and $W_4$.
For that reason we do not provide here an explicit expression for $W_0$ form factor.

\subsection{Vector Integrals}

\subsubsection{Two-Point Feynman Integrals}

\begin{equation}
  \boxed{J_{F}^{\mu}(0,r) = \Qv(r) \; T_1 \; r^\mu}
\end{equation}
\begin{flalign}
  T_1 = \frac{T_0}{2} = -\frac{1}{2\euv} + 1 &&
\end{flalign}

\subsubsection{Two-Point Axial Integrals}

\begin{equation}
  \boxed{J_{A}^{\mu}(p,q;0) = \frac{\Qv(k)}{\pn} \left( E_1 p^\mu + E_2 k^\mu + \frac{k^2}{2 \pn} E_3 n^\mu \right)}
\end{equation}
\begin{flalign}
  E_1 = &  \frac{1}{\euv} \frac{\ln(1-x)}{x} - \frac{2 \Li(x) + \ln^2(1-x)/2}{x}
  \\
  E_2 = & - \frac{1}{\euv} \frac{x + \ln(1-x)}{x^2} + \frac{2 \Li(x) + \ln^2(1-x)/2 - 2x}{x^2}
  \\
  E_3 = & - \frac{1}{\euv} \frac{(x-2)\ln(1-x) - 2x}{x^3} + \frac{4x + (x-2)(2 \Li(x) + \ln^2(1-x)/2)}{x^3} &&
\end{flalign}

\subsubsection{Three-Point Feynman Integrals}

\begin{equation}
  \boxed{ J_{F}^{\mu}(0,k,p) = \frac{\Qv(k)}{\abs{k^2}} \left( R_1 \; p^\mu + R_2 k^\mu \right) }
\end{equation}
\begin{flalign}
  R_1 & = - R_0 - 2 T_0
  &
  R_2 & = T_0
  &&
\end{flalign}
\begin{flalign}
  R_1^{PV} & = \frac{1}{\eir^2} + \frac{2}{\eir} + 4 - \frac{\pi^2}{6}
  &
  R_2^{PV} & = - \frac{1}{\eir} - 2
  &&
\end{flalign}
\begin{flalign}
  R_1^{NPV} & = \frac{2 - 2 I_0 - \ln(1-x)}{\eir} - 4 I_1 + 2 I_0 \ln(1-x) + \frac{\ln^2(1-x)}{2} - 4
  \\
  R_2^{NPV} & = - \frac{1}{\eir} + 2
  &&
\end{flalign}

\subsubsection{Three-Point Axial Integrals}

\begin{equation}
  \boxed{ J_{A}^{\mu}(0,k,p;0) = \frac{\Qv(k)}{\abs{k^2}} \frac{1}{\pn} \left( S_1 \; p^\mu + S_2 k^\mu + \frac{k^2}{2\sdot{k}{n}} S_3 n^\mu \right) }
\end{equation}
\begin{flalign}
  S_1 & = \frac{(2-x) R_0 - x(P_0 + S_0)}{2(1-x)}
  \\
  S_2 & = \frac{P_0 - R_0 + S_0}{2(1-x)}
  \\
  S_3 & = - \frac{(2-x) P_0 - x(R_0 - S_0)}{2(1-x)} &&
\end{flalign}
\begin{flalign}
  S_1^{NPV} = & \left(2 I_0 + \ln(1-x) + \frac{x \ln{x}}{1-x}\right) \frac{1}{\eir} + 4 I_1 - 2 I_0 \ln(1-x) \nonumber \\
              & + \frac{x \Li(1-x)}{1-x} - \frac{\ln^2(1-x)}{2}
  \\
  S_2^{NPV} = & \frac{\ln{x}}{1-x} \; \frac{1}{\eir} - \frac{\Li(1-x)}{1-x}
  \\
  S_3^{NPV} = & - \left(I_0 + \frac{\ln{x}}{1-x}\right) \frac{1}{\euv} + I_1 - I_0 \ln{x} + \frac{x \Li(1-x)}{1-x} - \frac{\ln^2{x}}{2} - \Li(1) &&
\end{flalign}

\begin{equation}
  \boxed{ J_{A}^{\mu}(0,k,p;k) = \frac{\Qv(k)}{\abs{k^2}} \frac{1}{\qn} \left( U_1 \; p^\mu + U_2 k^\mu + \frac{k^2}{2\sdot{k}{n}} U_3 n^\mu \right) }
\end{equation}
\begin{flalign}
  U_1 & = \frac{(2-x) R_0 + x(P_0 + U_0)}{2}
  \\
  U_2 & = \frac{U_0 - R_0 - P_0}{2}
  \\
  U_3 & = \frac{(2-x) P_0 + x(R_0 - U_0)}{2} &&
\end{flalign}
\begin{flalign}
  U_1^{NPV} = &   \frac{2 I_0 - x \ln{x} + (1 + x) \ln(1-x)}{\eir} + 4 I_1 - 2 I_0 \ln(1-x) \nonumber \\
              & + x \Li(1-x) - (1-x) \frac{\ln^2(1-x)}{2} - 2 x \Li(1) \\
  U_2^{NPV} = &   \frac{I_0 + \ln(1-x)}{\eir} + I_1 - I_0 \ln{x} + \Li(1-x) - \frac{\ln^2{x}}{2} \nonumber \\
              & + \frac{\ln^2(1-x)}{2} - 3 \Li(1) \\
  U_3^{NPV} = & - \frac{I_0 + (1-x)\ln{x} + x \ln(1-x)}{\euv} - I_1 + I_0 \ln{x} - x \Li(1-x) \nonumber \\
              & + \frac{\ln^2{x}}{2} - x \frac{\ln^2(1-x)}{2} + (1+2x)\Li(1) &&
\end{flalign}

\begin{equation}
  \boxed{ J_{A}^{\mu}(0,k,p;p) = \frac{\Qv(k)}{\abs{k^2}} \frac{1}{\qn} \left( W_1 \; p^\mu + W_2 k^\mu + \frac{k^2}{2\sdot{k}{n}} W_3 n^\mu \right) }
\end{equation}
\begin{flalign}
  W_1 & = \frac{(2-x) R_0 + x^2 E_0 - 2 x W_0}{2}
  \\
  W_2 & = - \frac{R_0 + x E_0}{2}
  \\
  W_3 & = \frac{R_0 + (2-x) E_0}{2} &&
\end{flalign}

\subsection{Tensor Integrals}

\subsubsection{Two-Point Feynman Integrals}

\begin{equation}
  \boxed{ J_{F}^{\mu\nu}(0,r) = \Qv(r) \left( T_2 r^\mu r^\nu + T_3 r^2 g^{\mu\nu} \right) }
\end{equation}
\begin{flalign}
  T_2 & = - \frac{1}{3 \euv} + \frac{13}{18}
      &
  T_3 & = \frac{1}{12 \euv} - \frac{2}{9}
  &&
\end{flalign}

\subsubsection{Three-Point Feynman Integrals}

\begin{equation} \label{eq:r36}
  \boxed{ J_{F}^{\mu\nu}(0,k,p) = \frac{\Qv(k)}{\abs{k^2}} \left( R_3 p^\mu p^\nu + R_4 k^\mu k^\nu + R_5 \{kp\}^{\mu\nu} + R_6 k^2 g^{\mu\nu} \right) }
\end{equation}
\begin{flalign}
  R_3 & = R_0 + \frac{3+2\eps}{1+\eps} T_0
  &
  R_4 & = - \frac{T_0}{2}
  \\
  R_5 & = - \frac{T_0}{2(1+\eps)}
  &
  R_6 & = \frac{T_0}{4(1+\eps)}
  &&
\end{flalign}
\begin{flalign}
  & R_3^{NPV} = \frac{-3 + 2 I_0 + \ln(1-x)}{\eir} + 4 I_1 - 2 I_0 \ln(1-x) - \frac{\ln^2(1-x)}{2} + 7
  \\
  & R_3^{PV} = \frac{1}{\eir^2} - \frac{\pi^2}{6}
  \\
  & R_4 = \frac{1}{2 \eir} - 1
  \\
  & R_5 = \frac{1}{2 \eir} - \frac{3}{2}
  \\
  & R_6 = - \frac{1}{4 \euv} + \frac{3}{4} &&
\end{flalign}

\begin{empheq}[box=\fbox]{multline}
    J_{F}^{\mu\nu\eta}(0,k,p)
    = \frac{\Qv(k)}{\abs{k^2}} \Big(
      H_1 \; p^\mu p^\nu p^\eta
    + H_2 \; \{ppk\}^{\mu\nu\eta}
    + H_3 \; \{pkk\}^{\mu\nu\eta}
    \\
    + H_4 \; k^\mu k^\nu k^\eta
    + H_5 \; k^2 \{p^\mu g^{\nu\eta}\}
    + H_6 \; k^2 \{k^\mu g^{\nu\eta}\}
    \Big)
\end{empheq}

\begin{flalign}
  H_1 = & \; R_0 + \frac{2 (11 + 12 \eps + 4 \eps^2) T_2}{(1+\eps)(2+\eps)}
  \\
  H_2 = & \; \frac{2 T_2}{(1+\eps)(1+2\eps)}
  \\
  H_3 = & \; \frac{T_2}{2+\eps}
  \\
  H_4 = & \; T_2
  \\
  H_5 = & \; - \frac{T_2}{2 (1+\eps)(1+2\eps)}
  \\
  H_6 = & \; - \frac{T_2}{4 + 2\eps} &&
\end{flalign}

\begin{flalign}
  H_1^{NPV} = & \frac{11/3 - 2 I_0 - \ln(1-x)}{\eir} - \frac{85}{9}
  \\
  H_2^{NPV} = & - \frac{1}{3 \eir} + \frac{11}{9}
  \\
  H_3^{NPV} = & - \frac{1}{6 \eir} + \frac{4}{9}
  \\
  H_4^{NPV} = & - \frac{1}{3 \eir} + \frac{13}{18} &&
\end{flalign}
\begin{flalign}
  H_5^{NPV} = & \frac{1}{12 \euv} - \frac{11}{36}
  \\
  H_6^{NPV} = & \frac{1}{12 \euv} - \frac{2}{9} &&
\end{flalign}

\subsubsection{Three-Point Axial Integrals}

\begin{empheq}[box=\fbox]{multline}
  J_{A}^{\mu\nu}(0,k,p;0)
  = \frac{\Qv(k)}{\abs{k^2}} \frac{1}{\qn} \Big(
    S_4 \; p^\mu p^\nu
  + S_5 \; \{pk\}^{\mu\nu}
  + S_6 \; k^\mu k^\nu
  \\
  + \frac{k^2}{2 \sdot{p}{n}} S_7 \; \{pn\}^{\mu\nu}
  + \frac{k^2}{2 \sdot{k}{n}} S_8 \; \{kn\}^{\mu\nu}
  + \left(\frac{k^2}{2 \sdot{k}{n}}\right)^2 S_9 \; n^\mu n^\nu
  + k^2 S_{10} \; g^{\mu\nu}
  \Big)
\end{empheq}
\begin{flalign}
  S_4 = & -\frac{(1+\eps) x^2}{2 (1+2\eps) (1-x)} (P_0+S_0) + \frac{-2+4x-x^2+\eps(-4+8x-3x^2)}{2 (1+2\eps) (1-x)} R_0 - (2-x) T_0
  \\
  S_5 = & \frac{(1+\eps) x}{2 (1+2\eps) (1-x)} (P_0-R_0+S_0) + T_0
  \\
  S_6 = & \frac{(1+\eps)}{2 (1+2\eps) (1-x)} (-P_0+R_0-S_0) - T_0
  \\
  S_7 = & \frac{(1+\eps x)}{2 (1+2\eps) (1-x)} (-P_0+R_0-S_0) - T_0
  \\
  S_8 = & \frac{(1+\eps) x}{2 (1+2 \eps) (1-x)} (P_0 - R_0 + S_0) + T_0
  \\
  S_9 = & \frac{2-4x+x^2+\eps(4-8x+3x^2)) P_0}{2(1+2\eps) (1-x)} + \frac{(1+\eps) x^2}{2(1+2\eps)(1-x)} (R_0 - S_0) - (2-x) T_0
  \\
  S_{10} = & \frac{P_0+S_0 - R_0}{4(1+2\eps)} &&
\end{flalign}

\begin{flalign}
  S_4^{NPV} = & \frac{-2 (1-x) I_0 + x^2 \ln{x}/(1-x) - (1-x) \ln(1-x) - x + 2}{\eir} \nonumber \\
              & - 4 (1-x) I_1 + 2 (1-x) I_0 \ln(1-x) + \frac{x^2 \Li(1-x)}{1-x} \nonumber \\
              & + \frac{(1-x) \ln^2(1-x)}{2} - \frac{x^2 \ln{x}}{1-x} + 2 x - 4
  \\
  S_5^{NPV} = & \frac{1}{(1-x)} \left( \frac{-x \ln{x} + x - 1}{\eir} - x \Li(1-x) + x \ln{x} - 2 x + 2 \right) &&
\end{flalign}
\begin{flalign}
  S_6^{NPV} = & \frac{1}{(1-x)} \left( \frac{\ln{x} - x + 1}{\eir} + \Li(1-x) - \ln{x} + 2 x - 2 \right)
  \\
  S_7^{NPV} = & \frac{1}{(1-x)} \left( \frac{\ln{x} - x + 1}{\euv} + \Li(1-x) - (2 - x) \ln{x} + 2 x - 2 \right)
  \\
  S_8^{NPV} = & \frac{1}{(1-x)} \left( \frac{-x \ln{x} + x - 1}{\euv} - x \; \Li(1-x) + x \ln{x} - 2 x + 2 \right)
  \\
  S_9^{NPV} = & \frac{-(1-x) I_0 - (1-2x) \ln{x}/(1-x) - x + 2}{\euv} \nonumber \\
              & - (1-x) I_1 + (1-x) I_0 \ln{x} + \frac{x^2 \Li(1-x)}{1-x} \nonumber \\
              & + \frac{(1-x) \ln^2{x}}{2} - \frac{x^2 \ln{x}}{1-x} + (1-x)\Li(1) + 2 x - 4)
  \\
  S_{10}^{NPV} = & \frac{1}{2} \left( -\frac{\ln{x}}{\euv} - \Li(1-x) + 2 \ln{x} \right)
\end{flalign}

\begin{empheq}[box=\fbox]{multline}
  J_{A}^{\mu\nu}(0,k,p;k)
  = \frac{\Qv(k)}{\abs{k^2}} \frac{1}{\qn} \Big(
    U_4 \; p^\mu p^\nu
  + U_5 \; \{pk\}^{\mu\nu}
  + U_6 \; k^\mu k^\nu
  \\
  + \frac{k^2}{2 \sdot{p}{n}} U_7 \; \{pn\}^{\mu\nu}
  + \frac{k^2}{2 \sdot{k}{n}} U_8 \; \{kn\}^{\mu\nu}
  + \left(\frac{k^2}{2 \sdot{k}{n}}\right)^2 U_9 \; n^\mu n^\nu
  + k^2 U_{10} \; g^{\mu\nu}
  \Big)
\end{empheq}
\begin{flalign}
  U_4 = & \frac{(1+\eps)x^2}{2+4\eps} (-P_0 + R_0 - U_0) - R_0 + (-2+x) T_0
  \\
  U_5 = & \frac{\eps x}{2 + 4 \eps} (- P_0 + R_0 - U_0) + T_0
  \\
  U_6 = & \frac{1 + 3\eps}{2 + 4\eps} P_0 + \frac{1 +\eps}{2 + 4\eps} (R_0 - U_0) - T_0
  \\
  U_7 = & \frac{-1 + x +\eps x}{2 + 4\eps} (P_0 - R_0 + U_0) - T_0
  \\
  U_8 = & \frac{(1 +\eps) x}{2 + 4\eps} (P_0 - R_0 + U_0) - P_0 + T_0
  \\
  U_9 = & \frac{(1 +\eps) x^2}{2 + 4\eps} (P_0 + R_0 - U_0) + P_0 + (-2 + x) T_0
  \\
  U_{10} = & \frac{1 - x}{4 + 8\eps} (P_0 - R_0 + U_0)
  &&
\end{flalign}
\begin{flalign}
  U_4^{NPV} = & \frac{-2 I_0 + x^2 \ln{x} - (1 + x^2) \ln(1-x) - x + 2}{\eir} - 4 I_1 + 2 I_0 \ln(1-x) \nonumber \\
              & - x^2 \Li(1-x) + \frac{(1-x^2) \ln^2(1-x)}{2} - x^2 \ln{x} + x^2 \ln(1-x) \nonumber \\
              & + 2 x^2 \Li(1) + 2 x - 4
  \\
  U_5^{NPV} = & - \frac{1}{\eir} + x \ln{x} - x \ln(1-x) + 2
  \\
  U_6^{NPV} = & \frac{-I_0 - \ln(1-x) + 1}{\eir} - I_1 + I_0 \ln{x} - \Li(1-x) + \frac{\ln^2{x}}{2} \nonumber \\
              & - \frac{\ln^2(1-x)}{2} - \ln{x} + \ln(1-x) + 3 \Li(1) - 2
  \\
  U_7^{NPV} = & \frac{(1-x) \ln{x} - (1-x) \ln(1-x) + 1}{\euv} - (1-x) \Li(1-x) \nonumber \\
              & - \frac{(1-x) \ln^2(1-x)}{2} - (2-x) \ln{x} + (2-x) \ln(1-x) \nonumber \\
              & + 2 (1-x)\Li(1) - 2
  \\
  U_8^{NPV} = & \frac{I_0 + (1-x) \ln{x} + x \ln(1-x) - 1}{\euv} + I_1 - I_0 \ln{x} + x \Li(1-x) \nonumber \\
              & - \frac{\ln^2{x}}{2} + \frac{x \ln^2(1-x)}{2} + x \ln{x} - x \ln(1-x) - (1+2x)\Li(1) + 2
  \\
  U_9^{NPV} = & \frac{-I_0 - (1-x^2) \ln{x} - x^2 \ln(1-x) - x + 2}{\euv} - I_1 + I_0 \ln{x} \nonumber \\
              & - x^2 \Li(1-x) + \frac{\ln^2{x}}{2} - \frac{x^2 \ln^2(1-x)}{2} - x^2 \ln{x} \nonumber \\
              & + x^2 \ln(1-x) + (1+2x^2) \Li(1) + 2x - 4
  \\
  U_{10}^{NPV} = & \frac{1-x}{2} \bigg( \frac{-\ln{x} + \ln(1-x)}{\euv} + \Li(1-x) + \frac{\ln^2(1-x)}{2} + 2 \ln{x} \nonumber \\
                 & - 2 \ln(1-x) - 2 \Li(1) \bigg)
  &&
\end{flalign}

\begin{empheq}[box=\fbox]{multline}
  J_{A}^{\mu\nu}(0,k,p;p)
  = \frac{\Qv(k)}{\abs{k^2}} \frac{1}{\qn} \Big(
    W_4 \; p^\mu p^\nu
  + W_5 \; \{pk\}^{\mu\nu}
  + W_6 \; k^\mu k^\nu
  \\
  + \frac{k^2}{2 \sdot{p}{n}} W_7 \; \{pn\}^{\mu\nu}
  + \frac{k^2}{2 \sdot{k}{n}} W_8 \; \{kn\}^{\mu\nu}
  \\
  + \left(\frac{k^2}{2 \sdot{k}{n}}\right)^2 W_9 \; n^\mu n^\nu
  + k^2 W_{10} \; g^{\mu\nu}
  \Big)
\end{empheq}

\begin{align}
  W_4 = & (-2 + x) R_0 + x W_0 + \frac{(1 + \eps) x^2}{1 + 2 \eps} \left( -2 E_0 - \frac{x}{1-x} E_1 - \frac{x(4-x)}{2(1-x)} E_2 - \frac{x^2}{2(1-x)} E_3 \right)
  \\    & + \frac{(2-x) (-2 + 2 x + x^2 + \eps (-4 + 4 x + x^2)}{2 (1 + 2 \eps) (1-x)} T_0
  \\
  W_5 = & \frac{1}{2} R_0 + \frac{(3 + 2 \eps) x}{2 + 4 \eps} E_0 + \frac{(1 + \eps) x^2}{(1 + 2 \eps) (1-x)} \left( E_1 + \frac{4-x}{2} E_2 + \frac{x}{2} E_3 \right)
  \\    & + \frac{2 + 4 \eps - 4 x - 6 \eps x + x^2 + \eps x^2}{2 (1+2\eps)(1-x)} T_0
  \\
  W_6 = & - \frac{1}{1 + 2 \eps} E_0 + \frac{(1 + \eps) x}{(1 + 2 \eps) (1-x)} \left( - E_1 + \frac{4-x}{2} E_2 - \frac{x}{2} E_3 \right) + \frac{x - \eps (2 - 3 x)}{2 (1 + 2 \eps) (1-x)} T_0
  \\
  W_7 = & \frac{(3 + 2 \eps) (-2 + x)}{2 + 4 \eps} E_0 + \frac{(3 + 2 \eps - x) x}{2 (1 + 2 \eps) (-1 + x)} E_1 + \frac{x (6 - 4 x + x^2 + \eps (4 - 2 x + x^2)}{2 (1 + 2 \eps) (-1 + x)} E_2
  \\    & + \frac{x^2 (1 + \eps x)}{2 (1 + 2 \eps) (-1 + x)} E_3 - \frac{1}{2} R_0 + \frac{2 - 2 x + x^2 + \eps x^2}{2 + 4 \eps - 2 x - 4 \eps x} T_0
  \\
  W_8 = & \frac{x}{1 + 2 \eps} E_0 + \frac{x + 2 \eps x + x^2}{2 + 4 \eps - 2 x - 4 \eps x} E_1 - \frac{x (2 + x + \eps (4 - 2 x + x^2)}{2 (1 + 2 \eps) (-1 + x)} E_2
  \\    & - \frac{(1 + \eps) x^3}{2 (1 + 2 \eps) (-1 + x)} E_3 + \frac{x (1 + \eps x)}{2 (1 + 2 \eps) (-1 + x)} T_0
  \\
  W_9 = & - \frac{x^2}{1 + 2 \eps} E_0 + \frac{(1 - \eps (-2 + x)) x^2}{(1 + 2 \eps) (-1 + x)} E_1 - \frac{x^2 (-2 - 2 x + x^2 + \eps (-4 + x^2)}{2 (1 + 2 \eps) (-1 + x)} E_2
  \\    & - \frac{x^2 (2 - 4 x + x^2 + \eps (4 - 8 x + 3 x^2)}{2 (1 + 2 \eps) (-1 + x)} E_3 + \frac{(1 + \eps) (-2 + x) x^2}{2 (1 + 2 \eps) (-1 + x)} T_0
  \\
  W_{10} = & \frac{1 - x}{1 + 2 \eps} E_0 + \frac{x}{2 + 4 \eps} E_1 - \frac{(-4 + x) x}{4 + 8 \eps} E_2 + \frac{x^2}{4 + 8 \eps} E_3 + \frac{-2 + x}{4 + 8 \eps} T_0
\end{align}

  \printbibliography

\end{document}